\documentclass{aa}
\usepackage{graphicx}
\usepackage{txfonts}
\usepackage{amsmath, amssymb, latexsym}
\ifpdf
   \DeclareGraphicsExtensions{.pdf,.jpeg,.png}
\else
   \usepackage[dvips]{graphicx}
\fi

\usepackage{morefloats}
\usepackage{xcolor}
\usepackage{setspace}
\usepackage{wrapfig}
\usepackage{bm}
\usepackage{fancybox} 
\usepackage{ifthen} 
\usepackage{hyperref}
\usepackage{graphicx}
\usepackage{caption}
\usepackage{subcaption}

\newif\ifrefer
\refertrue

\begin{document}

\title{Metal distribution in the ICM - a comprehensive numerical study of twelve galaxy clusters }

\titlerunning{The distribution of metals in the ICM}
\authorrunning{Harald H\"oller et.al.}

\author{Harald H\"oller\inst{1},  
Josef St\"ockl\inst{1},
Andrew Benson\inst{2},
Markus Haider\inst{1}, 
Dominik Steinhauser\inst{1}, 
Lorenzo Lovisari\inst{3},
Florian Pranger\inst{1}
}

\institute{
Institute of Astro- and Particle Physics, University of Innsbruck, \email{\href{mailto:harald.hoeller@uibk.ac.at}{harald.hoeller@uibk.ac.at}}
\and
Observatories of the Carnegie Institution for Science, Pasadena, CA 
\and
Argelander-Institut f\"ur Astronomie, Bonn
}

  \abstract{
   We present a simulation setup for studying the dynamical and chemical evolution of the intracluster medium (ICM) 
   and analyze a sample of 12 galaxy clusters that are diverse both
   kinetically (pre-merger, merging, virialized) and in total mass (M$_{\text{vir}} = 1.17\times 
   10^{14} - 1.06 \times  10^{15} \mathrm{M}_{\odot}$). We analyzed the metal mass fraction in the ICM as 
   a function of redshift and discuss radial trends as well as projected 2D metallicity maps. 
   The setup combines high mass resolution N-body simulations with the semi-analytical galaxy formation model \textsc{Galacticus} 
   for consistent treatment of the subgrid physics (such as galactic winds and ram-pressure stripping) 
   in the cosmological hydrodynamical simulations. The interface between \textsc{Galacticus} 
   and the hydro simulation of the ICM with \textsc{FLASH} is discussed with respect to observations of star formation rate histories, 
   radial star formation trends in galaxy clusters, and the metallicity at different redshifts. As a test for the robustness of 
   the wind model, we compare three prescriptions from different approaches. For the wind model directly taken from 
   \textsc{Galacticus}, we find mean ICM metallicities between $0.2-0.8\mathrm{Z}_{\odot}$ within the inner 1Mpc at $\mathrm{z}=0$. 
   The main contribution to the metal mass fraction comes from galactic winds. The outflows are efficiently mixed in the ICM, 
   leading to a steady homogenization of metallicities until ram-pressure stripping becomes effective at low redshifts. We find a very 
   peculiar and yet common drop in metal mass fractions within the inner $\sim200$kpc of the cool cores, which is 
   due to a combination of wind suppression by outer pressure within our model 
   and a lack of mixing after the formation of these dense regions. 
   } 
  
  \keywords{Hydrodynamics -- Methods: numerical -- Galaxy: abundances -- Galaxies: interactions -- Galaxies: evolution}

   \maketitle

\section{Introduction}\label{Introduction}

The study of galaxy clusters has benefited immensely from multiwavelength studies over the 
past few decades. First, photometric studies of rich clusters of galaxies have revealed an increase 
of blue spiral galaxies with higher redshifts~\citep{Oemler1974},
which indicates that there is an evolutionary 
sequence in these local clusters. A few years later, the influence of the density of the 
environment in which galaxies reside was established as a correlation between galaxy
morphology and galaxy number density~\citep{Dressler1980}. While these correlations 
are verifiable for relaxed galaxy clusters, kinetically complex systems pose 
a challenge for observational studies \citep[see e.g.][]{Pranger2013},
which leaves questions
on the kinetic state of these potentially merging or post-merger clusters open. Our best 
tracer for the kinetic state of these objects are radio and X-ray observations of 
the intracluster medium (ICM) as expounded by \citet{Sarazin1986} and studied, for example, by~\citet{Markevitch2000} and~\citet{Cassano2010}.
In addition to the X-ray surface brightness \citep[centroid-shift method, see, e.g.,][]{Poole2006,Maughan2008}
and the presence of radio halos, the distribution of metals seems to be a 
valid tracer for kinetic peculiarities within galaxy clusters, which was studied 
observationally \citep[see, e.g.,][]{Simionuescu2009} and in simulations \citep[e.g., ][]{Kapferer2009a}.

A number of different processes have been suggested to be responsible for the 
enrichment of the ICM with metals. Ram-pressure (RPS) stripping~\citep{Gunn1972a} is an 
interaction of infalling galaxies with the dense gas of cluster centers in which 
galactic material is removed from the gravitational potential of the galaxies and
deposited into the ICM. Shortly after the discovery of iron lines in the ICM~\citep{Mitchell1976}, 
galactic winds~\citep{DeYoung1978} were suggested as a process that might   be able to efficiently 
transport metals into the intracluster medium. Mass outflow from supersonic jets as found 
in radio galaxies and quasars was studied numerically by \citet{DeYoung1986}, and observations
\citep[e.g., ][]{Willis1985} showed significant spectral Fe features associated with AGNs.
While these mechanisms are of galactic origin,~\citet{Gerhard2002} found
an isolated compact HII region within the Virgo cluster, suggesting that there  
might be a fractional contribution from intracluster light to the metals found 
in the ICM. 

In this paper we study cosmological simulations of galaxy clusters with respect to 
the dynamics and the chemical evolution of the ICM. We consider galactic winds and 
ram-pressure stripping as enrichment processes. 
The paper is structured as follows: In Sect. 
~\ref{SimulationSetup} we present our unique simulation setup, for which we combined 
a sophisticated subgrid model by including the well-tested semi-analytical
model \textsc{Galacticus} into a grid based treatment of the ICM with the \textsc{FLASH4} 
hydro code. 
The details of the interface between these two codes are described in Sect.~\ref{Sec:Feedback}, where 
we describe three different wind models and the ram-pressure stripping algorithm. 
We present the actual sample of 12 galaxy clusters we analyzed in Sect.~\ref{GalaxyClusterSample}. 
Here we also motivate the choice of objects that were included for the purpose of this work. 
In Sect.~\ref{secResults} we display and describe the results of the 12 cluster simulations 
as well as the comparison between the three galactic wind models depicted in Sect.~\ref{Subsec:Winds}. 
Section~\ref{Sec:CompObs} is dedicated to a comparison of our results with observations and explores the potential of the simulation setup presented here for future studies in collaboration 
with X-ray surveys. 
We summarize the results and give a brief overview of the development roadmap in Sect.~\ref{Discussion}. 

\section{Simulation setup}\label{SimulationSetup}

Because we wish to explore feedback processes that couple galaxies and the ICM on cosmological scales, 
we constructed a set of consecutive simulations that incorporate 
cosmological structure formation, a semi-analytical galaxy formation model, and 
a hydrodynamical treatment of the ICM. In the following sections we present this unique setup 
in some detail. We note that all codes are publicly available, are well-tested
within the communities, and are continuously developed and refined.  

In the discussion we do not review all the implementations in detail, 
but focus on software and parameter choices as well as certain aspects of the software packages that are of crucial importance. For the specifics of the publicly available codes we refer 
to the appropriate code papers. A detailed description of the setup we have developed and assembled 
can be found in~\citet{HoellerThesis2013}.

\subsection{IC and N-body simulation of structure formation}\label{secNbody}

The sample of galaxy clusters depicted in Sect.~\ref{GalaxyClusterSample} is generated by adopting 
cosmological parameter values taken from WMAP7 data~\citep{Larson2011} with the initial 
condition generation codes \texttt{NGenIC} (which is part of the \texttt{GADGET-2} 
package, see~\citet{Springel2005-2}) and \texttt{MUSIC}~\citep{Hahn2011}.
The values used are $\Omega_\text{c}=0.266$, $\Omega_{\Lambda}=0.733$, $\Omega_\text{b}=0.045$, $H_0=70.3$, 
$\sigma_8=0.809$ and $n_\text{s}=0.966$ in combination with the the matter 
transfer function from~\citet{Eisenstein1998}.

For our sample we generated a large number of unconstrained cosmological volumes between 48Mpc/h and 
102Mpc/h in size, each with $512^3$ dark matter particles. In these unconstrained simulations 
we identified galaxy clusters using a halo finder (see Sect.~\ref{HaloFinderMergerTree}) and cut out 
a smaller volume of 20Mpc/h with the heaviest halo in the center to facilitate handling of the data. 
The sample presented in Sect.~\ref{GalaxyClusterSample} is a balanced mixture of 
different masses and kinetic states of galaxy clusters. 
In addition to the mass resolutions, which are just given by the relation between simulation volume and 
particle number, one major input parameter in the N-body simulations executed with 
\texttt{GADGET-2} is the softening length $\epsilon$. In particular, test runs have shown that the  
number of stable, lower mass halos found is very sensitive to this parameter. 
We refer to~\cite{Rodinov2005}, who 
calculated the softening needed because of the peak density in the dark matter 
simulation. In our case these peak densities reach up to $\sim 1.5 \times 10^{23}$g/cm$^3$ , yielding a 
(physical) value of $\epsilon \leq 0.5$kpc/h. With this the values
of the number 
of halos (and galaxies) in our simulation converged. 

We output the 200 \texttt{GADGET-2} snapshots equidistantly spaced in time that we 
use as input for the remainder of our analysis; this is already more than necessary for convergence of the 
galaxy properties, as shown in~\cite{Benson2012Convergence}.

\subsection{Halo finder and merger tree}\label{HaloFinderMergerTree}

\begin{figure}
\centering
\includegraphics[scale=0.4]{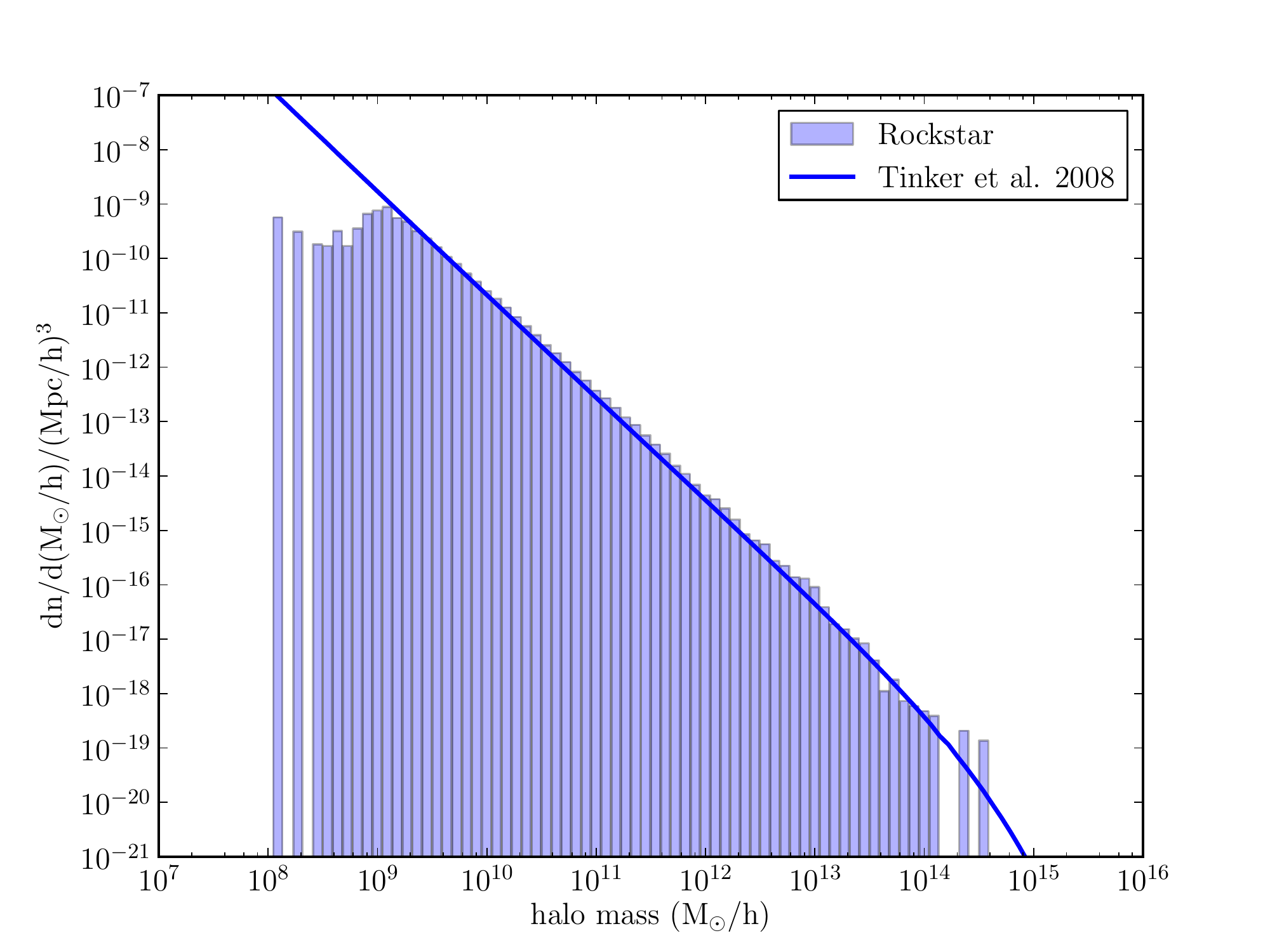} \\
\caption{Halo mass function at $\mathrm{z}=0$ from a simulation using $512^3$ particles, each of 
mass $6.1 \times 10^7M_{\odot}$ (histogram), compared with the mass function of~\cite{Tinker2008} (line).} \label{figHaloMasFunc}
\end{figure}

The identification of bound objects in the structure formation simulation is the basis 
for modeling galaxy properties in our semi-analytical galaxy formation model. 
The merger history of halos and subhalos  determines the evolution of the tracked objects. 
To identify halos we used the software \texttt{Rockstar}~\citep{Behroozi2011}, which has proven 
very reliable in the comparison project of~\cite{Knebe2011Halos}.  
In Fig.~\ref{figHaloMasFunc} we show the halo mass function of our highest resolution 
run, which yields a completeness down to $10^9M_\odot$. 
Merger trees were generated from \texttt{Rockstar} output using the algorithm 
\texttt{Consistent Trees} ~\citep{Behroozi2013}, in which
we enforced a mass-resolution lower limit of $\geq$ 1000 particles per halo. 
The merger trees obtained from \texttt{Consistent Trees} were converted 
into the appropriate input format for \textsc{Galacticus} using the publicly available 
converter \texttt{rockstar2galacticus}\footnote{
\tiny{\texttt{https://bitbucket.org/markushaider/rockstar2galacticus}}}. 

\subsection{Semi-analytical galaxy formation model} \label{secSAM}

As described in \cite{Benson2012}, {\sc Galacticus} functions by first constructing a merger tree and then solving the set of differential equations that describe how galaxies form in that merger tree. 

We constructed merger trees by simply reading their definition from the output of \texttt{rockstar2galacticus}. Each merger tree consists of a set of dark matter halos, some of which will be isolated halos, others of which are subhalos hosted within isolated halos. The merger-tree definition output by \texttt{rockstar2galacticus} specifies for each halo the descendent (i.e., the halo it becomes at the subsequent snapshot), its host (i.e., the larger halo within which it exists if it is a subhalo), and a variety of other properties, including the mass, position, velocity, dimensionless spin parameter, and half-mass radius. {\sc Galacticus} uses this information to construct the merger tree and to initialize the masses, positions, velocities, angular momenta, and density profiles of all halos in the tree. After it is constructed, galaxy formation proceeds in this tree following {\sc Galacticus'} usual methodology.

Merger trees extracted from N-body simulations require some additional considerations compared with the description of merger-tree evolution given in \cite{Benson2012}:
\begin{itemize}
 \item Subhalos are tracked in the simulation until they fall below the detected threshold of the group-finding algorithm. This provides some information about when galaxies in subhalos should be considered to merge with another galaxy. {\tt Galacticus} utilizes this information to determine both the time at which mergers will take place and the halo with which the merger occurs (including the possibility of subhalo-subhalo mergers). Since it is possible that merging will not occur immediately when the subhalo is lost from the simulation, we allowed for a delay between the time at which the subhalo was last seen and the time of merging. This delay was computed using the algorithm of \cite{boylan-kolchin_dynamical_2008}, with the orbital parameters of the subhalo taken at the time at which it was last detected in the simulation.
 \item In merger trees derived from N-body simulations (as opposed to those constructed using extended Press-Schechter algorithms, for example),
subhalos can sometimes escape their parent halos and again become isolated systems. {\sc Galacticus} detects such events in merger trees and, 
at the appropriate time, detaches the subhalo from its host, promoting it to be again an isolated halo. This means that the halo can again 
begin to accrete new material from the intergalactic medium (if it continues to grow in mass).
 \item Another possibility in N-body simulation merger trees is for a subhalo to jump between branches of the tree between subsequent 
snapshots (i.e., the subhalo moves between host halos that are not part of the same tree branch). {\sc Galacticus} also detects this type of 
event in merger trees and moves the subhalo between hosts at the appropriate time. This will affect to which halo any outflowing material 
from the subhalo is deposited, for example.
\end{itemize}


\begin{figure}
\centering
\includegraphics[scale=0.4]{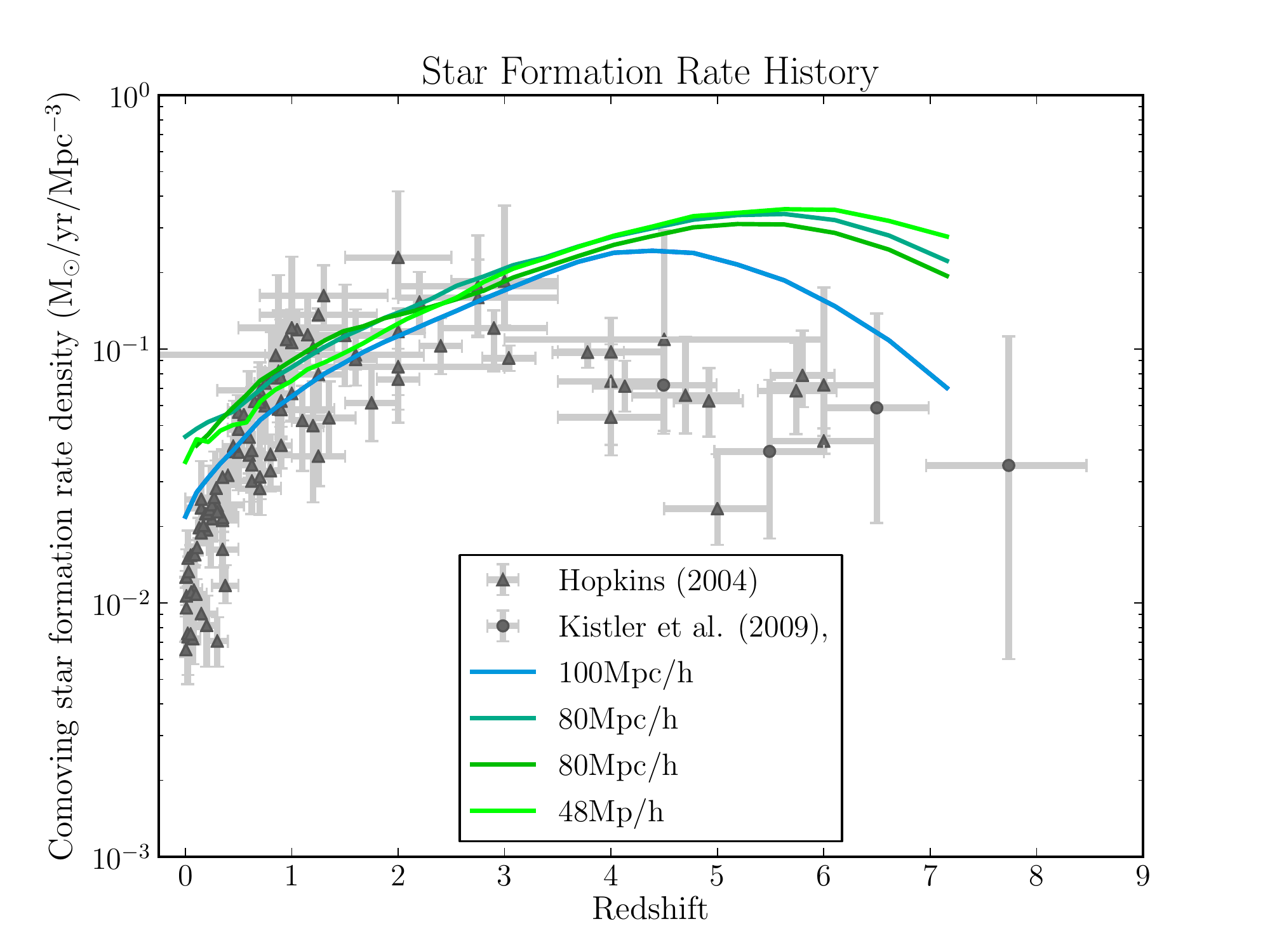} \\
\caption{Star formation histories for different simulation volumes. The larger the volumes, the better the 
agreement with the observational data from~\citep{Hopkins2004} and~\citep{Kistler2009}. }  \label{galacticusSFR}
\end{figure}


\subsection{Hydrodynamical simulation}

We used \textsc{FLASH4} for the hydrodynamical treatment of the ICM in cosmological 
simulations. The Riemann solver was the unsplit HLLD implemented in \textsc{FLASH4} together 
with a vanLeer slope limiter and Balsara prolongation for the correct treatment of the 
refinement boundaries.
The 20Mpc/h volume was discretized into three nested grids of $256^3$ cells, each of which results in a
resolution of $\sim$20kpc/h for the inner 5Mpc/h in whose center the cluster is located at z=0.

Because we adopted post-processed data from the \texttt{GADGET-2} dark matter simulation for 
the semi-analytical model and the feedback processes into the subsequent hydrodynamical simulation, we decided not to use the particle unit in \textsc{FLASH4,} but instead generated dark matter 
density fields from mapping the \texttt{GADGET-2} snapshots directly onto the grid. 
We mapped the dark matter particles with the Delaunay tessellation field interpolation 
code \texttt{DTFE}~\citep{Schaap2000a}. The strength of this approach compared with 
cloud-in-cell-like algorithms is that the resolution of 
the grid data stays ahead of the adaptive mass resolution that
is ensured by using a particle-based dark matter (DM) simulation like ours 
with \texttt{GADGET-2}.
 
In particular, this avoids small-scale variations in the distribution of the DM particles compared with the 
\texttt{GADGET-2} run that are caused by the grid-based restrictions on the gravitational potential and 
an inconsistency of the \textsc{FLASH} CiC formulation at refinement boundaries.
Another advantage gained by employing these read-in density
fields is the performance improvement in \textsc{FLASH4} without the particle unit.  

The adjustment of the metallicities $X_{i}$ in the hydrodynamical simulation is 
implemented by making use of a special variable type (\texttt{SPECIES}) that becomes
advected with the fluxes, but with the additional implication that the sum over all 
metal mass-fractions $X_i$ equals 1. 
We initialized these \texttt{SPECIES} with the 
primordial abundances of H and He and did not model any metallicity enrichment other than 
galactic feedback, meaning that we did not model intracluster supernovae (SNe) at the current stage,
\begin{equation}
\qquad X_{i} = \frac{\rho_i}{\sum \rho_i} \, , \qquad  \sum X_i = 1
.\end{equation}
We did not consistently compute the ratio of H and He, but reduced its mass fraction by equal 
parts when enriching a cell because we are not interested in the details of H/He abundances.
For the disambiguation of enrichment that is caused by galactic winds and
ram-pressure stripping (Sect. ~\ref{Sec:Feedback}), we introduced distinct fields
to be able to trace the evolution of either separately. At this moment we are unable 
to track individual elements consistently within our subgrid model {\sc Galacticus}, hence
metal mass fractions always refer to total metallicities in this work except when explicitly 
noted otherwise. The details of implementing the interfaces between SAM and 
hydrodynamical simulation can be found in~\cite{HoellerThesis2013}. For the algorithms 
that govern the outflowed metal masses from nodes within {\sc Galacticus}, we refer 
to~\cite{benson2010} and~\cite{Benson2010a}.

Cooling is one physical process that is directly affected by the enrichment of the intracluster medium
with heavier elements from the subgrid physics. We include cooling tables
from~\cite{Sutherland1993a}, which are given for a number of temperatures and
metallicities between which we linearly interpolated in log-space. The tables include
cooling functions over a range of $10^4 - 10^{8.5}$K, but we truncated the cooling below
T=$10^6$K  to avoid artificially dense and cool regions during the formation process 
of the galaxy cluster. This lower cooling limit is not even remotely reached when the actual
merger-shock-heated cores have formed in the simulations. Moreover,
we detected problematic behavior only 
in a run with the highly effective wind model \textbf{B\textup{,}} which we describe in Sect.~\ref{Subsec:Winds}. 
Observations of cooling flows~\citep{Fabian1994} indicate that their 
occurrence is not only related to the kinetic state of the
galaxy cluster and additional heating processes such as AGN feedback~\citep{Fabian2012},
but also to the ICM metallicity~\citep{Allen1998}. In Sect.~\ref{secResults} we discuss the 
finding of cool cores in our simulations with respect to the ICM metallicities.

\section{Feedback models} \label{Sec:Feedback}

The main strength of our setup is the ability to model galactic feedback processes in a 
cosmological simulation for a large number of objects (typically $\sim 10^4$). We treat the input 
of mass, momentum and energy from galactic winds and ram-pressure stripping into the ICM 
by adopting numerical and analytical prescriptions. In the following sections we discuss the 
details of these feedback models. 

\subsection{Galactic winds} \label{Subsec:Winds}

The details of mass, momentum, and energy outflow of galactic winds is observationally poorly constrained because of 
the extremely low densities and high temperatures of these flows. However, there is consensus that starburst winds, 
driven by kinetic energy and momentum from stellar winds and supernovae, play a major role not only 
in the evolution of galaxies themselves - as shown in numerical studies such as~\cite{Cole2002} 
or~\cite{SpringelHernquist2003} - but especially for the enrichment of the gas in the Universe 
with heavier elements~\citep{Adelberger2003,Steidel2010}. 
Multiwavelength observations in the optical \citep[e.g., ][]{Heckman1990}, UV~\citep{Meurer1995,Delgado1998} and
X-ray bands \citep{Su2010,Strickland2009} have indicated that
there is a strong correlation
between the strengths of starbursts and kinematically confirmed galactic outflows or 
diffuse ionized gas in the close vicinity of these galaxies. 

To model galactic wind feedback from the \textsc{Galacticus} output into the intracluster 
medium of the hydrodynamical simulation, we adapted two models derived from numerical studies. For the 
first model (\textbf{model A}) we refer to~\cite{Leitherer1999} and~\cite{Veilleux2005}, who presented a very 
simple wind prescription derived from simulations with \texttt{Starburst99} . This model yields a
constant mass-loading factor for solar metallicity and constant outflow and mechanical 
luminosity integrated over 40 Myr scales,  
\begin{eqnarray} \label{LeithererModel}
\nonumber       \qquad \dot{M}_{\ast \textbf{A}} & = & 0.26 \left[ \text{SFR}/M_{\odot}yr^{-1} \right] \, M_{\odot}yr^{-1}, \\
\nonumber       \qquad \dot{E}_{\ast \textbf{A}} & = & 7 \times 10^{41} \left[ \text{SFR}/M_{\odot}yr^{-1} \right] \text{erg}s^{-1}, \\
                \qquad \dot{p}_{\ast \textbf{A}} & = & 5 \times 10^{33} \left[ \text{SFR}/M_{\odot}yr^{-1} \right] \text{dyne.}
\end{eqnarray}
The expected residing time within a grid cell of a galaxy with typical $\sim1000$km/s $\cong 10^{-6}$kpc/yr 
velocity in the innermost of our grids is $\geq$20 Myr. Therefore, for the vast majority of objects, 
not more than one cell is occupied during this 40 Myr integration time. Because the resolution of the spatial 
grid is not high enough to resolve galactic morphologies (and 
orientations), we added mass, momentum, and pressure from Eq. \eqref{LeithererModel} cell-wise. 
Implicitly, therefore, we assumed that the outflowed material mixes and equilibrates
with the surrounding ICM on the time scales mentioned above. 
This wind model does not provide a parametrization with the galaxies' masses. From analytical 
and numerical analysis we know that the gravitational binding energy of the host 
strongly influences the efficiency of the galactic outflow. We adopted the numerical 
results from~\cite{Recchi2013}, which imply an averaged wind efficiency as a function 
of the total baryonic content of the galaxy $M_{\text{b}}$,
\begin{equation} \label{windEffRecchi}
 \text{Eff}_{\textbf{A}} =  \left\{
\begin{array}{c l l}     
     & 1                                                & \text{for} \quad \log(M_{\text{b}}) < 7, \\
     & 3.8 - 0.41 \cdot \log(M_{\text{b}})              & \text{for} \quad 7 \leq  \log(M_{\text{b}}) \leq 9.4, \\
     & 0                                                & \text{for} \quad \log(M_{\text{b}}) > 9.4.
\end{array} \right.
\end{equation}
The mass-loading factor of constant 1 for objects of mass $\log(M_{\text{b}}) < 7 $ 
most likely underestimates the actual wind in this range, but the 
mass-resolution limit of our SAM is approached in this same regime. 
A second efficiency parameter describes the pressure contrast of the wind with 
respect to the outer confining pressure of the ICM. Currently, we only consider isotropic 
gas pressure for the wind suppression and no ram pressure from velocity differences 
between the infalling galaxies and the intracluster medium. When the ICM pressure exceeded the wind pressure that we 
obtain from Eq.~\eqref{LeithererModel}, the flux from the 
galaxies into the ICM was set to zero.  
We also adopted from~\cite{Recchi2013} the resulting metallicity of the galactic outflow 
with constant solar value $Z_{\odot}$ , for which we refer to~\cite{Lodders2003} with a 
value of $Z_{\odot} = 0.0149$.  

The second prescription implemented (\textbf{model B}) is taken from~\cite{Hopkins2012}, which is another 
systematic study on galactic winds making use of \texttt{Starburst99}. Their results  
show a sublinear relation between star formation rate and mass flux and a wind efficiency 
with steeper dependency at a higher level for less massive galaxies,  
\begin{equation} \label{HopkinsModel}
\qquad  \dot{M}_{\ast \textbf{B}} = 3.0 \left[ \text{SFR}/M_{\odot}yr^{-1} \right]^{0.7} \, M_{\odot}yr^{-1} .
\end{equation}
Because this wind model does not provide for energy and momentum fluxes, we calculated the internal energy of 
the outflowed material for a $10^7$K gas with $\dot{E}_{\ast \textbf{B}} = 10^7 k_B (\gamma-1)$ and 
a corresponding gas pressure of $\dot{p}_{\ast \textbf{B}} = \sqrt{2 \dot{M}_{\ast \textbf{B}} 
\dot{E}_{\ast \textbf{B}}}$, assuming an ideal gas with $\gamma = 5/3$: 
\begin{equation} \label{windEffHopkins}
\qquad  \text{Eff}_{\textbf{B}} = 32.2 - 2.6 \log(M_{\text{b}}). 
\end{equation}
In addition, we added the pressure contrast efficiency to this
prescription to take into 
account wind suppression in high-density regimes at the galaxy cluster centers.

Below we introduce a variation of this as \textbf{model B'}, where the wind efficiency
is set to be equivalent to model A:
\begin{equation} \label{windEffHopkinsModified}
\qquad  \text{Eff}_{\textbf{B'}} \equiv \text{Eff}_{\textbf{A}}.
\end{equation}

The galactic wind tracer of \textsc{Galacticus} taken directly from the semi-analytical 
model composes our \textbf{model C}. {\sc Galacticus} tracks gas that has been ejected from galaxies, 
but has yet to re-equilibrate and become part of the hydrostatic hot atmosphere surrounding each galaxy. 
We refer to this material as outflowed gas. The evolution of the mass of this component is given by
\begin{eqnarray}
\nonumber \qquad  \dot{M}_{\rm outflowed} & = & \beta_{\rm disk} \dot{M}_{\star,{\rm disk}} + \beta_{\rm spheroid} \dot{M}_{\star,{\rm spheroid}} - \\
&& - \delta M_{\rm outflowed} / \tau_{\rm dyn}.
 \label{eq:outflowedReservoir}
\end{eqnarray}
Here, the first two terms describe supernovae-driven outflows from the disk and spheroid components of galaxies. 
These outflow rates are proportional to the star formation rates, $\dot{M}_\star$, with the coefficient of proportionality given by
\begin{equation}
\qquad \beta = \left(V_{\rm outflow} / V \right)^{\alpha_{\rm outflow}},
\end{equation}
where $V$ is the circular velocity of the component at its scale radius, and we adopted $V_{\rm outflow}=250$km/s and 100km/s 
for disk and spheroid, and $\alpha_{\rm outflow}=3.5$ for both components. The final term in Eq. ~\eqref{eq:outflowedReservoir} 
describes  the re-incorporation of the outflowed material into the hydrostatic hot atmosphere. 
Here, $\tau_{\rm dyn}$ is the dynamical time of the halo, and $\delta = 5$.

Because we introduced source terms into the equations of mass, momentum, and energy conservation, 
our numerical scheme is no longer strictly conservative. Therefore
we limited the density, 
pressure, and internal energy contrast per time step to 1\% of the cell values for stability reasons. 
In addition to the numerical motivation, the recycling of matter via the 
subgrid model is not perfectly consistent because we did not model accretion of mass onto the galaxies in 
form of negative source terms. For studying ICM metal enrichment at the current 
grid resolution, however, the details of this interface dynamics are not relevant.

\subsection{Ram-pressure stripping} \label{Subsec:Ramp}

Ram-pressure stripping in galaxy clusters contributes not only to the surpression of 
galactic star formation rates (SFR) as a strong function of cluster-centric radius, but also 
deposits partially enriched material into the ICM along the infalling trajectories,
as shown in surveys by~\cite{Lewis2002} and numerical simulations by~\cite{Quilis2000}, for instance. 
Both effects have to be taken into account for our study of the metallicity distribution in galaxy clusters.  
With an SFR density interpretation of the morphology density relation~\citep{Gunn1972a} 
in galaxy clusters, the suppression of galactic winds will not
only be caused by external pressure, but also by an ever-decreasing intrinsic star formation rate. 
The contribution from these supernovae and stellar-wind-driven outflows will decrease with 
cluster-centric radius, while ram-pressure stripping is acting in dense environments and high
relative velocities, as we mainly find inside the virial radii of galaxy clusters.  

To model galactic disk ram-pressure stripping we applied the
criterion of Gunn\&Gott \citep{Gunn1972a} by comparing the disk-restoring force at a given radius 
with the ram pressure exerted on the galaxy moving through the ICM:
\begin{equation}
\qquad p_{\text{ram}} = \rho_{\text{ICM}} v_{\text{rel}}^2.
\end{equation}
The surface densities $\sigma_{\text{gas}}$ and $\sigma_{\text{stellar}}$ are taken 
directly from the \textsc{Galacticus} models. 
\begin{equation}
\qquad F_{\text{rest}}(r) = 2 \pi G \sigma_{\text{gas}}(r) \sigma_{\text{stellar}}(r)
.\end{equation}
Galactic disks in \textsc{Galacticus} are modeled as having exponential surface density profiles. 
Because the density decreases monotonically, we therefore simply iterated over radius and stripped all material outside the 
stripping radius $r_{\text{strip}}$ where $p_{\text{ram}} > F_{\text{rest}}(r)$. The material 
lost thus amounts to 
\begin{equation}
M_{\text{strip}} = \int\limits_{r_{\text{strip}}}^{\infty} r \frac{e^{(-r/r_{\text{disk}})}}{r_{\text{disk}}^2} \, dr 
 = \frac{(r_{\text{strip}}+r_{\text{disk}})}{r_{\text{disk}}} M_{\text{disk}} \, e^{(-r_{\text{strip}}/r_{\text{disk}})}. \label{Eq:strippedMass}
\end{equation}
Because ram-pressure stripping is not an instantaneous effect, we determined the actual fluxes 
into the ICM by defining a ram-pressure stripping time scale that we took 
from~\cite{Steinhauser2012}, who found stripping timescales between 100 and 250Myr.  
For the total amount of gas and metals stripped, and thus deposited into the ICM, 
the exact number of this time scale is not very relevant because the trajectories of infalling galaxies will result 
in much longer paths, along which they experience ram pressure. 

At this point of our setup development we did not model the effect that disk ram-pressure stripping has 
on the galaxies. However, we are working on an implementation where we take this process into account by directly feeding information about the ram-pressure force extracted from the hydrodynamical simulation into \textsc{Galacticus}. 
In \textsc{FLASH} we keep track of the lost material on a per-galaxy basis. This mass of lost material is allowed to decrease on the free-fall time scale of the halo, allowing for re-accretion of gas for objects that were stripped before \citep[see, e.g.,][]{BensonBower2011}. Our
approach furthermore assumes that galactic disks do not react to 
the outer stripping by radial pressure equilibration. In other words, the stripping 
radius is a sharp outer boundary and the disk can only grow beyond this radius via accretion. 
\begin{table*}
\centering
\begin{tabular}{l*{8}{l}r}\label{12clusters}
        &        & box                  &               & cutout center         & $M_{\text{vir,BCG}}$  & $R_{\text{vir,BCG}}$  & DM mass res           & r$_{200}$     & M$_{200}$             \\
cluster & IC gen & [Mpc/h]              & seed          & x-y-z [Mpc/h]         & [$M_{\odot}$]         & [Mpc/h]               & [$M_{\odot}$]         & [Mpc/h]       & [$M_{\odot}$]         \\
\hline
C01     & \texttt{MUSIC}  & 80           & 13671         & 38-47-66              & 5.44E14               & 2.39                  & 2.82E8                & 1.24          & 4.37E14               \\
C02     & \texttt{MUSIC}  & 80           & 30596         & 21-16-64              & 1.06E15               & 2.99                  & 2.82E8                & 1.48          & 7.54E14               \\
C03     & \texttt{MUSIC}  & 80           & 30596         & 22-16-62              & 2.09E14               & 1.53                  & 2.82E8                & 0.99          & 2.24E14               \\
C04     & \texttt{MUSIC}  & 48           & 1502          & 35-33-20              & 2.99E14               & 1.90                  & 6.1E7                 & 0.92          & 1.81E14               \\
C05     & \texttt{MUSIC}  & 80           & 7964          & 62-62-39              & 1.17E14               & 1.44                  & 2.82E8                & 0.73          & 8.7E13                \\
C06     & \texttt{MUSIC}  & 64           & 22109833      & 33-16-11              & 2.08E14               & 1.75                  & 9.7E7                 & 0.87          & 1.57E14               \\
C07     & \texttt{NGenIC} & 100          & 27036         & 89-49-70              & 1.45E14               & 1.71                  & 5.52E8                & 0.79          & 1.13E14               \\
C08     & \texttt{MUSIC}  & 80           & 17794         & 25-29-42              & 2.69E14               & 1.89                  & 2.82E8                & 0.95          & 1.99E14               \\
C09     & \texttt{MUSIC}  & 80           & 25922         & 25-48-52              & 5.29E14               & 2.37                  & 2.82E8                & 1.24          & 4.36E14               \\
C10     & \texttt{MUSIC}  & 80           & 32706         & 66-39-23              & 5.45E14               & 2.41                  & 2.82E8                & 1.23          & 4.26E14               \\
C11     & \texttt{MUSIC}  & 48           & 8736          & 19-14-24              & 2.11E14               & 1.79                  & 6.1E7                 & 0.90          & 1.69E14               \\
C12     & \texttt{NGenIC} & 100          & 20905         & 34-76-35              & 2.49E14               & 1.86                  & 5.52E8                & 0.98          & 2.15E14               \\
\end{tabular}
\caption{Cluster sample. For traceability we list the initial conditions generators (Col. 2), the box sizes (Col. 3), and the random seeds (Col. 4). With 
the cosmological parameters listed in Sect.~\ref{secNbody} it is possible to reproduce the N-body simulations. We specify the (approximate) 
position of the clusters on which we centered the 20Mpc/h cutout in Col. 5.  } \label{Table:clusterSample}
\end{table*}

The metallicity profiles of the disks were modeled using results taken from~\cite{Ferguson1998}, 
where the mean O/H ratio is given by 
\begin{equation}
\qquad \log (O/H) = -2.8 - 0.65  \log(R_{25}) .
\end{equation}
For the reference solar abundance ration of $O/H=0.429$ we refer to~\cite{Asplund2009}. 
It is probably a fairly crude assumption to model all disks with log-linear profiles, because observations have shown a wide variety of radial metallicity gradients~\citep[see, e.g.,][]{Shaver1983,Zaritsky1994}. Particularly in more recent works~\citep{Bresolin2012,Scarano2013} observations give remarkable evidence for an omnipresent break at the corotation 
radius, which suggests that a log-linear profile might be underestimating the amount of metals 
in the outer parts of galactic disks.

\section{Galaxy cluster sample}\label{GalaxyClusterSample}

Our sample of 12 clusters was selected in such a way as to permit an analysis across a range of masses and 
states of virialization. In Table~\ref{Table:clusterSample} we present the main parameters used for generating 
the initial conditions as well as the main characteristics 
of the clusters, such as r$_{200}$ and M$_{200}$. The differences in the virialized masses of the 
most massive halos, the brightest cluster galaxies (BCG) $M_{\text{vir,BCG}}$ 
and the M$_{200}$ values are first indicators 
of the degree of virialization of these objects. The difference in mass resolutions is solely caused by the 
differences in volumes of the initial N-body simulations. We used the same cosmological parameters 
and number of DM particles for all runs as described in Sect.~\ref{secNbody}. 

\begin{figure*}
\centering
\includegraphics[scale=0.6]{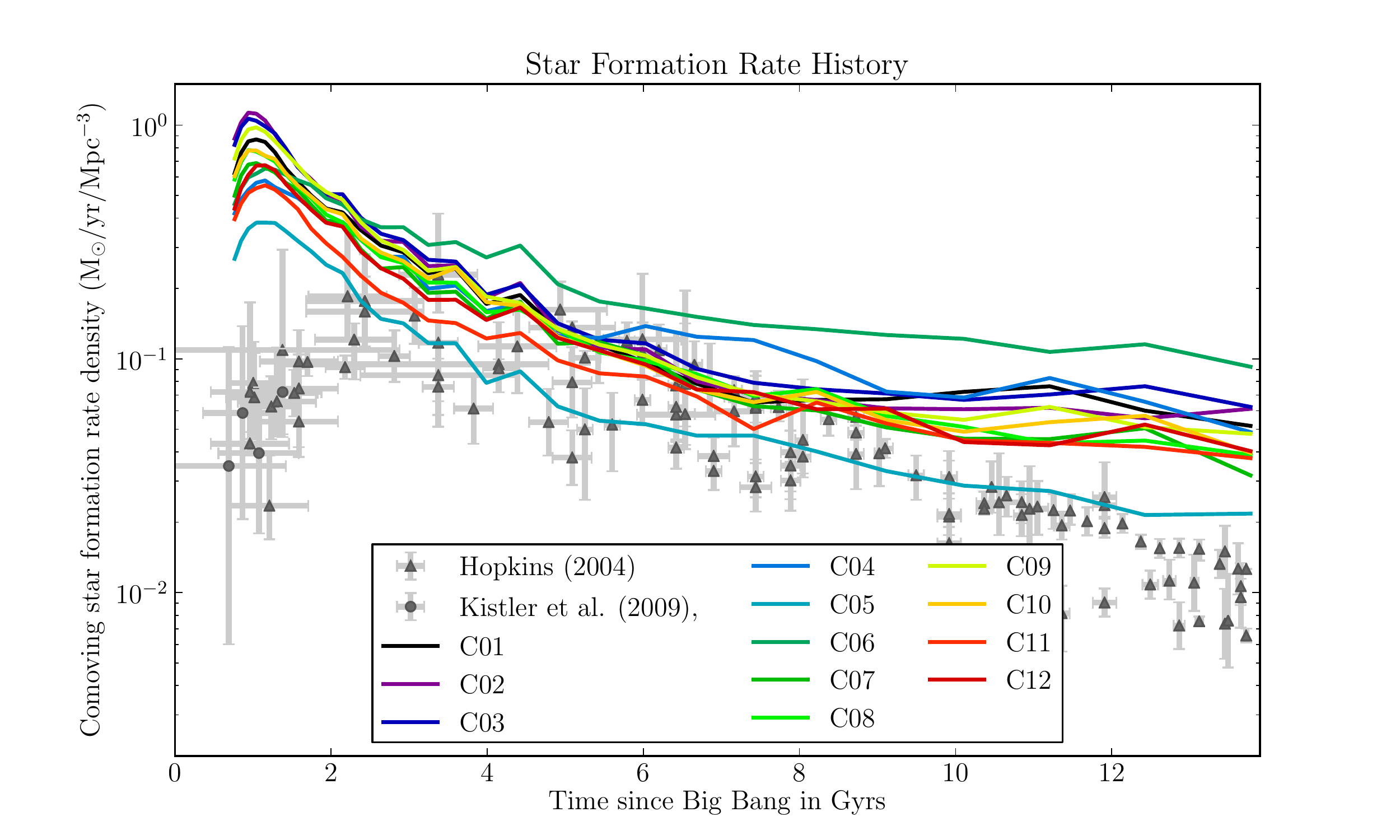} \\
\caption{SFR histories of all 12 clusters in Gyr.}  \label{SFRhistories}
\end{figure*}

The first and foremost physical quantity that we require to model the galactic winds 
and the enrichment of the ICM are the star formation rates. In Fig.~\ref{SFRhistories} we 
plot the SFR histories of all 12 clusters. It is clear that not only can the integrated star formation rates 
differ enormously between the least massive clusters (e.g., C05) and more massive ones (such as C01 or C02), but 
that the SFR at $z=0$ can also differ by almost an order of magnitude. In Sect.~\ref{secResults} we
discuss the structure formation histories of these objects also with respect to the relative 
peaks in the SFR history curves. In many cases, these peaks directly correspond to merger 
events. To interpret the plot, we recall that the 20Mpc/h volumes are 
not directly comparable with the observational data because we derived $\Omega_\text{c}$ values 
inside these boxes between $\Omega_\text{c}\sim0.7-0.9,$ and in this respect these 
volumes are no longer representative for the cosmos.  Specifically, they are highly 
overdense regions by virtue of having been selected to contain massive clusters.
One direct consequence of this volume cut is that the star formation rates plotted are 
significantly higher than the observational comparison data from~\cite{Hopkins2004} 
and~\cite{Kistler2009}. This tendency is also shown in Fig.~\ref{galacticusSFR} 
, we show that the degree of accordance with observational data increases with 
simulation volume. 

Furthermore, {\sc Galacticus} probably overestimates the SFRs at low 
redshifts, especially within the virial radius of the galaxy clusters, because of 
the lack of a generic treatment of disk ram-pressure stripping.

\begin{figure*}
\centering
\includegraphics[scale=0.6]{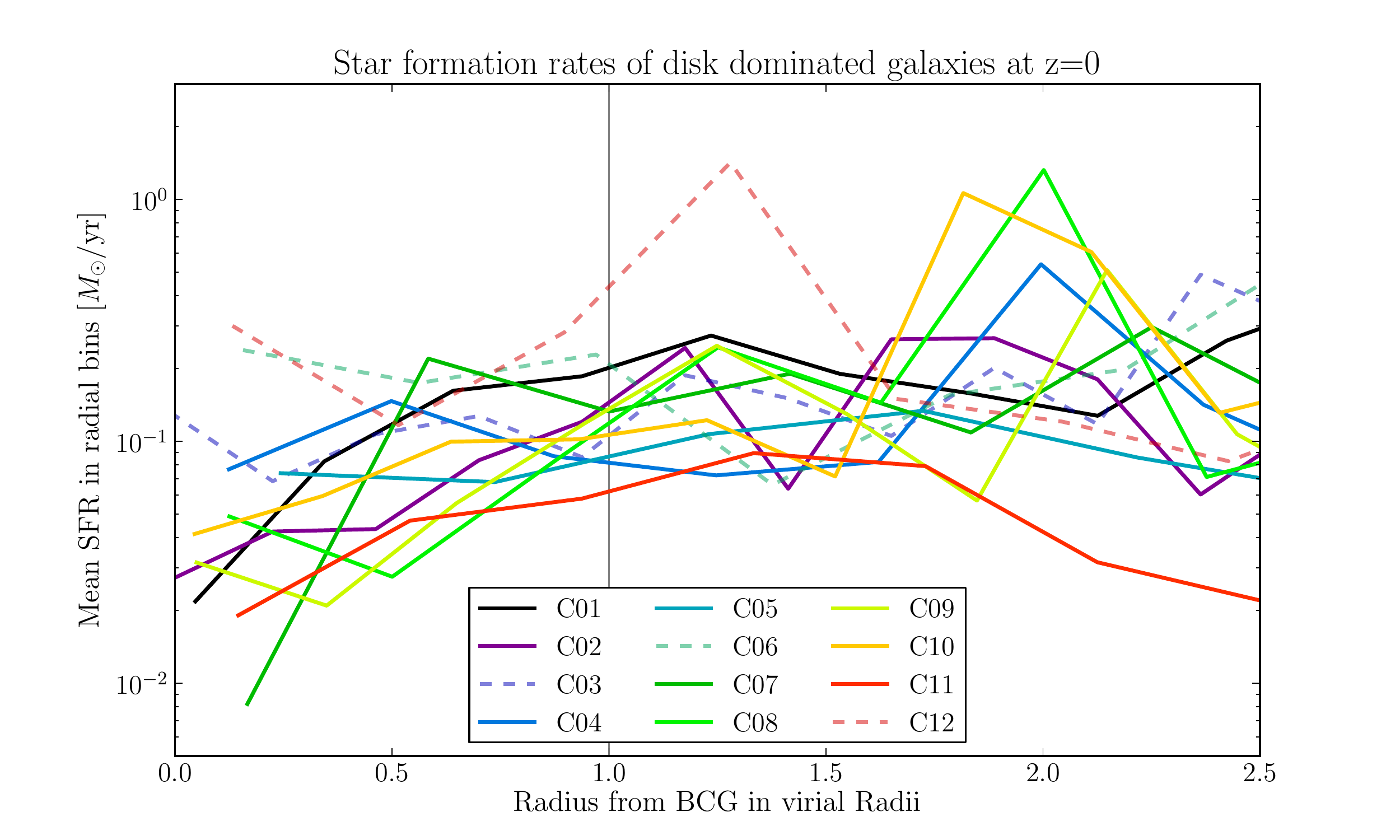} 
\caption{Radially binned mean star formation rates from \textsc{Galacticus}. The dotted, semitransparent lines 
belong to clusters that are in the least virialized states (see Sect.~\ref{secResults}).}  \label{binnedSFR1}
\end{figure*}

\begin{figure*}
\centering
\includegraphics[scale=0.6]{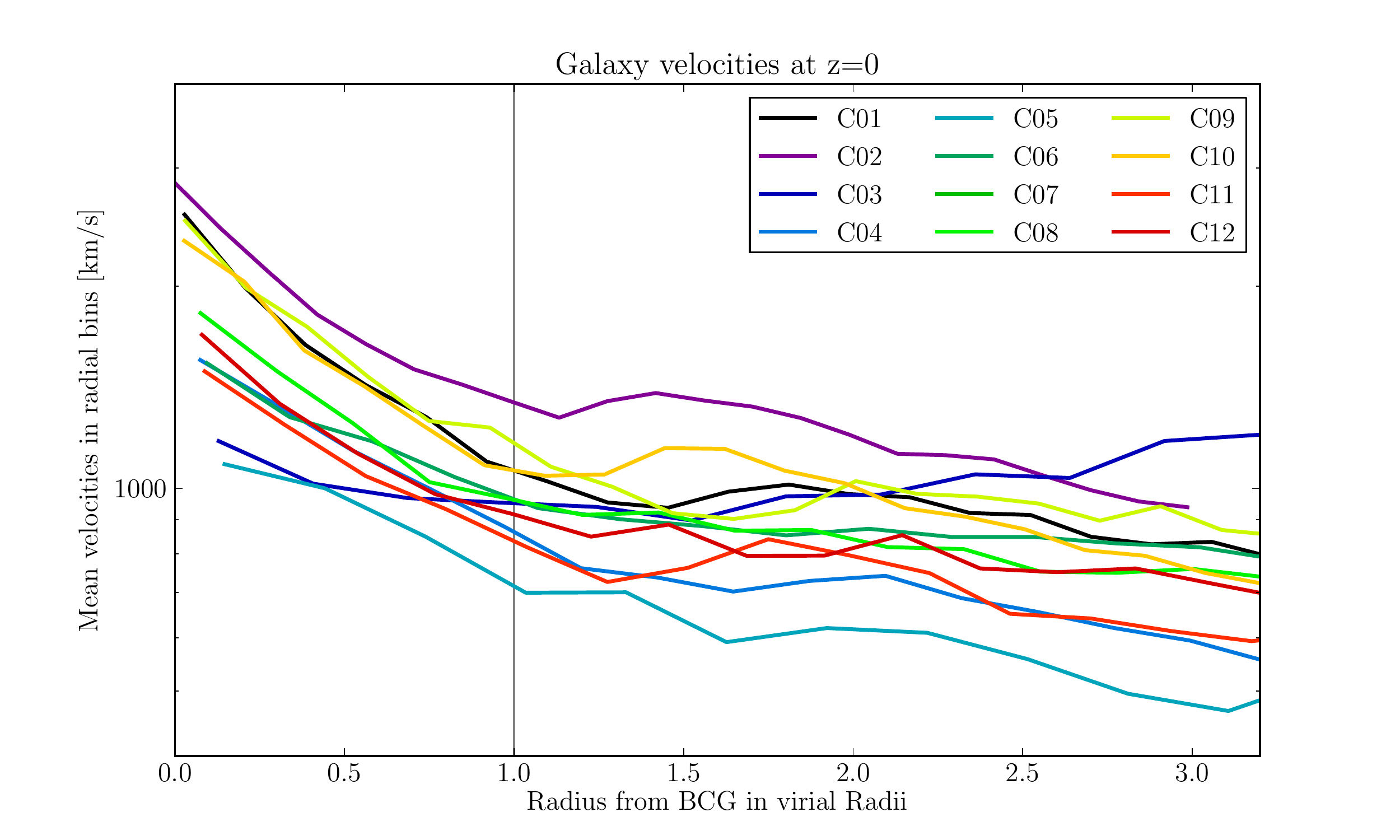}
\caption{Radially binned mean velocities}  \label{binnedVelo1}
\end{figure*}

In Fig.~\ref{binnedSFR1} we plot the radially binned mean 
star formation rates of all samples of galaxies and our \textit{disky galaxies,} 
which we define as those with a nonzero disk mass. The quenching of the SFR inside the 
virial radius is mainly due to the hot-halo ram-pressure stripping.
This is modeled in \texttt{Galacticus} using the results
from~\cite{Font2008} and affects all halos that are subhalos of another halo. Within the BCG virial radius, 
all infalling galaxies are satellites and hence experience stripping of their hot halo while 
falling into the galaxy cluster.
The radially binned mean velocities are plotted in Fig.~\ref{binnedVelo1}  
and support the expected increase in efficiency of ram-pressure stripping with decreasing 
cluster centric radius.  

\begin{figure*}
\centering
\includegraphics[scale=0.6]{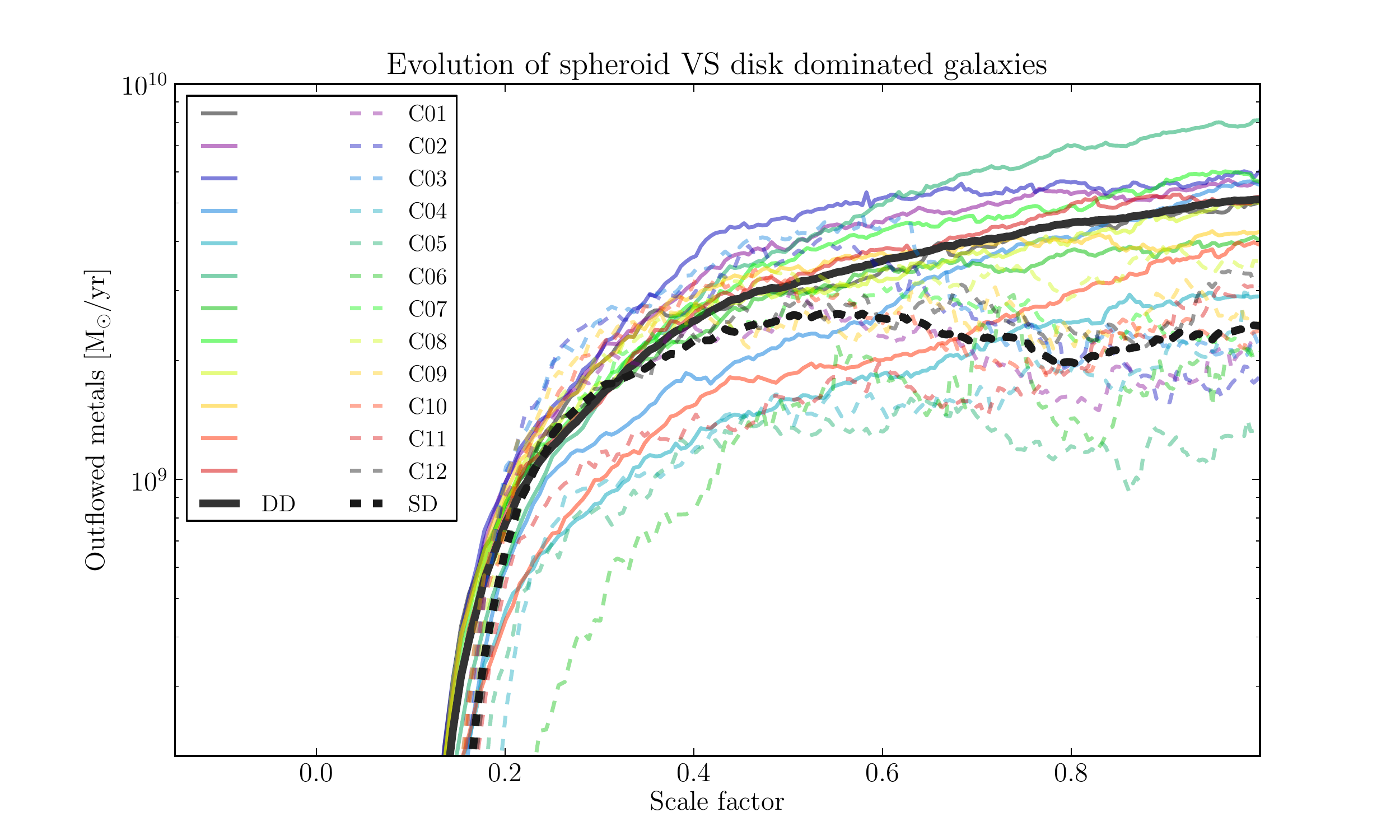} \\
\caption{ Evolution of spheroid-dominated (SD, dotted) and disk-domitaned (DD, solid lines)
galaxies with respect to the outflowed metals. The thicker, dark lines denote 
the mean values of all 12 clusters. \label{outflowedHisto}}
\end{figure*}

In Fig.~\ref{outflowedHisto} we illustrate the different evolution of disk-dominated (DD) 
and spheroid-dominated (SD) galaxies in our simulations. The classification of these two types 
is straightforward. We define a bulge-to-total-mass ration $B/T$:
\begin{equation}
\qquad B/T = \frac{M_{\rm \star, sph} + M_{\rm gas, sph}}{M_{\rm \star, sph} + M_{\rm gas, sph}+M_{\rm \star, disk} + M_{\rm gas, disk}},
\end{equation}
and identify $B/T \leq 0.5$ as disk-dominated (DD) and $B/T > 0.5$ as spheroid-dominated (SD), while 
the morphological classification is based on the plotted time.
Our data clearly show that disky galaxies primarily contribute to metals ejected and mixed 
into the ICM. This seems to contradict ~\cite{Arnaud1992}, who found for a sample of local 
galaxy clusters a correlation between total cluster gas mass (and iron mass) and luminosity of 
ellipticals and S0s, but no correlation between cluster gas mass and spiral luminosity.
The authors concluded that the iron mass in the ICM is probably directly proportional to the stellar
mass in cluster ellipticals and S0s, and that therefore only ellipticals were involved in the metal 
enrichment of the ICM. In our simulation data the gas mass as function of galaxy luminosity 
and morphology agrees perfectly with the results presented in~\cite{Arnaud1992}, 
see Table~\ref{Tab:Arnaud}. However, from that finding we do not deduce that only ellipticals 
(and S0s) were involved in metal enrichment, but refer to the generally established idea of a 
morphological galaxy transformation in clusters \citep[see, e.g.,][]{Vogt2004}. 
Spiral galaxies that contributed to the ICM metal content in the past might have been 
transformed and, at the present epoch, be classified as S0s or ellipticals.

\begin{table}
\centering
\begin{tabular}{l*{2}{l}r}
                               & Observational           & Simulation sample \\
\hline
Gas mass and SD                 & $1.1 \pm 0.25$ (E)       & $1.19$ ($B/T \geq 0.6$) \\
                               & $1.5 \pm 0.25$ (E+S0)    & $1.26$ ($B/T \geq 0.5$) \\
Metal mass and SD               & $1.0$ (E+S0)             & $1.13$ \\
\end{tabular}
\caption{Comparison of simualtion results with data from~\cite{Arnaud1992} within the 
inner 3 Mpc of the galaxy clusters.}  \label{Tab:Arnaud}
\end{table}

We discuss the necessity of generating and analyzing samples that are both narrower 
in kinetic states and larger in number for statistically significant statements in 
Sect.~\ref{Discussion}. The 12 clusters presented here illustrate 
the potential of the simulation setup and the wide spectrum of objects that can 
be covered and studied, however. 

\section{Results of cosmological hydrodynamical cluster simulations}\label{secResults}

The results presented in the subsections below are based on simulations made with wind model \textbf{C}. 

\subsection{C01} \label{Subsec:C01}

The source C01 has a major subcluster merger event at $\mathrm{z} \sim 0.7$ and another smaller merger 
at $\mathrm{z} \sim 0.3$. In the graphical Appendix~\ref{Sec:App}, we show the projected metallicities, 
DM density, temperature, and baroclinic vorticity (a quantity proportional to $\nabla p \times \nabla \rho$ 
that displays shock-induced turbulence, see Fig.~\ref{Fig:BVMall}(a)) at $\mathrm{z}=0$. In temperature and vorticity, 
the shock structures are visible. The elongated DM profile is another indicator of past mergers. 
C01 is a virializing, cool-core cluster with a peak temperature of $\sim 2 \times 10^{8}$K
at a cluster-centric radius $\mathrm{r} \sim 0.2  $ Mpc, and a core temperature of $\sim 1.5 \times 10^{7}$K. In Figs.~\ref{Fig:radWind}
and~\ref{Fig:radRPS} (please note the different scales), we plot the radial profile of 
the metallicities due to galactic winds and disk 
ram-pressure stripping. Because C01 is a massive cluster, winds are effectively suppressed toward the 
center, and the level of RPS metal contribution is comparatively high.  
Interestingly, the sum of both contributions is almost constant out to the virial radius with 
values in the range $Z\sim 4-6 \times 10^{-3}$ , except for the cool core where the 
metal mass-fraction obviously drops significantly below $1/3 \mathrm{Z}_{\odot}$. 

\begin{figure}
\includegraphics[scale=0.4]{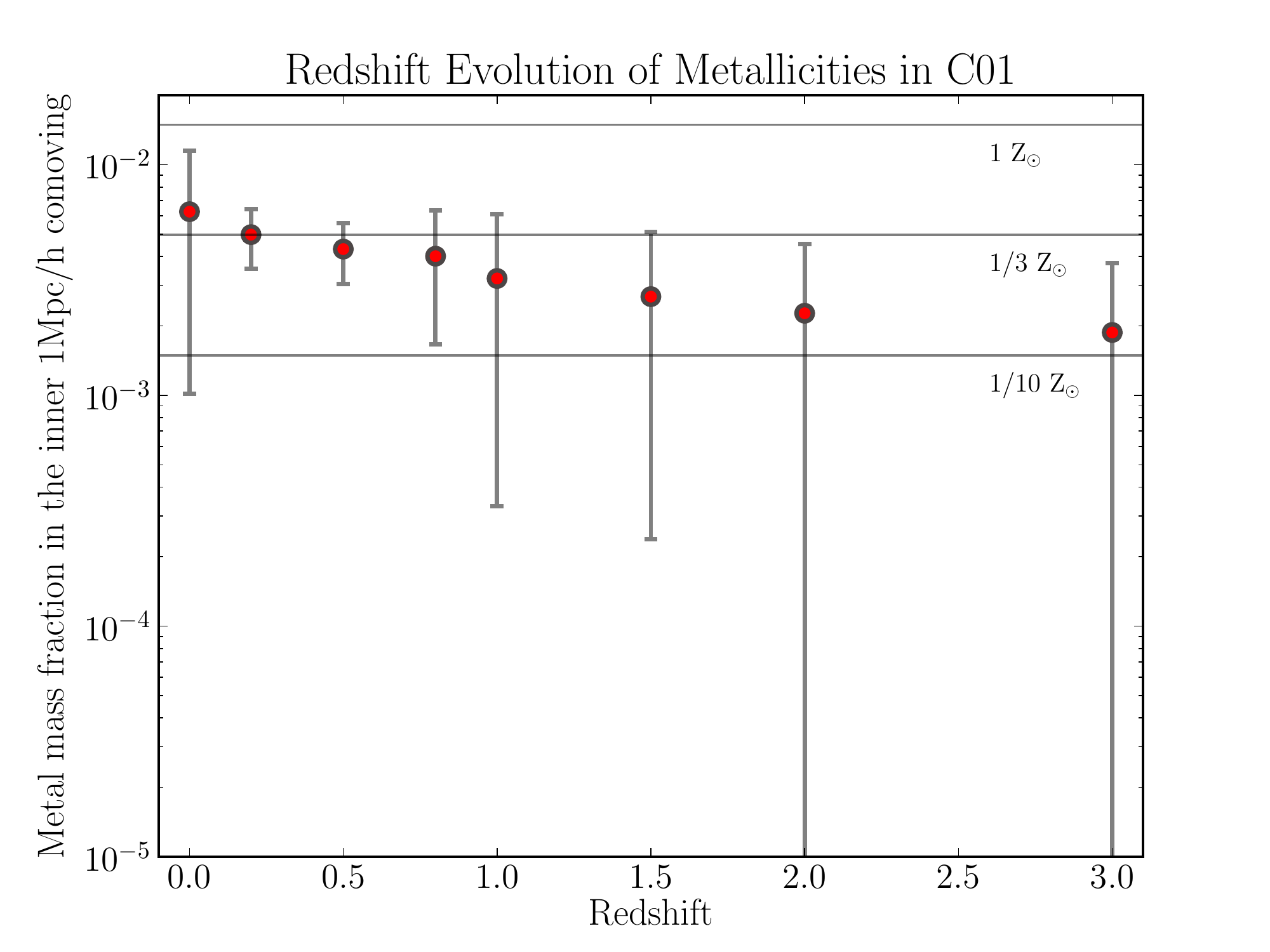}
\caption{Metallicity evolution C01 within a comoving 1 Mpc/h sphere around the 
density maximum. Plotted are mean values and the variance. } \label{Fig:metEvoC01}
\end{figure}

In Fig.~\ref{Fig:metEvoC01} we plot the evolution of metallicities within the inner comoving 1 Mpc/h. 
The lowest metallicity of the cool core is at the same level as the mean metallicity at a redshift $\mathrm{z}>1,$
which also coincides with the time of its formation. 
Furthermore, Fig.~\ref{Fig:metEvoC01} shows a steady \textit{homogenization} of metallicities 
toward $\mathrm{z}=0.2,$ but the variance (error bars) again
increases at $\mathrm{z}=0,$ which is due 
to the onset of RPS. 

The simulation of C01 with the wind model \textbf{B} ran into numerical problems with a cooling catastrophe 
already at redshift $\mathrm{z}\sim1.2$. Analysis shows that this wind prescription yields super-olar 
metallicities for the major portion of the ICM already at such high redshifts. 
This suggests not only that the formation of cool cores is strongly 
correlated with the ICM metallicities, but also that the wind prescription from~\cite{Hopkins2012} 
probably yields too high metallicities in the ICM. In Sects.~\ref{Subsec:Comp} and~\ref{Discussion} we discuss 
more arguments to support that claim, and we describe how we adapted the wind model for a more detailed analysis.

\subsection{C02} \label{Subsec:C02}

The source C02 is the most massive cluster in the sample (virial mass of $1.06 \times 10^{15}\mathrm{M}_{\odot}$) and also has the highest peak galactic velocities in Fig.~\ref{binnedVelo1}. 
The cluster experiences a strong merger at $\mathrm{z} \sim 0.3$ and there are two big groups (pre-merger) within 2 Mpc/h (one 
of which is cluster C03). The large DM halo in Fig.~\ref{Fig:DMall}(b) appears to be more elongated than it actually is because of a 
projection effect of an infalling subcluster along the $x$-direction. 
The relatively high level of SFR at $\mathrm{z}=0$ in Fig.~\ref{SFRhistories} does not show a significant correlation
with the wind metallicity levels in the radial plot~\ref{Fig:radWind} toward the center. C02 also
shows the steepest decline in wind metallicities within its virial radius, but it is
among the highest wind metallicity objects of Fig.~\ref{Fig:Windall}(b).   
The ram-pressure stripped metals in Fig.~\ref{Fig:radRPS} 
are as expected amongst the absolute highest values in the sample. 
In Fig.~\ref{Fig:BVMall}(b) we show the baroclinic vorticity, which is dominated by the pre-merger shocks 
of the infalling cluster C03 (coming from the lower right), but there is a considerable level of baroclinity 
also around the center, which traces the previous merger at $\mathrm{z} \sim 0.3$. This event obviously was not able 
to disrupt the cool core, which we also find in C02, which likewise shows the significant drop in metallicity within the 
central $\sim 150$kpc.

\subsection{C03}

Cluster C03 is not just a different final cut of the same volume as C02, but the whole data analysis 
presented in Sect.~\ref{SimulationSetup} was made individually for a shifted 20 Mpc/h side-length volume of 
the same 80 Mpc/h box. With this, we are able to test resolution effects of the nested grid approach. 
 
The source C03 is a configuration centered on an infalling smaller subcluster of C02, see Fig.~\ref{Fig:DMall}(c). 
The strong hydrodynamical shocks caused by the ongoing approach are seen in temperature and baroclinic vorticity 
(Figs.~\ref{Fig:Tempall} and~\ref{Fig:BVMall}). The projected metal mass-fraction coming from galactic 
winds in Figs.~\ref{Fig:Windall}(c) shows an s-like shape that is not seen as clearly in (b). This is mostly a projection effect, however. This special feature arises from the second infalling subcluster, which 
had a recent close fly-by between C02 and C03 with a trajectory spiraling toward the main cluster core. 
Because C03 is not an actual isolated galaxy cluster, this untypical behavior extends to the radial SFR, the radial 
wind metallicities plots, and the radial mean velocities (Figs.~\ref{binnedSFR1},~\ref{Fig:radWind},
and~\ref{binnedVelo1}). 
While all virialized or semivirialized objects in Fig.~\ref{Fig:Coreall} show ring-like structures 
in wind metals in the core cluster region, C03 as one of the \textit{untypical} clusters presents a large cloud 
of dragged-along high-metallicity gas. All the relaxed systems show wind suppression 
in the high-density (and thus high-pressure) region of the cluster centers, as expected.

\subsection{C04}

The cluster C04 shows a major-merger event at $\mathrm{z} \sim 0.6$ and ongoing wafting of the central 
halo for the rest of the simulation. This cluster has a relatively low mass, showing also one
of the smallest velocity dispersions in Fig.~\ref{binnedVelo1}. The contribution 
to the metal masses by disk ram-pressure stripping is generally low for C04, which is 
seen both in the radial metallicity profile in Fig.~\ref{Fig:radRPS} and 
in the projected metal mass-fraction in Fig.~\ref{Fig:RPSall}(d). The RPS metals 
show much structure and a strong peak in the radial plot, which is due to \textit{recent} stripping, 
especially in the very center of the cluster, while the overall (total and relative) contribution to the metallicity 
by ram-pressure stripping is significantly smaller than for cluster C01, for example. 
Especially in the zoomed-core total metal (wind + RPS) projection plots~\ref{Fig:Coreall}, this cluster 
presents itself interestingly inhomogeneous with a high-metallicity region in the top left corner that 
is caused by two infalling groups that are already very close to the actual cluster
at $\mathrm{z}=0$, as seen in the dark matter projection~\ref{Fig:DMall}(d).
C04 shows a slightly cool core that is dislocated from the DM density peak by $\sim 220 $ kpc, where 
 the temperature also peaks at $\sim 8 \times 10^{7}$K.   

\begin{figure}
\includegraphics[scale=0.4]{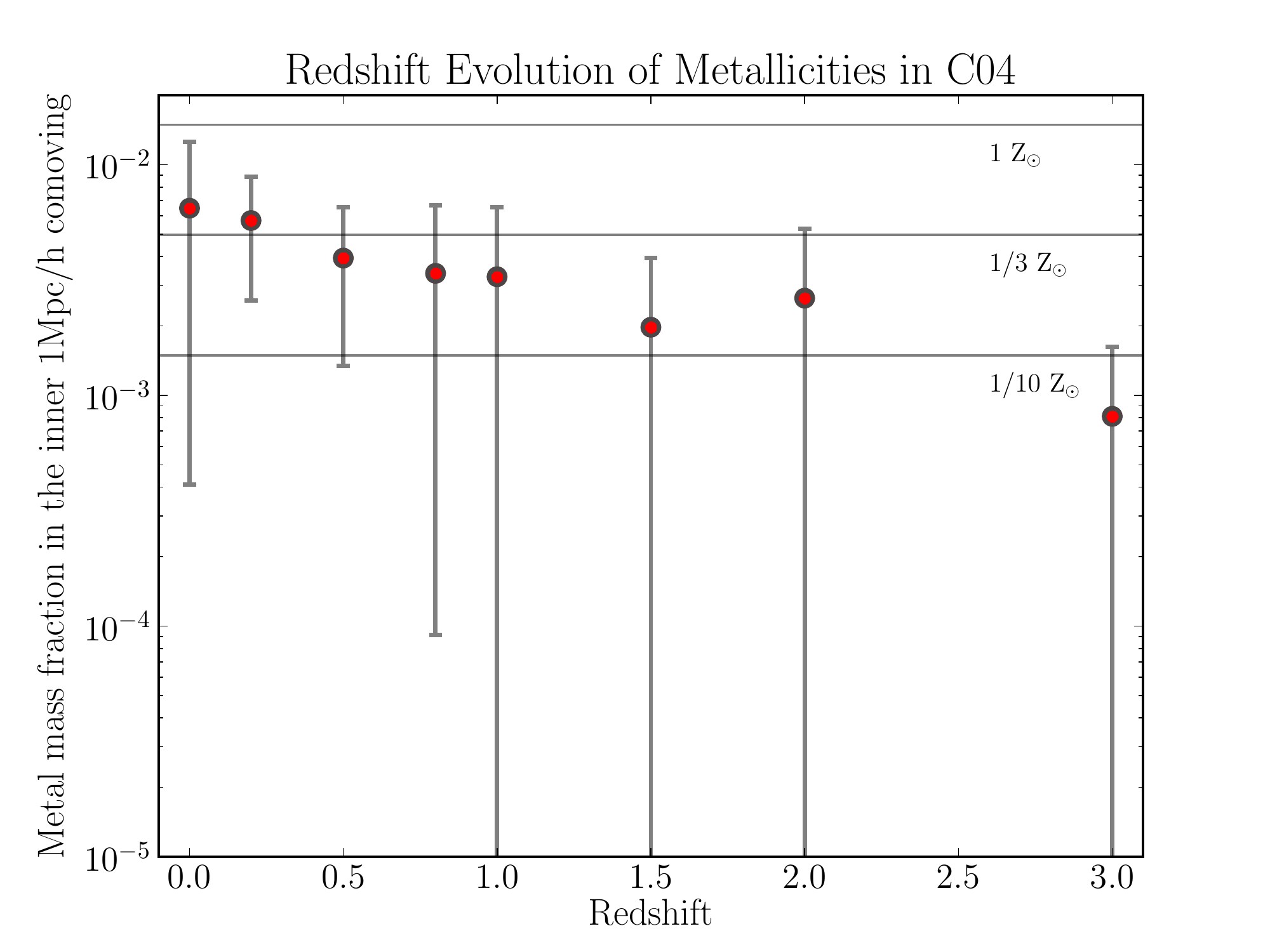}
\caption{Metallicity evolution of C04 within a comoving 1 Mpc/h sphere around the
density maximum. Plotted are mean values and the variance. } \label{Fig:metEvoC04}
\end{figure}

In Fig.~\ref{Fig:metEvoC04} we plot the evolution of the central metallicity levels 
with time. The metallicity of C04 drops at about $\mathrm{z}=1.5$ because of filament 
infall of poorly enriched gas. Figure~\ref{Fig:metEvoC04} also
shows an increased level 
of variance (i.e., inhomogeneity) at $\mathrm{z}=0$ due to localized RPS peaks. 

\subsection{C05}

The cluster C05 has the lowest mass in our sample. It experiences a triple merger at $\mathrm{z}\sim 0.3-0.2$ 
and relaxes to become semivirialized by $\mathrm{z} = 0$. There is a central cool core where the 
temperature drops from $\sim 3 \times 10^{7}$K to $\sim 8 \times 10^{6}$K. 
The metallicity maps in Fig.~\ref{Fig:Windall}(e) show several extraordinarily high metallicity clumps 
within a rather low metallicity cluster environment.
  
The overall modest metallicity values are probably related to the lowest integrated star formation rates in the sample 
(see Fig.~\ref{SFRhistories}).  


\subsection{C06}

The cluster C06 shows a merger event at $\mathrm{z} \sim 0.2$ and ongoing wafting thereafter until $\mathrm{z}=0$. 
Nevertheless, there is a slight but extended cool core with a temperature contrast of approximately 0.4 
within 250 kpc. Although C06 has the highest integrated star formation rate, the level of  
metal mass-fraction contribution from galactic winds is medium
for most of the plotted 
volume in Fig.~\ref{Fig:radWind}. There is a striking drop of metallicity toward the 
unvirialized cluster core, whose direction strongly correlates with that of the elongated DM 
halo, as seen in Figs.~\ref{Fig:Windall}(f) and~\ref{Fig:DMall}(f). 
Figure~\ref{Fig:Windall}(f) shows that there are several high-metallicity regions outside the actual cluster that originate 
from infalling galaxies with recent high SFR episodes, while the very core remains depleted 
of wind metals. The 20 Mpc/h volume in whose center C06 resides
is generally excessively busy and dense with a total matter density of $\Omega_c = 0.95, $ which provides 
a good indication of the shape of the volume-averaged SFR history. The mass of   
C06 within $r_{200}$ of $1.5 \times 10^{14}M_{\odot}$ is similar to 
that of C11 and C04, which also manifests itself 
in very similar and rather low velocity dispersions (see Fig.~\ref{binnedVelo1}). 
This means that disk ram-pressure stripping in C06 is much less significant than in high-mass clusters. 
But we found  a high-metallicity RPS peak close to the center, which is caused by recent stripping events. 
The radial wind-metallicity profile of this galaxy cluster is  \textit{untypical,}  which shows the unique central depletion of wind metals that is also illustrated in the projected total metal 
mass-fraction in Fig.~\ref{Fig:Coreall}(f). This extended region of low metallicities is caused by the 
$\mathrm{z} \sim 0.2$ merger of two early formed cool cores that
have a metallicity level of high redshift, as 
discussed for previous clusters.  

\begin{figure*}
\centering
\includegraphics[scale=0.6]{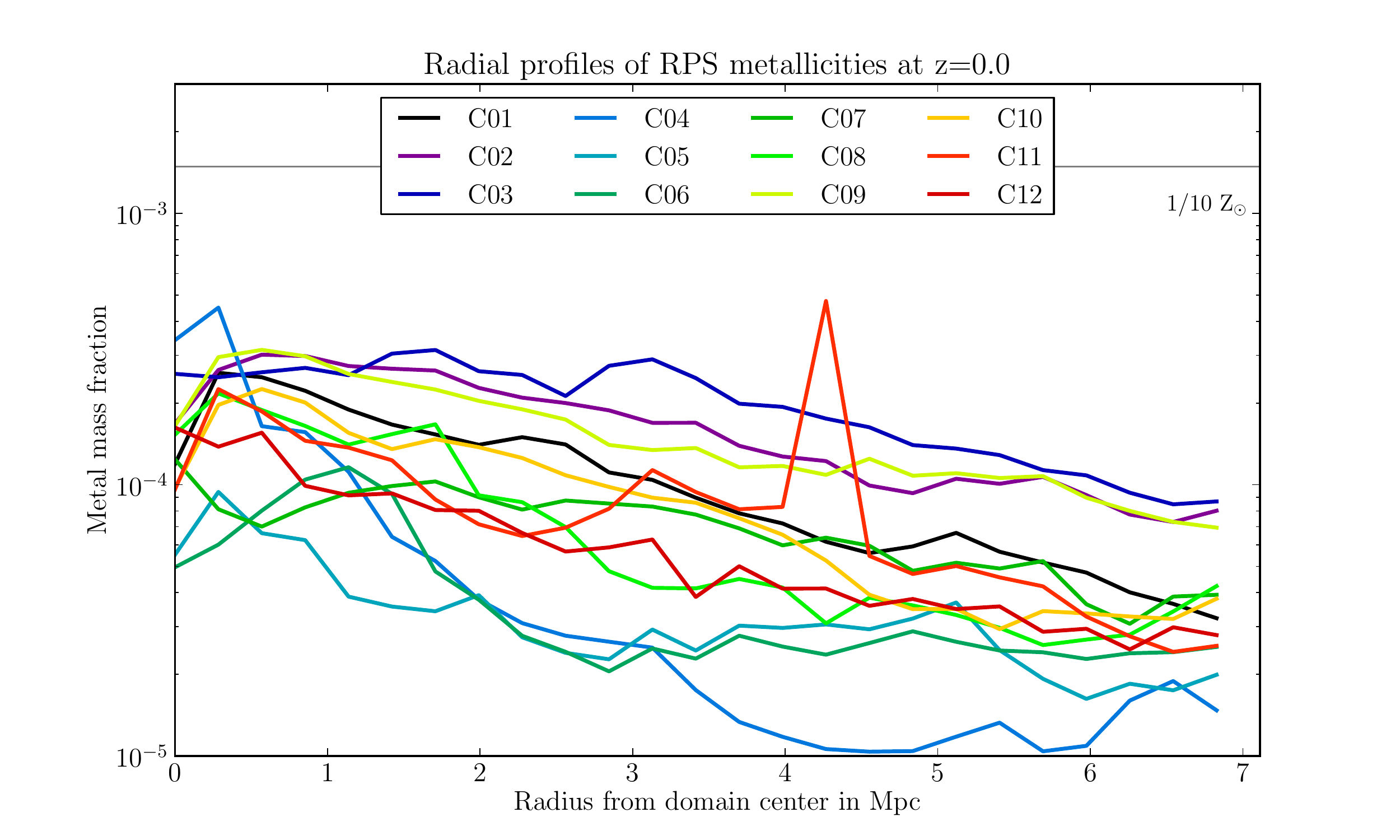} \\
\caption{Radial profiles of metal mass-fractions from RPS.}  \label{Fig:radRPS}
\end{figure*}

\subsection{C07}

\begin{figure}
\includegraphics[scale=0.4]{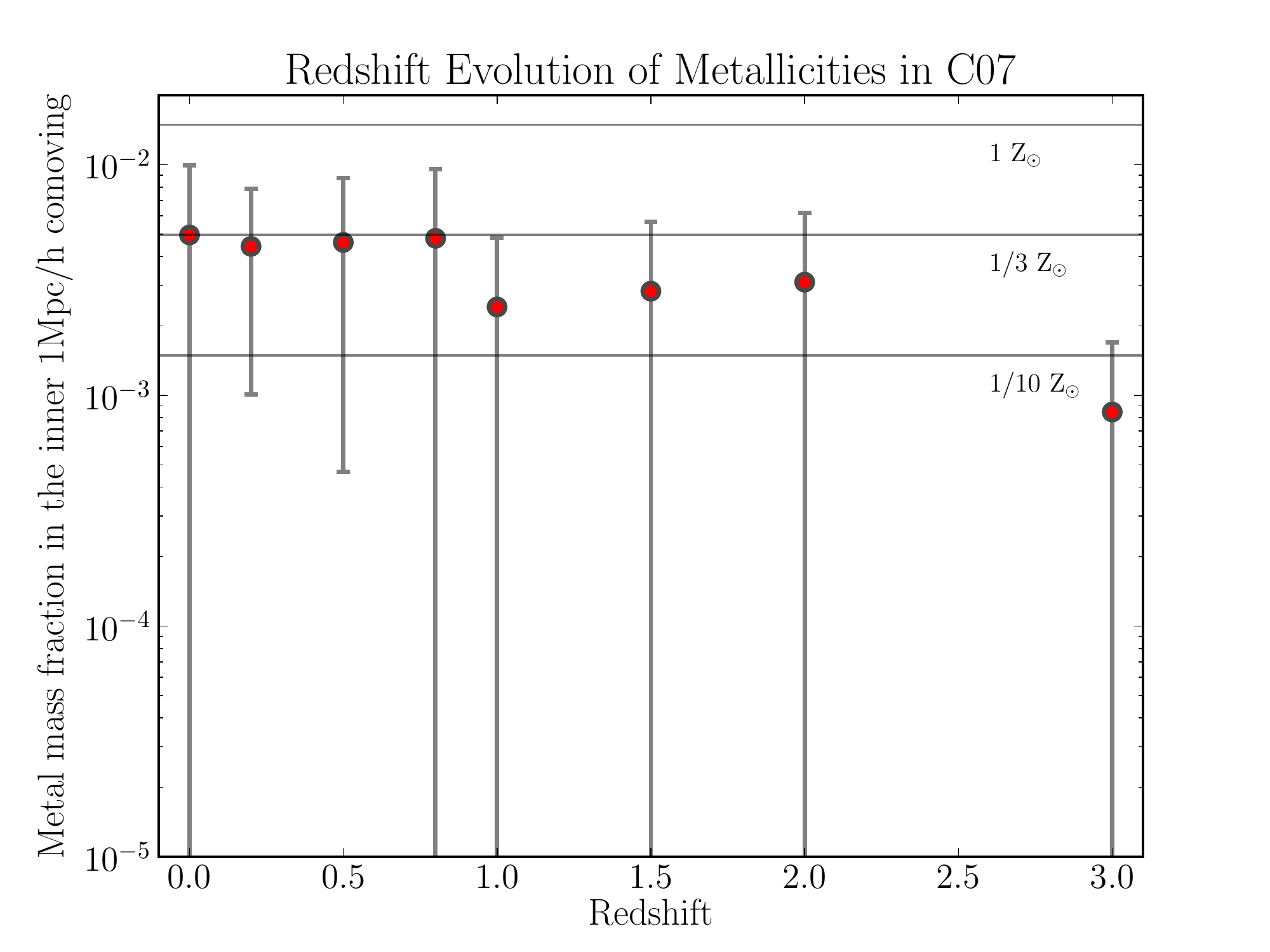}
\caption{Metallicity evolution of C07 within a comoving 1 Mpc/h sphere around the
density maximum. Plotted are mean values and the variance. } \label{Fig:metEvoC07}
\end{figure}

The configuration C07 is a pre-merger case of several subclusters at a distance of 
$\sim 1.5 $ Mpc/h, evident in the baroclinic vorticity plot with a butterfly-like 
shock structure. Although the measured velocities of the galaxies in C07 are similar to 
relaxed systems of similar total mass, no extended \textit{static} ICM halo has developed yet, 
and the contribution to metals by disk ram-pressure stripping is smaller than for the virialized cases,
while we find much structure in the RPS metallicity projection in Fig.~\ref{Fig:RPSall}(g). This 
suggests that the stripping events took place at low redshift.  
Because configuration C07 is not an actual galaxy cluster, a number of cool cores 
coincide with the DM peaks with temperature contrasts of $0.5-0.3$.  
The redshift evolution of metallicity in Fig.~\ref{Fig:metEvoC07} shows a peculiar peak at 
$\mathrm{z} = 0.8$ that is similar to the final mean metallicity of $1/3\mathrm{Z}_{\odot}$ at 
$\mathrm{z} = 0.0$. 
 
The DM structure in Fig.~\ref{Fig:DMall}(g) in the lower left  corner seems close in the 
$x$-projection, but is at a distance of about 6 Mpc/h. 

\subsection{C08}

Cluster C08 shows a major-merger event at $\mathrm{z}\sim1$ and after this, smooth and steady 
mass accretion. DM density, gas density, and pressure peak at the very center of the cluster, 
which also hosts a steep cool core. The baroclinity values are moderate, as expected for clusters with no recent 
significant merger events. 
In Fig.~\ref{Fig:radWind} the radial wind metal mass-fraction profile is plotted, showing as in the other 
cool-core cases a significant drop of metallicity at the very center. 
The peak at distance 6.6 Mpc is caused by a combination of a recent
starburst and an already very rarefied ICM in these outer parts of the cluster surroundings. 
Although the radial profile suggests a typical metallicity profile attributable to galactic winds, the projection 
plot in Fig.~\ref{Fig:Windall}(h) reveals an extended high-metallicity cloud in the upper center 
of the $x$-projection. Furthermore, it has a very localized metallicity peak caused by galactic winds 
within the inner Mpc of the cluster (not a projection effect), a peculiarity that is not seen in any 
other of the 12 configurations. This supersolar central metallicity peak is seen even more clearly in the 
zoomed-core total metallicity projection~\ref{Fig:Coreall}(h).

\subsection{C09}

The cluster C09 shows major-merger events at $\mathrm{z} \sim 1$ and $\mathrm{z} \sim 0.7$ and is virialized 
at redshift zero. For this object the radial trend in SFR (Fig.~\ref{binnedSFR1}) is among the steepest within 
the sample. Furthermore, it has a very strong cool core where the temperature drops almost an 
order of magnitude in the inner 200 kpc and as in the other cool-core cases on this length scale 
we also see a drop in metallicities. Because of its early formation as a galaxy cluster, it has one of 
the strongest contributions from RPS to the metallicities, but also a drop in this quantity toward the cool core (Fig. ~\ref{Fig:radRPS}).

\subsection{C10}

\begin{figure}
\includegraphics[scale=0.4]{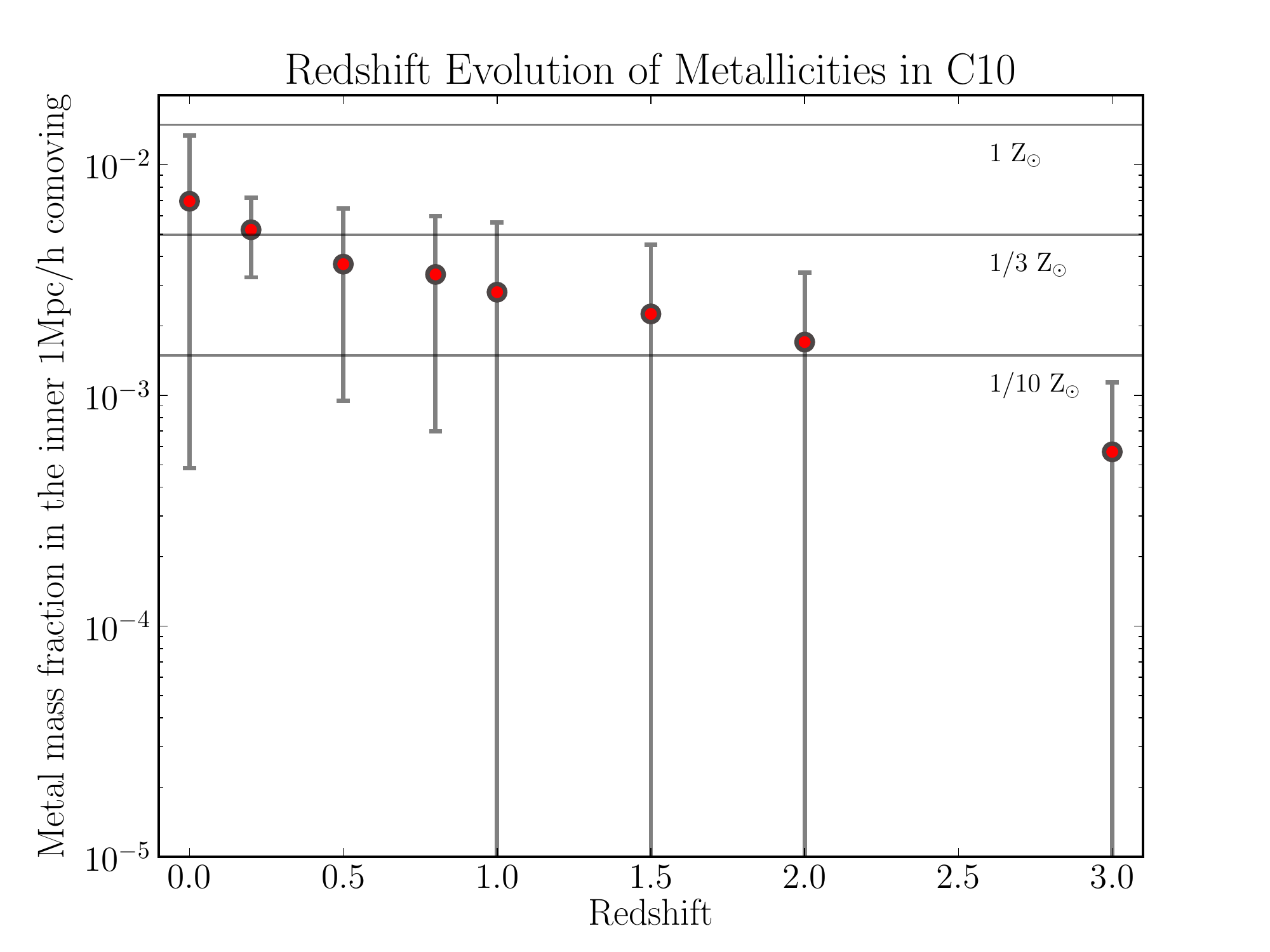}
\caption{Metallicity evolution of C10 within a comoving 1Mpc/h sphere around the
density maximum. Plotted are mean values and the variance. } \label{Fig:metEvoC10}
\end{figure}

The DM plot in Fig.~\ref{Fig:DMall}(j) shows cluster C10 as a seemingly relaxed system, but the 
very high baroclinic vorticity values witness the very strong merger event at 
$\mathrm{z} \sim 2$ with ongoing wafting until $\mathrm{z}=0$. This very energetic event 
was obviously unable to prohibit the formation of a cool core,
however. 
There is a peculiarity in temperature because of its kinetic state, as the temperature peak 
is more like an extended plateau between $0.2-1 $ Mpc. Figure~\ref{Fig:Tempall}(j) reveals not only this large high-temperature region, but also the infall of a lower-temperature 
object from the left with a distinct bow shock in front of it. In Fig.~\ref{Fig:metEvoC10} we plot 
the metallicity evolution in the inner comoving Mpc/h. This cluster  also shows a steady homogenization 
of metallicities until the data point at $\mathrm{z}=0.2$ and an increased variance at $\mathrm{z}=0$.  

\subsection{C11}

The cool-core cluster C11 also experiences a minor merger between $\mathrm{z} \sim 0.2-0.1$. 
With its virial mass of $2.11 \times 10^{14} \mathrm{M}_{\odot}$, C11 belongs to the lower mass 
clusters, which is also visible in the galaxy velocities in Fig.~\ref{binnedVelo1}. However, its metallicity 
values within the virial radius are the highest in the sample, although they show the typical central 
drop in the cool core. The total metallicity projection in Fig.~\ref{Fig:Coreall}(k) is remarkably 
symmetric for this virialized cluster. 
This may be also related with the generally very low star formation rate at $z=0$, seen both in the SFR history 
plot in Fig.~\ref{SFRhistories} and in the radial SFR trend in
Fig.~\ref{binnedSFR1}, which  suggests that the enrichment 
has occurred at an earlier epoch and the mixing within the ICM has effectively homogenized the metals.  

\subsection{C12}

\begin{figure*}
\centering
\includegraphics[scale=0.6]{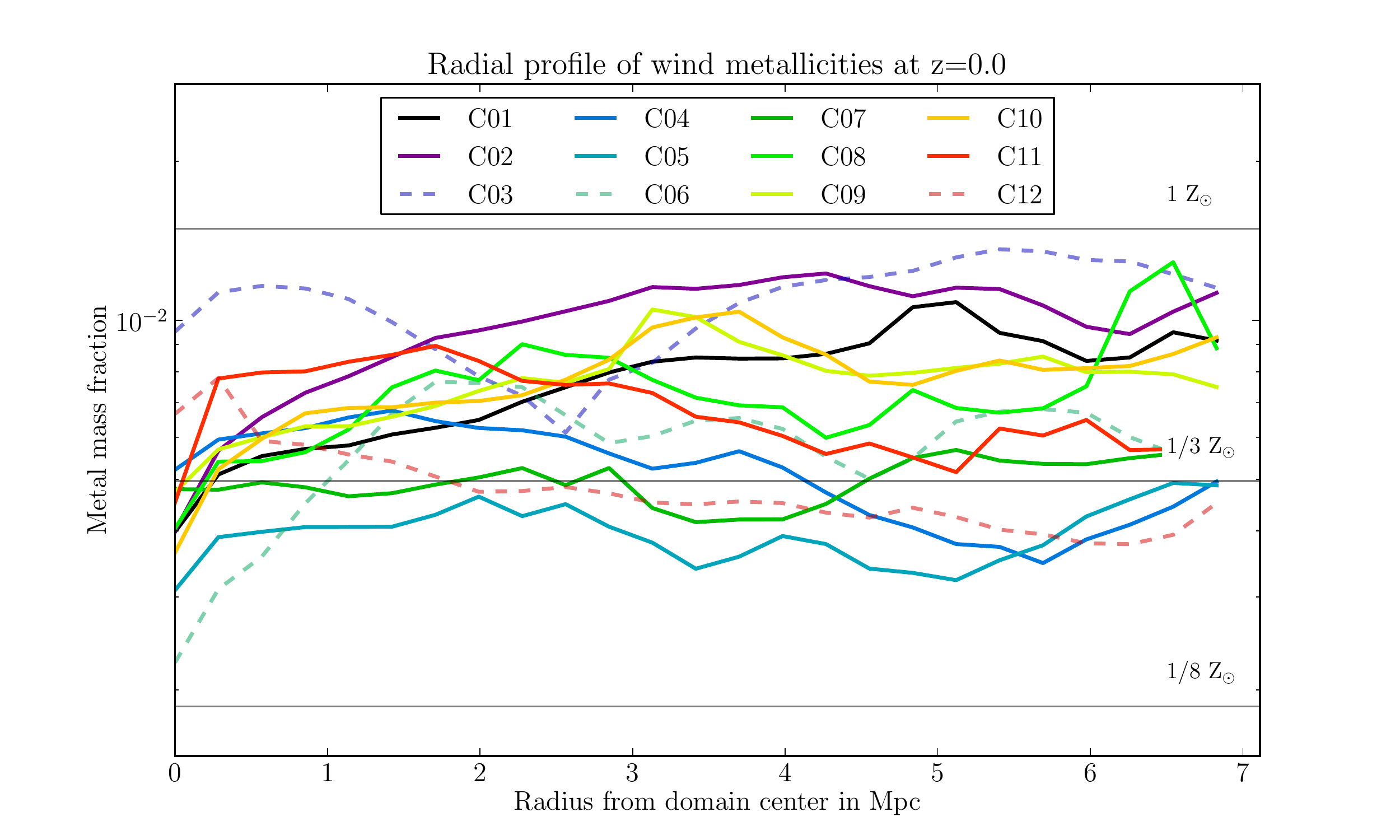} 
\caption{Radial profiles of metal mass fractions from galactic winds. The dotted cases are classified 
as \textit{untypical} because of their kinetic states. }  \label{Fig:radWind}
\end{figure*}

The structure in C12 shows an ongoing multiple merger and is hence a very unvirialized object that 
also shows the most untypical radial SFR profile in Fig.~\ref{binnedSFR1}, as well as a peculiar radial 
wind metallicity profile in Fig.~\ref{Fig:radWind}. That the multiple merger 
is still in a very early stage explains why the vorticity is rather localized and the 
temperature peak is both high and comparatively extended, Figs.~\ref{Fig:BVMall}(l) and~\ref{Fig:Tempall}(l).  
There are two pre-merger cool cores dislocated from the shock-heated temperature maximum.

\subsection{Comparison of wind models} \label{Subsec:Comp}

\begin{figure}[h!]
\centering
\includegraphics[scale=0.28]{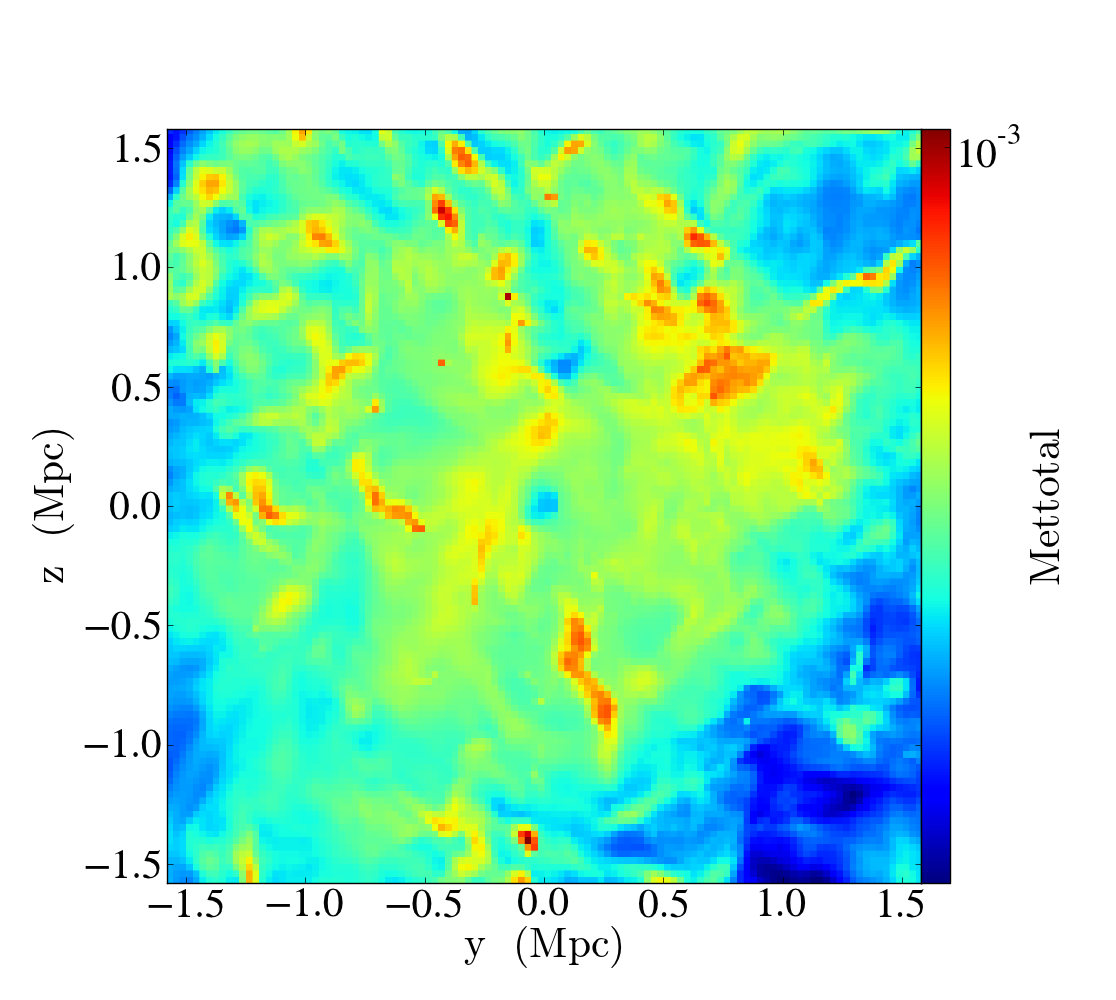}
\includegraphics[scale=0.28]{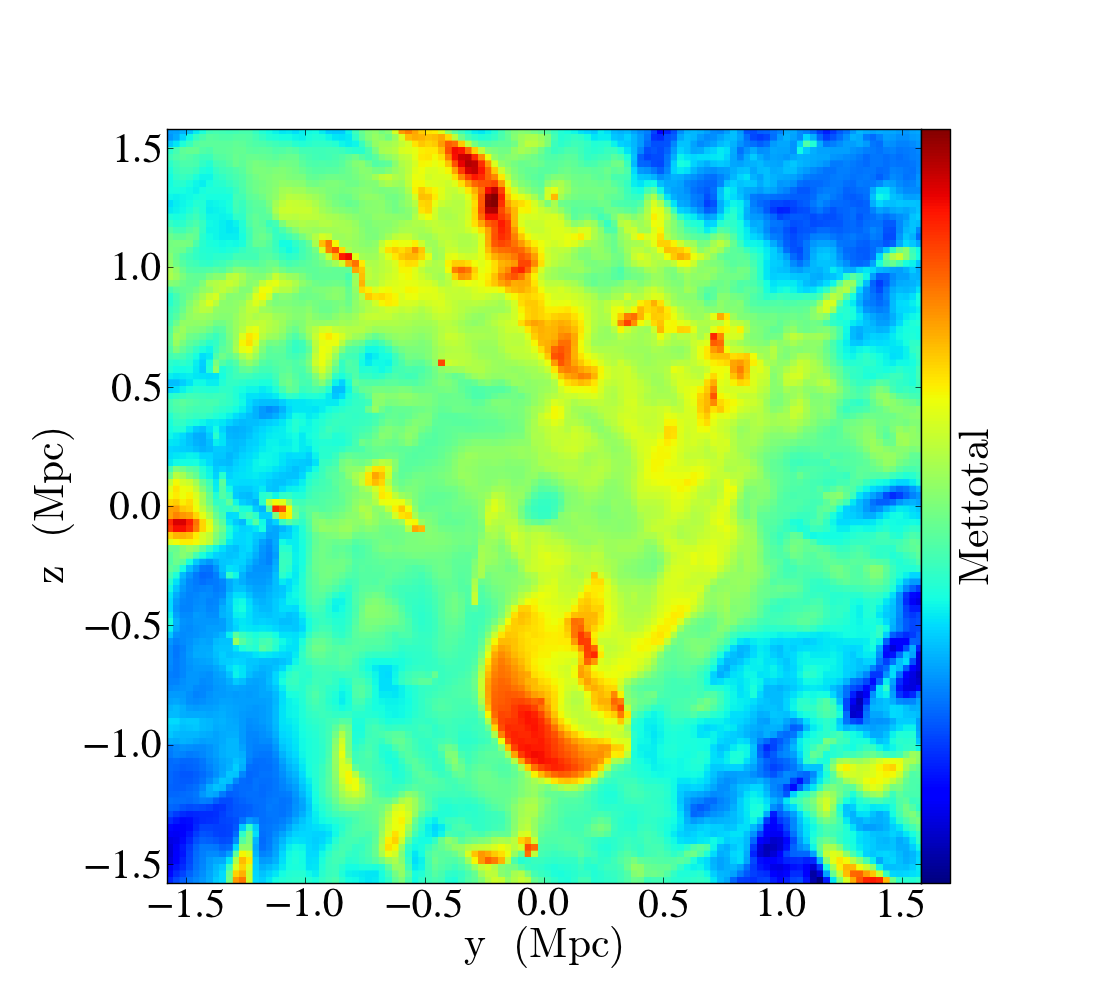}
\includegraphics[scale=0.28]{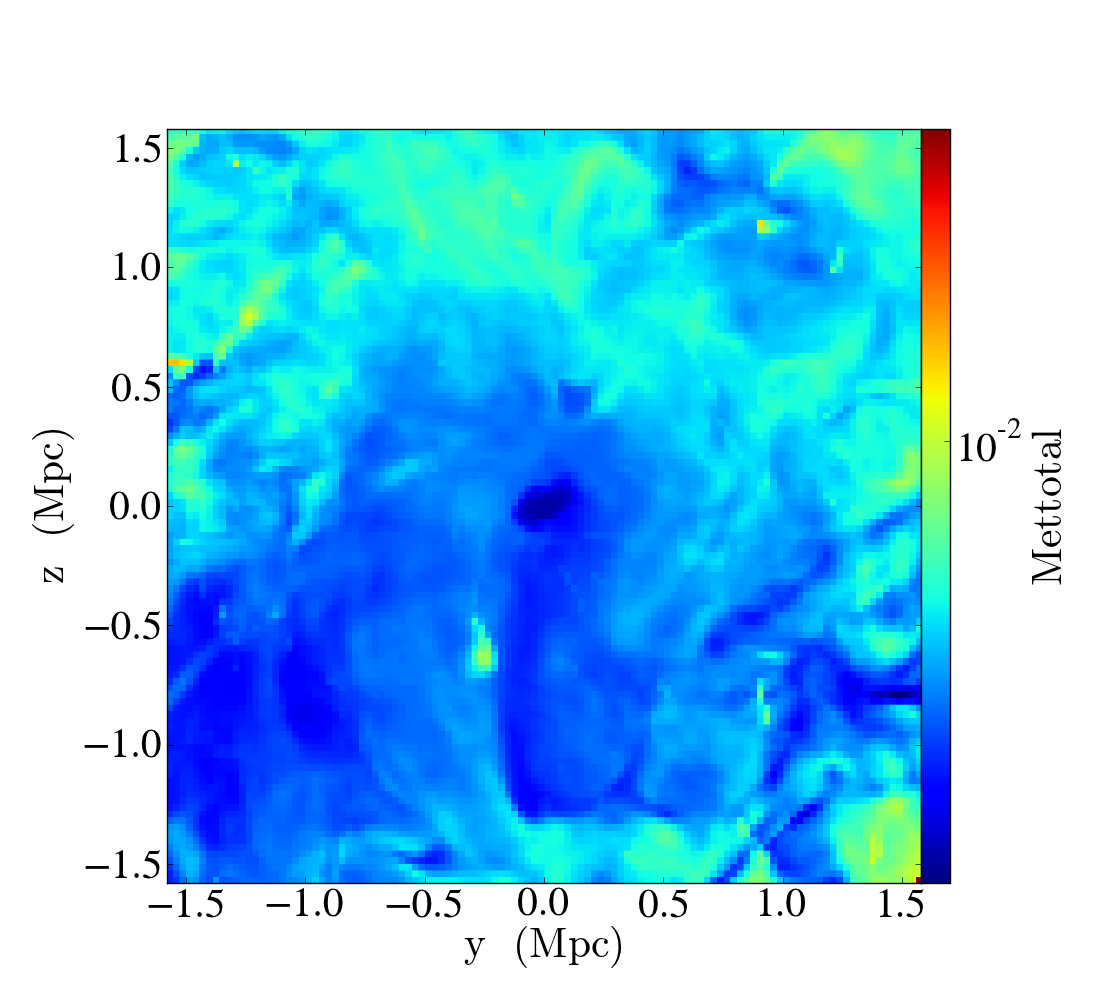}
\caption{Individually color-scaled (\texttt{yt} auto scaling) of X-ray flux-weighted metallicities with 
models \textbf{ A}, \textbf{B'}, and \textbf{C} in $x$-projections. } \label{Fig:modelsProj}
\end{figure}

\begin{figure}
\includegraphics[scale=0.4]{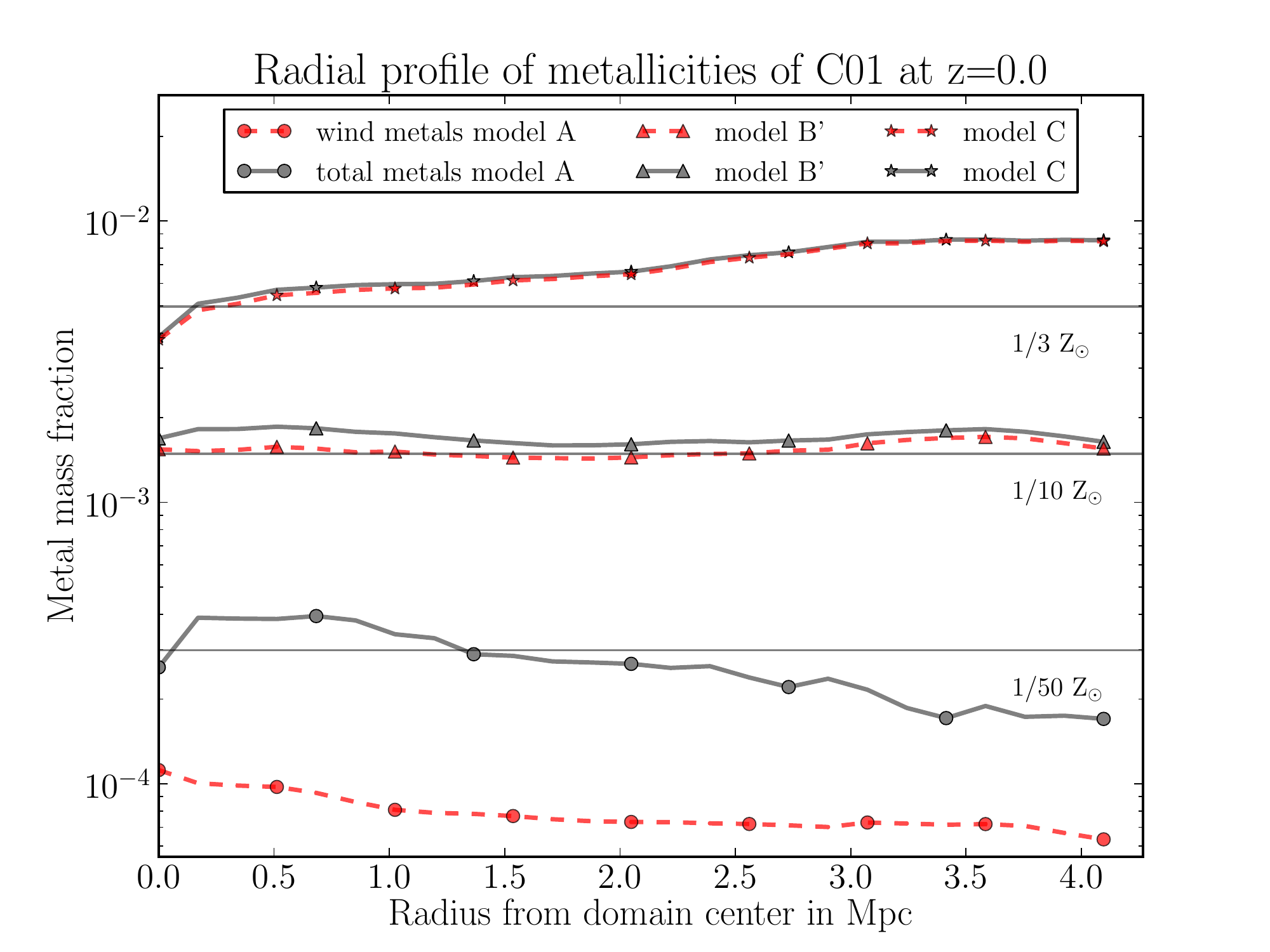}
\includegraphics[scale=0.4]{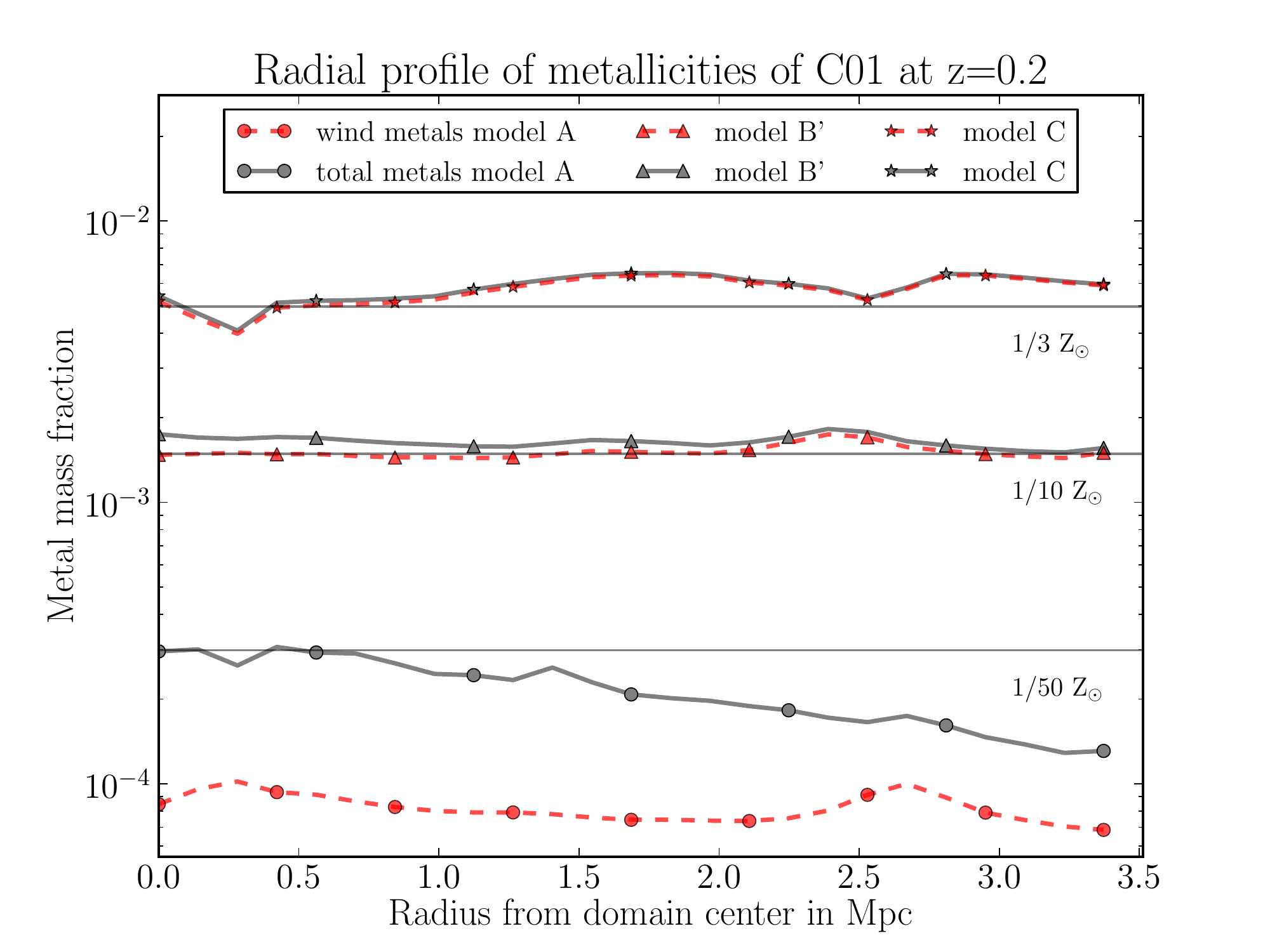}
\includegraphics[scale=0.4]{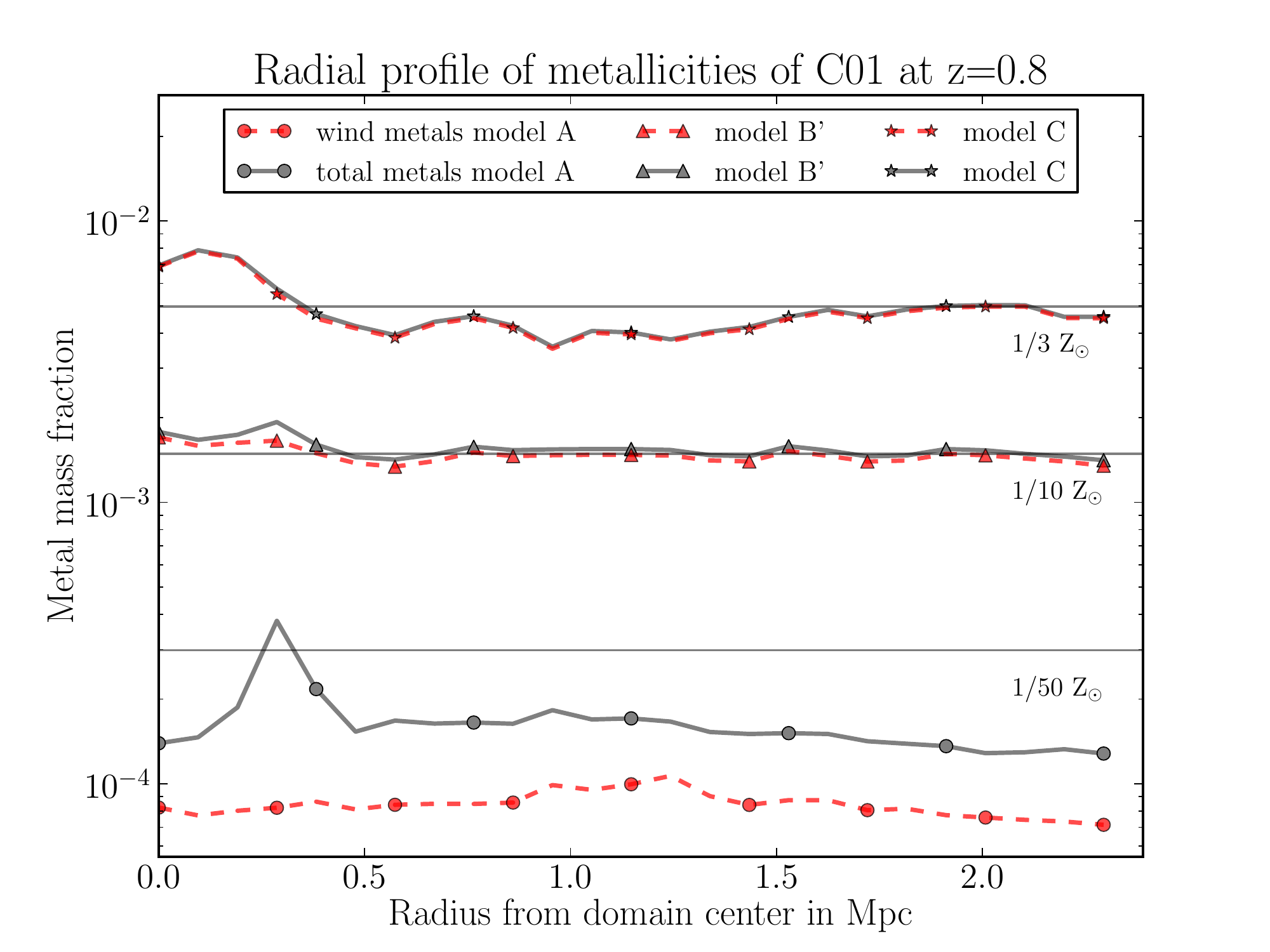}
\caption{Radial metallicity profile of C01 at showing the
contributions from the three wind models at different redshifts. } \label{Fig:metRadEvoC01}
\end{figure}

As mentioned in Sect.~\ref{Subsec:C01}, cluster C01 was simulated with the three different wind models 
we described in Sct.~\ref{Subsec:Winds}. Model \textbf{B} has very high efficiency values (Eq.~\eqref{windEffHopkins}) 
for low-mass galaxies, yielding outflows of up to several thousand $\mathrm{M}_{\odot}/\mathrm{yr}$ for 
starburst episodes. In combination with the assumption of solar metallicity of these winds, first 
simulations (without cooling) also showed ICM metallicities of almost constant value 
$\mathrm{Z}_{\odot}$  at a redshift of about $\mathrm{z} \sim 2$. Observations of ICM 
metallicities and galactic winds suggest that these values are hardly physical. Furthermore, 
the simulation of C01 with wind model \textbf{B} encountered numerical problems with 
a cooling catastrophe. In Fig.~\ref{Fig:metRadEvoC01} we plot the radial profiles of the metal mass fraction of C01 from 
winds and the total for the three models. Here we ran model \textbf{B} with 
the efficiency factor from model \textbf{A}, see Eq.~\eqref{windEffRecchi}, from now on denoted as model \textbf{B'}. 

Clearly, there are not only severe differences 
in the overall metallicity levels, but the shapes of the curves also differ significantly.  
The diversities are striking when the central projected metallicities are plotted in individual 
color scaling, see Fig.~\ref{Fig:modelsProj}. For this plot, the actual scales are not relevant because we only wished to 
compare relative differences. If the models were reproducible by simply 
scaling the levels of metallicities, these projections should look very much alike, which they do not. 
One reason is of course the fractionally different contribution by ram-pressure stripping - the less 
efficient the wind model, the more small-scale structures are
visible  - but also when regarding the wind 
contribution alone (not plotted), the projections show a few common but also distinct features. 
As briefly reported in the cluster description, the cool-core metallicities roughly agree with the central mean metallicities of their formation times. The simulations 
with wind models \textbf{A} and \textbf{B'} have much lower metallicities, and the cool core forms much later and is shallower as well. The wind metallicity profiles of these two runs definitely show
no drop in wind metals toward the center. However, the total metallicity shows a slight dip because the contribution 
from RPS metals in these two cases is larger in fraction. 
This suggests that after the cool core has formed, 
there is virtually no way to enrich it more by any process that we modeled in the 
setup. We discuss this peculiar feature in Sect.~\ref{Subsec:Results}. 

Without introducing any additional parameters  to reproduce observed ICM metallicities,
clearly only model \textbf{C} reaches values similar to those found in studies by~\cite{Tamura2001},
~\cite{Durret2005}, and~\cite{Simionuescu2009}, for example. Without question, one could additionally design the two wind 
prescriptions from~\cite{Hopkins2012} and~\cite{Veilleux2005} specifically to achieve higher metallicities. 
But at this point we were more interested in the robustness of these two wind models 
that originate in galaxy-scale simulations without specific modifications. There obviously is a wind model with an efficiency between those of \textbf{B} and \textbf{B'} that could reproduce expected ICM metallicities. 
We will study this in a future paper.

\section{Comparison with observations} \label{Sec:CompObs}

\subsection{Observations of ICM metallicities} \label{Subsec:Obs} 
Since the first X-ray observations of galaxy clusters, we know that the 
ICM is enriched by heavy elements. Several mechanisms have been proposed 
to explain how these metals were ejected into the ICM. Understanding whether the enrichment occurred 
before or after the cluster formation is also crucial to constrain the importance of each 
mechanism. Galactic winds, for example, are thought to be more important at high redshift because the 
star formation peaked around redshift $z \sim 2-3$~\citep{Hopkins2004}, while disk ram-pressure 
stripping plays a major role in the regions of high density. Recently,~\cite{Werner2013} 
analyzed a set of 84 Suzaku pointings of the  Perseus cluster and derived the metallicity profile 
in eight different directions up to the virial radius. They found a 'remarkably uniform 
iron abundance, as a function of radius', which suggests that the enrichment occurred before and 
during the formation of the cluster. In addition to the metallicity peak associated with the BCG, 
these data agree very well with the qualitative radial trend we found in most
profiles for relaxed clusters within our sample. As in~\cite{Werner2013},   previous studies by \cite{Matsushita2011},~\cite{Urban2011}, and ~\cite{Leccardi2008}, for instance, also showed either flat or 
slightly radially increasing metallicities in ranges typically
between $1/10-1\mathrm{Z}_{\odot}$. However, while observational high-temperature 
metal data mainly refer to Fe measurements, the total metallicities within this work 
can be considered to be O dominated. Therefore a complete quantitative comparison will only 
be possible with single-element tracking, which is currently being implemented into 
the presented simulation setup and which will be introduced in future investigations.
 While there are only few observational single-element studies because of the stringent 
flux limitations, analyses by~\cite{Tamura04} and ~\cite{Matsu07} also found 
comparatively flat trends in O.
Another definite limitation in comparing the radial plot of Fig.~\ref{Fig:radWind} with the 
observational profiles is that 
in the simulation data we did not separate X-ray detectable ICM and gas that does not 
significantly contribute to the X-ray flux (which is proportional to $\rho^2 \sqrt{T}$). 

Because mixing within the ICM homogenizes the distribution of all metal species, the degree 
of structure seen in the metallicity is a promising tracer for the enrichment history. 
Radial profiles fail in detecting the local inhomogeneities of the metals, and 
detailed 2D metal maps are thus required. Metallicity maps, in fact, are the perfect 
tools for studying the efficiency of the different processes. Although we 
know that the metallicity distribution in galaxy clusters shows significant 
inhomogeneity even in relaxed clusters~\citep[e.g.,][]{Lovisari2011}, we 
expect a more clumpy pattern for disturbed systems.  One clear 
example of this is the distribution of metals in the famous merging 
cluster A3667~\citep{Lovisari2009}. The metal peak was detected to be 
between the two main substructures and probably caused by galactic winds 
associated with the new star formation triggered by the merging 
activity.  This is not the only example, and other works showed 
that there are  maxima in the metallicity distribution that are not 
associated with the cluster center~\citep[e.g.,][]{Durret2005,Simionuescu2009,dePlaa2010,Kirk2011}. 
A direct comparison between observations and simulations is now required  
not only for single objects, but also from the statistical point of view 
for a full understanding of the ICM enrichment. While most of the work 
until know focused on the iron distribution, a complete view of the metal-enrichment processes can be only obtained by additionally studying the 
distribution of other elements (e.g., O, Si, and S), as done by~\cite{Sanders2004} for the 
Perseus cluster. Clearly, this requires a large number of photons 
(i.e., very long observations) and some prediction from the simulations.
In 2015 ASTRO-H~\citep{AstroH2012} will be launched. One of its science goals is
determining the metal distribution in galaxy clusters.

\subsection{Comparing the models with observations of galactic winds} 

\begin{figure*}
\centering
\includegraphics[scale=0.6]{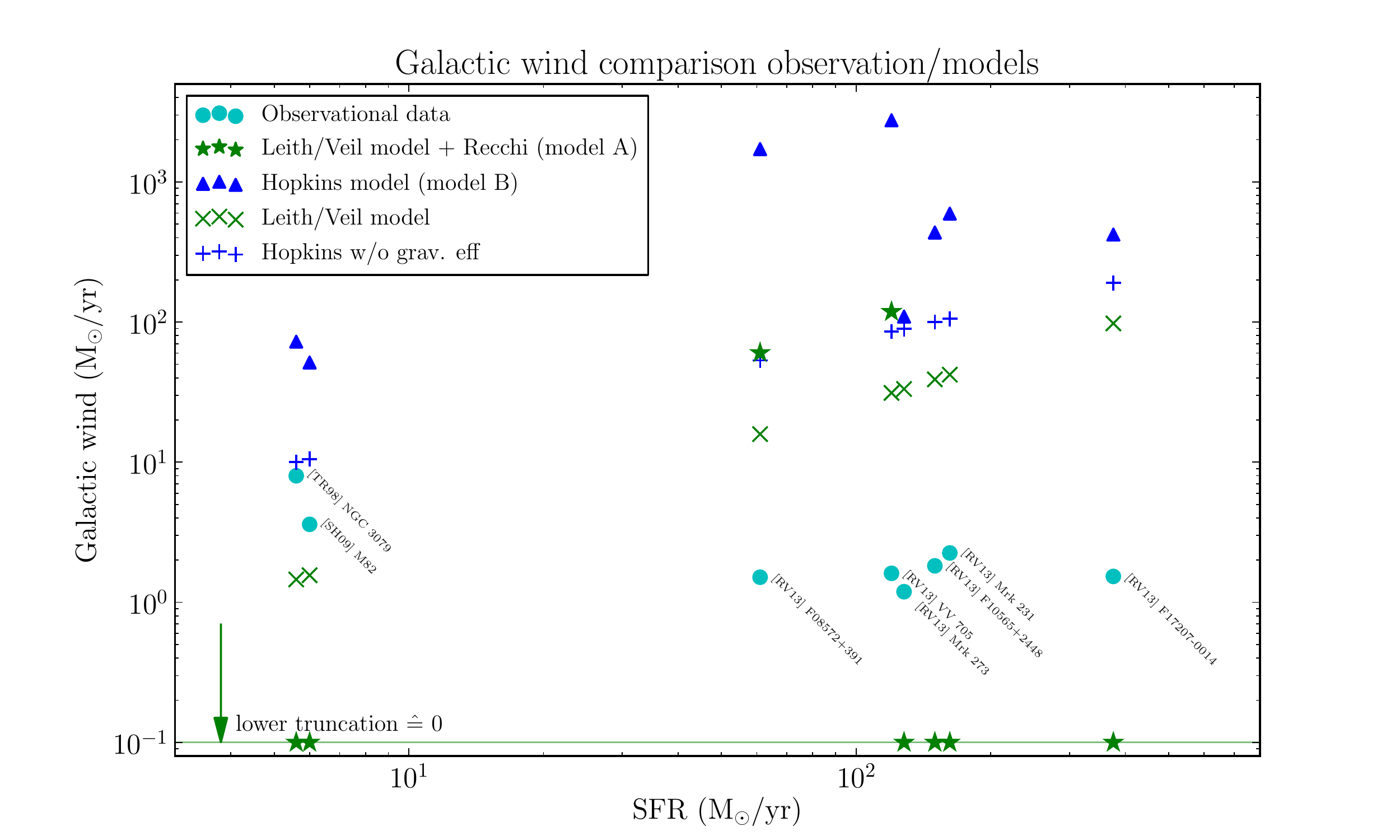}
\caption{Comparison of wind observations with models. Data on star formation rates and
galactic winds taken from~\citet{Rup+Veil13},~\citet{Strickland2009}, ~\citet{Trotter98}, mass estimates
taken from~\citet{Dasyra06} and $S_{\text{0.5}}$ fits for baryonic mass determination taken from~\citet{Catinella12}.  }
\label{Fig:WindObs}
\end{figure*}

In Fig.~\ref{Fig:WindObs} we compare observational estimates of galactic winds with the model 
predictions of the prescriptions presented in Sect.~\ref{Subsec:Winds}. While the observational 
data are very sparse, model \textbf{B} seems to overestimate the 
galactic wind at a given SFR and baryonic mass for all the objects. More data are definitely 
needed, especially for lower-mass systems and high SFRs - the galaxies that 
are thought to contribute a major fraction of ICM metals.
Because the gravitational efficiency in models \textbf{A} and \textbf{B'} is zero for objects with 
a baryonic masses higher than $10^{9.4}M_\odot$ and the sample is strongly biased on the high mass end, 
there are some objects with total wind suppression. The two data points 
from~\cite{Trotter98} and~\cite{Strickland2009} seem to agree best with the~\cite{Hopkins2012}
prescription without the gravitational efficiency parameter. 

\subsection{Simulation results} \label{Subsec:Results}

\begin{figure*}
\centering
\includegraphics[scale=0.6]{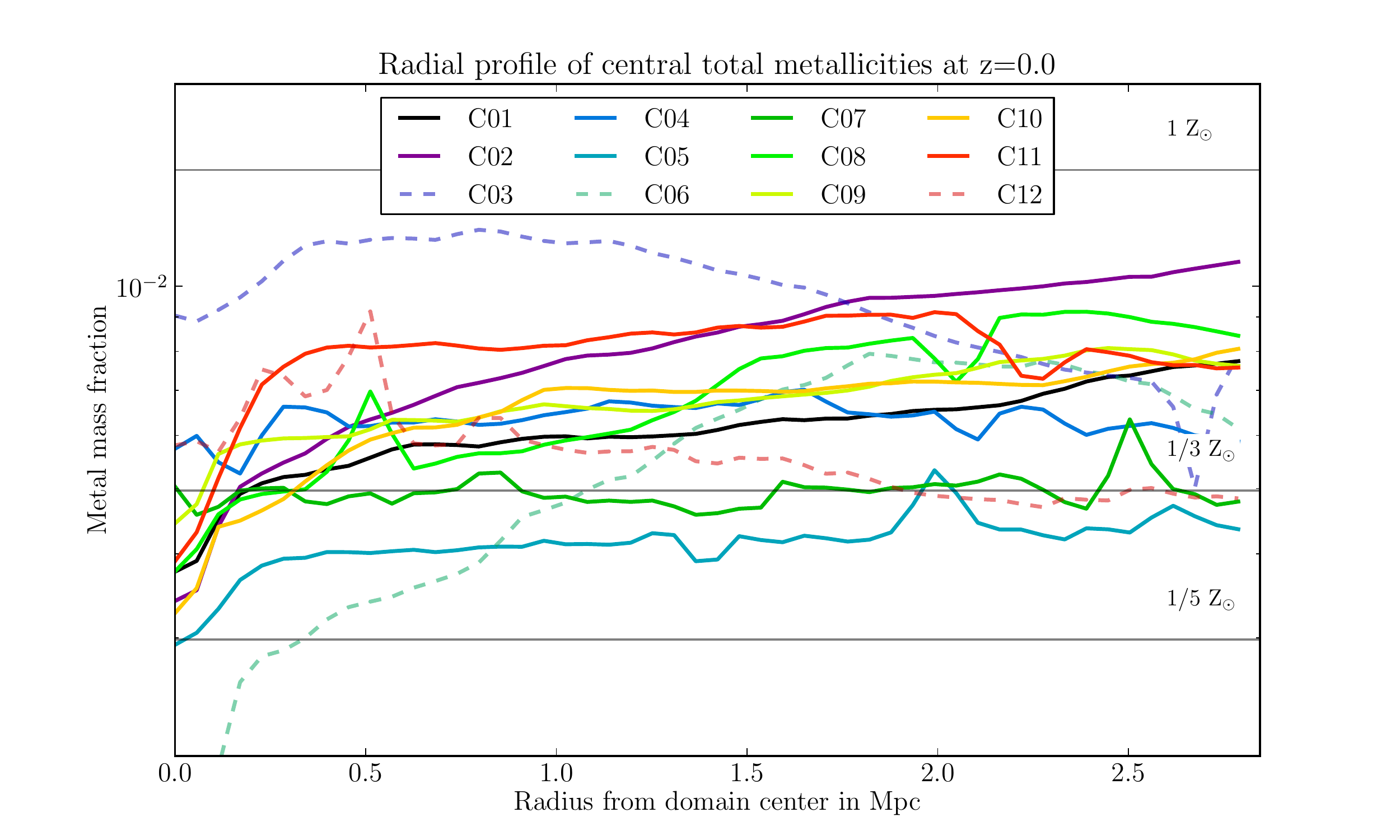}
\caption{Radial profiles of the total metal mass-fractions. The dotted cases are classified
as \textit{untypical} because of their kinetic states. }  \label{Fig:radTot}
\end{figure*}

As already mentioned in the previous section, the overall level of metals in the ICM in 
the simulations with wind model \textbf{C} agree very well with observations (see Fig.~\ref{Fig:radTot}). 
There is a distinct difference between our results and a number of measurements of central metallicities, however. Usually, 
observers find a peak in central metallicity, especially in cool-core clusters~\citep[see, e.g.,][]{David2001,Sanders2004,Sanders2007}. But there are also observations that
show the opposite trend~\citep{Sanderson2009} at the very center of cool-core clusters, which provide a 
striking indication of a bimodality between CC and non-CC clusters. These observations were fitted with a single-temperature 
model for their spectra, which is considered less  well suited than multitemperature profiles 
for cool-core clusters, see~\cite{Boute2000}.  

Our results showed an interestingly narrow distribution of core metallicities ($0.2-0.4\mathrm{Z}_{\odot}$) 
when only virialized and semivirialized systems were considered.
This result was also found observationally by~\cite{deGrandi2009}. 
Our analysis showed striking evidence that the lack of mixing within and in the vicinity of the cool core 
causes the metal mass-fraction to \textit{freeze} at the mean level of the time when the cool cores form 
(compare Figs.~\ref{Fig:metEvoC01},~\ref{Fig:metEvoC04},~\ref{Fig:metEvoC07}, and ~\ref{Fig:metEvoC10}). 
While the region depleted of wind metals can be larger than the extent of the cool core itself, the 
depleted volume from ram-pressure-stripped metals is smaller in size and similar to or smaller than 
the cool core. This is probably best seen for cluster C11 by
comparing Figs.~\ref{Fig:RPSall}(k) and~\ref{Fig:Windall}(k).

However, this finding should not be overinterpreted, because many numerical studies have shown a tendency 
to form cool cores in cluster simulations without introducing additional heating processes, such as 
AGN feedback. In this sense, the cool cores presented in this study are probably artificially deep and generally 
more prone to form than in models with additional heating prescriptions. It is remarkable, however, 
how well the core values agree with the mean metallicity values at the redshifts when they form (usually $\mathrm{z}\sim 1.5 -0.8$). 
There is no currently modeled physical process that would explain an increase  in metallicities toward the 
cool cores. The metal mass-fraction would not show any gradient if mixing were completely perfect; the less efficient the 
mixing, the lower the metallicity in the core, one would assume. An increase in core metal mass-fraction can only 
be explained if the metals are produced in the center. With the current semi-analytical model \textsc{Galacticus}, 
we do not see significant SFR in the BCGs that would yield an appropriate amount of SN-driven metal outflows within the 
model presented in Sect.~\ref{Subsec:Winds}. The early formation of a dense, high-pressure core suppresses galactic winds 
in this region. 
The length scale over which the ICM metallicity drops in Fig.~\ref{Fig:radTot} is in the order of 200 kpc. This length scale
is clearly associated with the BCG, a giant elliptical galaxy that (at the current setup status) is treated 
equally with all other galaxies. However, the characteristic length scales of a cD galaxy are significantly larger 
than the central resolution of the hydrodynamical simulation (20 kpc/h), and the gas residing in this object should 
be considered a part of the ICM, while in our model it is not transported into the ICM reservoir because of the mentioned 
wind suppression.  
While this weakness of our simulation setup will be removed in future revisions, we found this interesting 
mechanism of \textit{metallicity freezing} in the cluster cores, which present themselves almost unmixed with 
the surrounding ICM. This result also supports the hypothesis of~\citep{Boehringer2004}, who analyzed iron-to-silicon 
ratios in CC clusters. They claimed that `the central excess seen in cooling core clusters is most probably due to enrichment 
by SN type Ia in the cD galaxies in the recent past, during which the central region of the cluster was not disturbed'.  
While we see no indications that turbulent mixing is only suppressed recently, but rather from the time of 
CC formation on, we will study this phenomenon in the future in detail and also apply single-element tracking in 
the feedback processes.   

Moreover, although the inner 2.5 Mpc as plotted in Fig.~\ref{Fig:radTot} accounts for less 
than 1/1000 of the total simulated volume, there might be a radial bias toward higher metallicities in more massive 
systems because the drag of gaseous material other than DM is only taken from the $28 \textrm{ Mpc}^3$ box, and the 
inflow over the boxes boundaries is definitely underestimated. Comparisons of radial density profiles
have shown, however, that this effect is noticeable only for the highest-mass objects, such as C01 and C02 in this sample.

In Sect.~\ref{Subsec:Obs} we showed that to study the kinetic state of a galaxy cluster 
and the efficiency of the diverse enrichment processes, it is desirable to compare 2D metallicity 
maps with simulations. The richness of structures in metal mass-fraction is not shown to 
its full extent because we used common color maps (see Fig.~\ref{Fig:Coreall}) in a rather diverse sample of objects. 
The X-ray flux-weighted metallicity projections nevertheless show a wide spectrum of metal distributions. 
Although we are aware of the small sample and the degeneracy of the projection direction, we wish 
to point out a few common but distinct features. This work should be considered as the basis for
more systematic and more homogeneous choices of cluster samples for statistical analysis in 
conjunction with observational data. 

Apart from in situ subgrid model simulations such as that by~\cite{Tornatore2007}, clusters in all 
kinetical states show structures and inhomogeneities of metallicities on galactic scales. Metallicity 
peaks are directly associated with starburst episodes or ram-pressure stripping events of single galaxies 
modeled by the SAM.  
\begin{figure}
\centering
\includegraphics[scale=0.28]{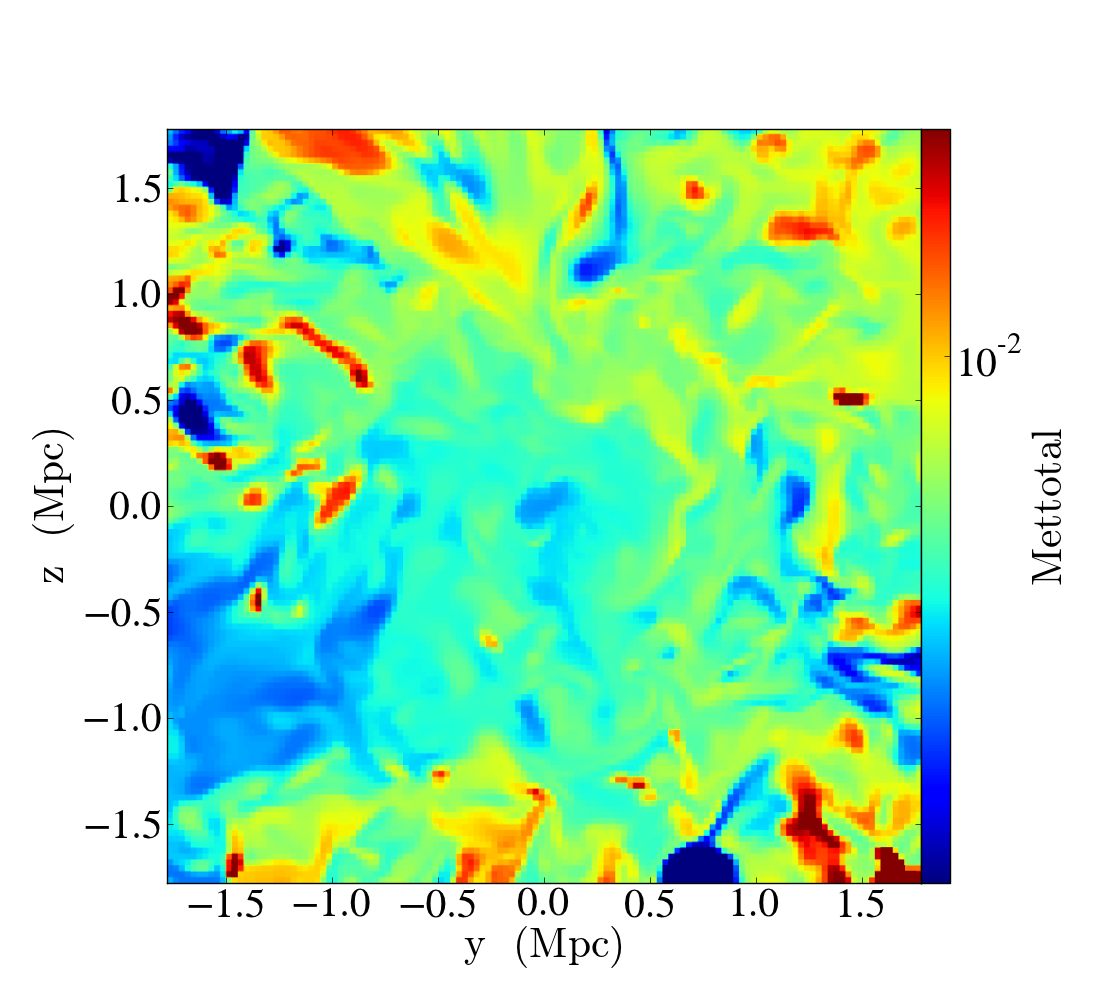}
\caption{Central $(y,z)$-slice of the total metal mass-fraction of cluster C01 at $\mathrm{z}=0$ 
showing galactic-scale inhomogeneities within the ICM. The color range 
is equivalent to the interval $0.13-1.3 \mathrm{Z}_{\odot}$. }  \label{Fig:C01SliceMet}
\end{figure}
\begin{itemize}
\item
High-mass relaxed systems show efficient mixing of galactic wind metals within the shock-heated 
ICM volume. Clusters C01, C09, and C11 show a remarkably symmetric spatial metal distribution on large scales  
(Figs.~\ref{Fig:Coreall}(a),(i), and (k)). However, this degree of homogeneity is to be understood 
as relative to the other investigated systems. The central region of C01 is plotted in Fig.~\ref{Fig:C01SliceMet} 
in a $(y,z)$-slice. It shows much small-scale inhomogeneities and structures - in this case 
using a slightly different color scale of $0.13-1.3 \mathrm{Z}_{\odot}$
-  that trace recent stripping events 
and wind mass-loss of single objects. 
There is a radial break in large-scale homogeneity at the edge of the shocked ICM; the infalling gas shows much
higher metallicity contrasts (orders of magnitude). A strong metallicity depletion of the 
cool core is a another common feature of these clusters. 

\item The cluster C02 has undergone a major-merger event at $\mathrm{z} \sim 0.2$, the cluster is not virialized. The 
metallicity projection suggests two planes of symmetry, along and perpendicular to the merger direction.  
Although not a direct observable, the RPS metals in C10 (Fig.~\ref{Fig:RPSall}(b)) show much structure, 
an indication that the shock-induced turbulence has only recently started to efficiently mix the material.  

\item The lowest-mass clusters display also the lowest level of metal mass-fraction, while C05, a relaxed system, 
again shows a high degree of radial symmetry, while the pre-merger object C07 presents itself much more 
structured and inhomogeneous, see Fig.~\ref{Fig:Coreall}(g). 

\item The two pre-merger subclusters in C06 reveal unusually large volumes of metallicity depletion, while the 
projection in Fig.~\ref{Fig:Coreall}(f) shows much structure and comparatively high metallicity values outside these regions. 

\item The multiple-merger system C12 displays an interesting bimodality where the upper left part in the 
$x-y$ projection is smoother than the inhomogeneous lower right subcluster(s). The elongated supersolar metallicity 
peak at the center coincides with a heavy shock from the ongoing merger. 

\item The low-metallicity, cool infalling structure in the left upper part of the metal mass-fraction projection of 
C10, Fig.~\ref{Fig:Coreall}(j), \textit{pushes} a higher metallicity bow-shock region in front of it. 

\item As also seen in the radial profiles, all cool-core clusters show a decrease in metallicity at their centers. The 
spatial extent of these low metal mass-fraction regions is smaller in RPS metal contribution than in wind metals. 

\end{itemize}

\section{Summary and outlook}\label{Discussion}

We presented a study on the temporal and spatial distribution of metals in the intracluster medium within 
a heterogeneous sample of 12 cosmological simulations of galaxy clusters. We applied a novel simulation 
setup that combines a solid and comprehensive subgrid model by including the well-tested semi-analytical 
model \textsc{Galacticus} in grid-based hydrodynamical simulations. 
The strength of this setup is to robustly model the histories of a large number of objects (tens of thousands 
of galaxies) based on their merger histories within a high-resolution hydrodynamical treatment
of the ICM, which facilitates studying the dynamics and chemical evolution. 

\noindent The main results can be summarized as follows: 
\begin{itemize}

\item The combination of the semi-analytical galaxy formation model \textsc{Galacticus} and a grid-based hydrodynamical 
cosmological simulation has proven reliable in reproducing a number of observables, such as the star formation histories, 
a radial trend in SFR within the galaxy cluster environment, and the overall level of metallicities in the ICM.

\item The diverse galaxy cluster samples show common and distinct features in radial trends and 2D metallicity 
projections. Relaxed objects display efficiently mixed metal mass-fractions within the shock-heated regions. Kinetically 
peculiar clusters display significantly more structure and individual characteristics.  
2D metallicity maps are more adequate for a detailed analysis of feedback process efficiencies and for tracing 
the kinetic state of galaxy clusters. 

\item Galactic winds dominate the overall contribution to ICM metals, while RPS adds at low redshifts and shows a 
significant radial trend. The radial course in total metal mass-fraction shows less spread at smaller radii, and 
cool-core clusters are metallicity depleted. The analysis showed that the mean metal mass-fraction 
values at the time of the formation of the CCs coincides with the final level within the central few hundred 
kpc. Turbulent mixing of ICM material within the CCs is heavily suppressed in our simulations. 

\item The comparison of wind models showed that without additional parameters and/or
models, only wind model \textbf{C} can reproduce metallicity values similar to those seen in X-ray observations. The formation 
time of the cool cores strongly depends on the ICM metallicities; depth and size of the metallicity dip 
in CCs is smaller the later they form.

\item  The main weakness of the simulation setup in the current status is that it neglects the ISM within the BCG, which is the most likely cause for the metal mass-fraction drop in the cluster cores. 
Results suggest that this material must be considered as part of the ICM in future studies. 

\end{itemize} 

The next step is the implementation of single-element 
tracking into our feedback models. This way, the comparison with observational data will be even more 
effective and reliable in analyzing the feedback process efficiencies as a function of space and time.
 We are furthermore working on an implementation of {\sc Galacticus} in which ram-pressure 
stripping is computed from real relative velocities and external ICM densities. This next 
release will reproduce even more realistic star formation rates within galaxy 
clusters and also model star formation quenching from disk ram-pressure stripping.
  
We will analyze the formation of cool cores when AGN feedback is included in the computations. Furthermore, the gas that is
currently \textit{captured} in 
the BCG by suppression of the transport process will be considered part of the ICM. In combination with the 
single-element tracking, we aim to be able to test the hypothesis of~\cite{Boehringer2004}, who claimed that 
the central metals originate more likely from supernovae type Ia in these central elliptical galaxies.

\begin{acknowledgements}
The authors acknowledge the UniInfrastrukturprogramm des BMWF
Forschungsprojekt Konsortium Hochleistungsrechnen, the
Forschungsplattform Scientific Computing at LFU Innsbruck and the doctoral school -
Computational Interdisciplinary Modelling FWF DK-plus (W1227).
The software used in this work was in part developed by the DOE NNSA-ASC OASCR Flash 
Center at the University of Chicago.
We  thank Matteo Bianconi, Asmus B\"ohm, Benjamin B\"osch, and Sabine Schindler for 
many valuable discussions and their perspective from observational (X-ray) astronomy. 
We furthermore acknowledge the anonymous referee, who
helped to substantially improve the paper.  
\end{acknowledgements}

\bibliography{cluster,thesis}
\renewcommand{\bibsection}{\section{References}}
\setlength{\bibhang}{1.24cm}
\setlength{\parindent}{3cm}
\setlength{\bibsep}{0cm}
\bibliographystyle{dcu}
\setcitestyle{authoryear,round,citesep={;},aysep={,},yysep={;}}
\gdef\harvardand{\&}

\begin{appendix}
\section{Graphical Appendix} \label{Sec:App}

\begin{figure*}[h!]
\centering
\begin{subfigure}[b]{0.25\textwidth}
\includegraphics[width=\textwidth]{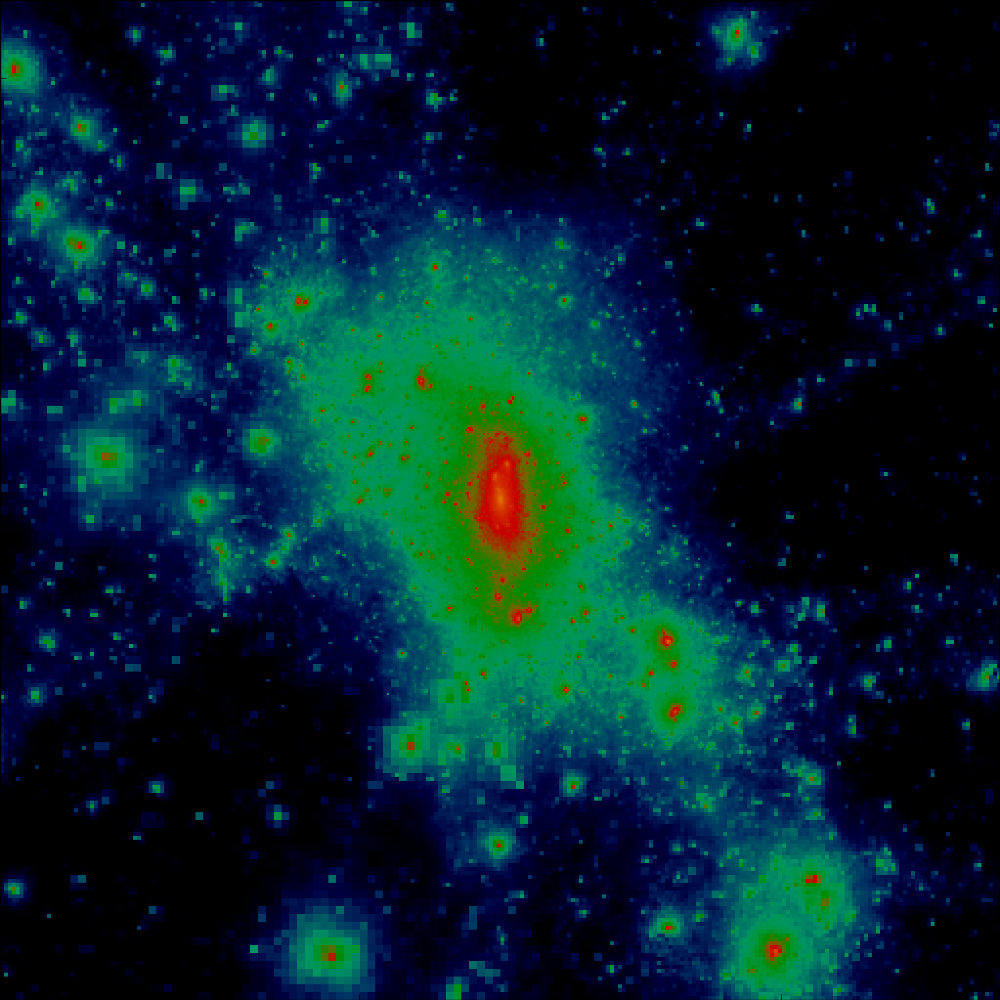} 
\label{Fig:DMC01} \caption{C01}
\end{subfigure}
\begin{subfigure}[b]{0.25\textwidth}
\includegraphics[width=\textwidth]{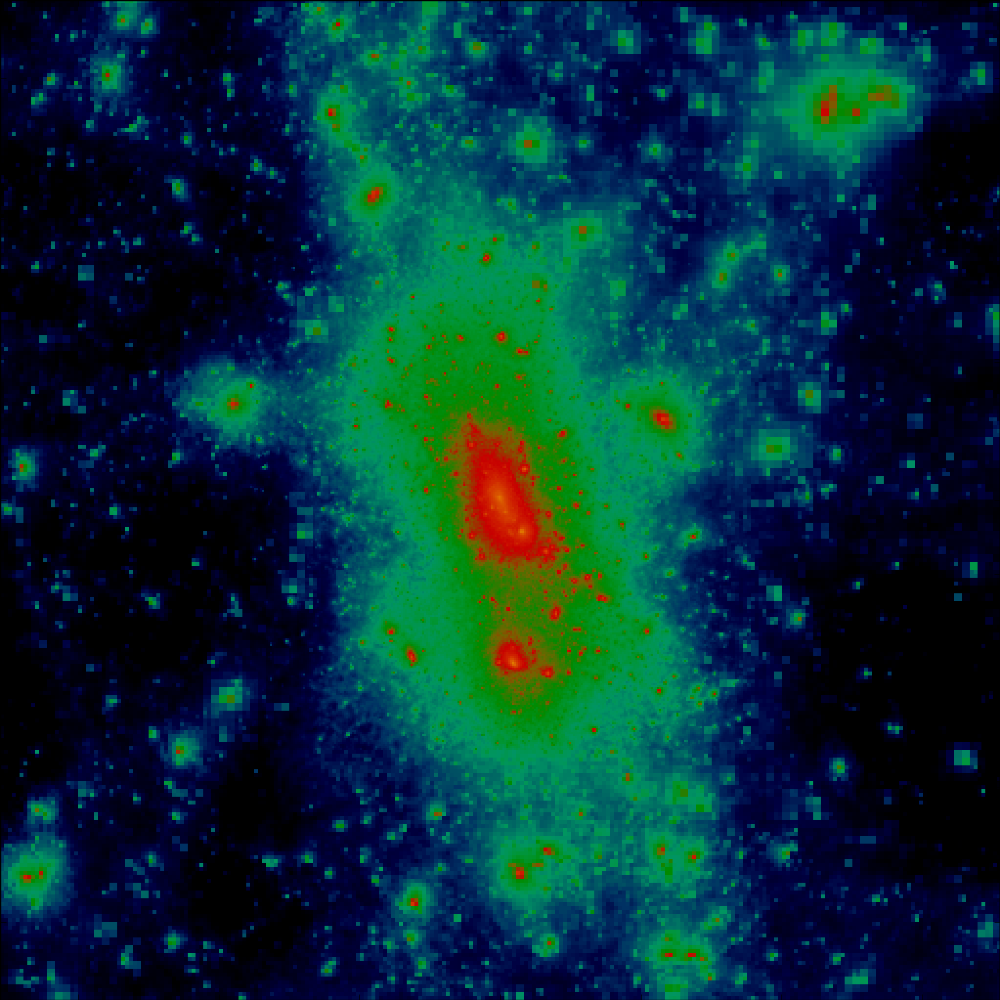} 
\label{Fig:DMC02} \caption{C02}
\end{subfigure}
\begin{subfigure}[b]{0.25\textwidth}
\includegraphics[width=\textwidth]{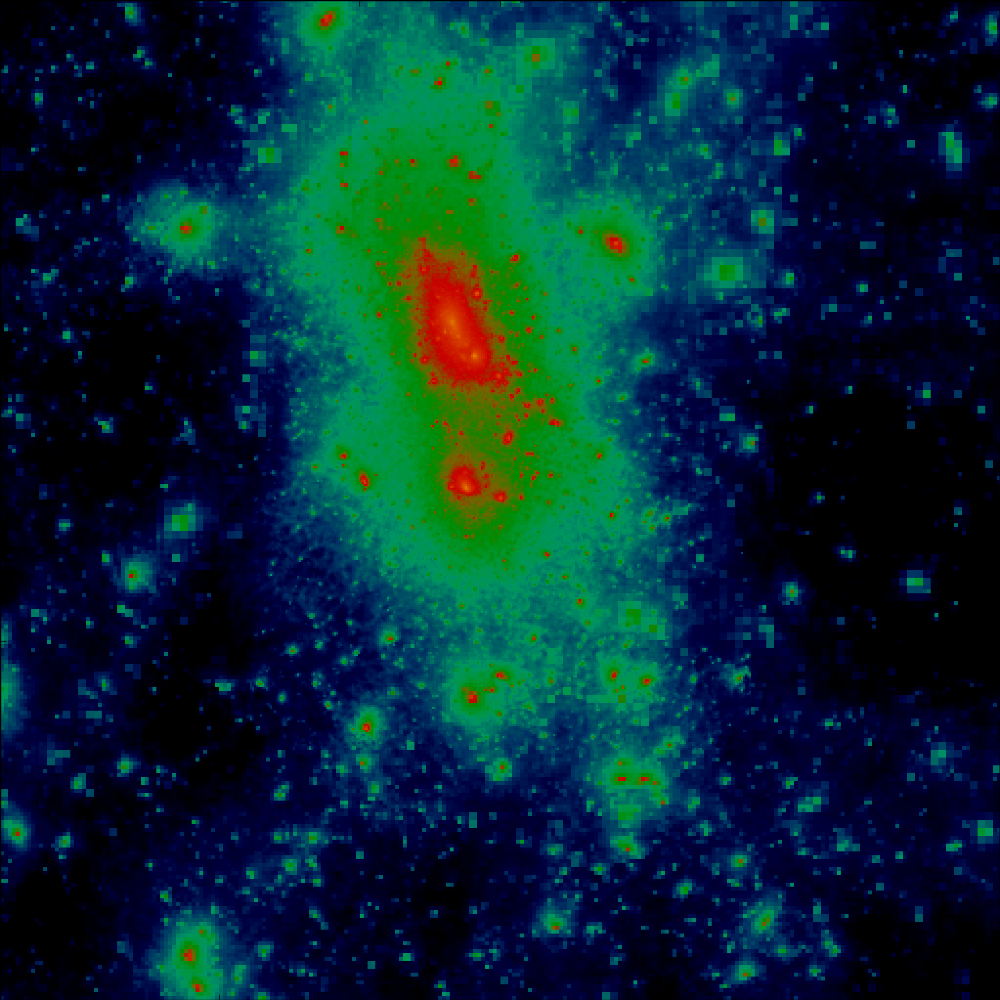} 
\label{Fig:DMC03} \caption{C03}
\end{subfigure}

\begin{subfigure}[b]{0.25\textwidth}
\includegraphics[width=\textwidth]{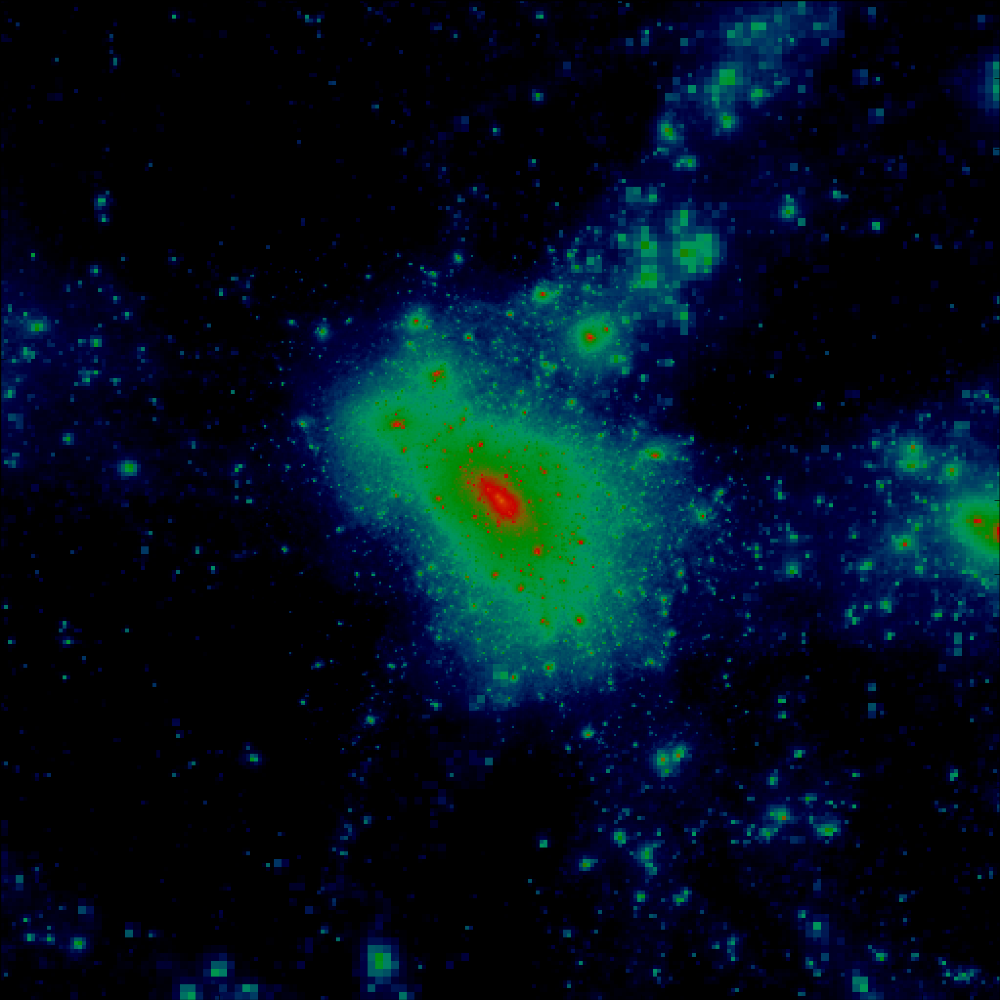}
\label{Fig:DMC04} \caption{C04}
\end{subfigure}
\begin{subfigure}[b]{0.25\textwidth}
\includegraphics[width=\textwidth]{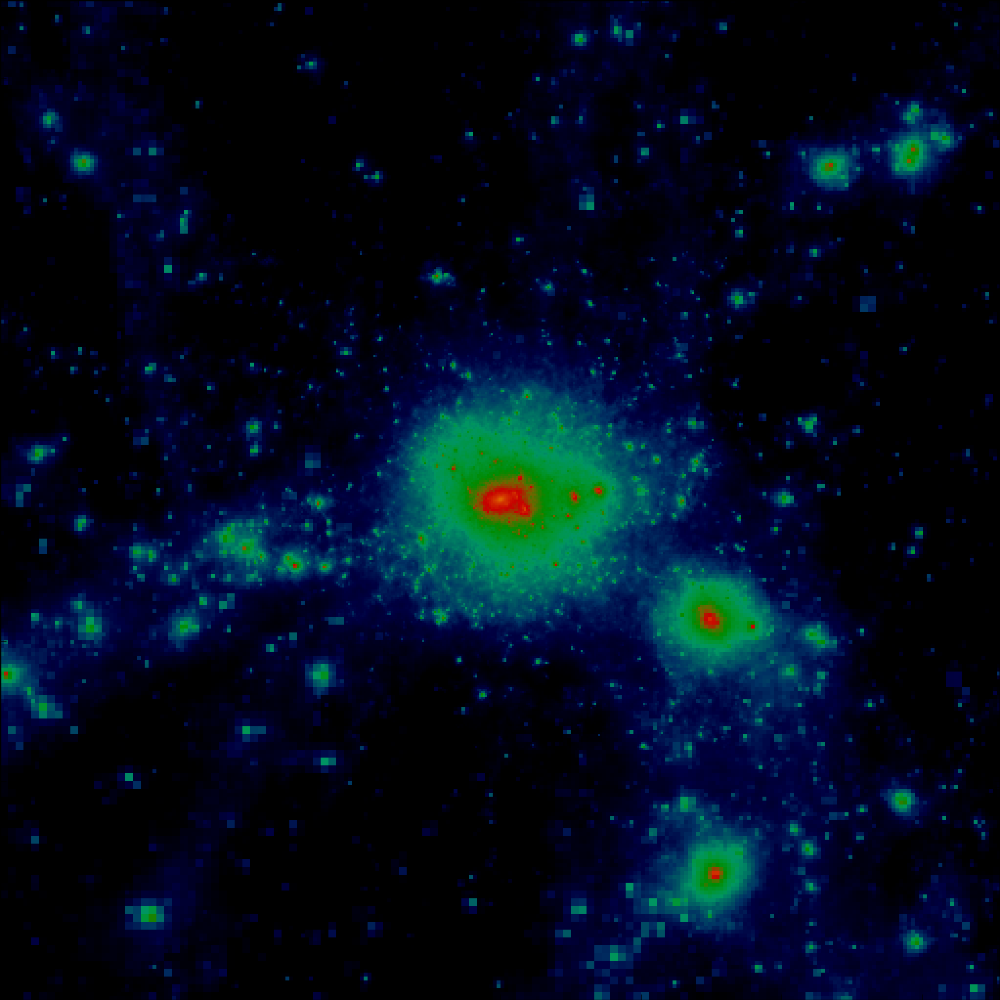}
\label{Fig:DMC05} \caption{C05}
\end{subfigure}
\begin{subfigure}[b]{0.25\textwidth}
\includegraphics[width=\textwidth]{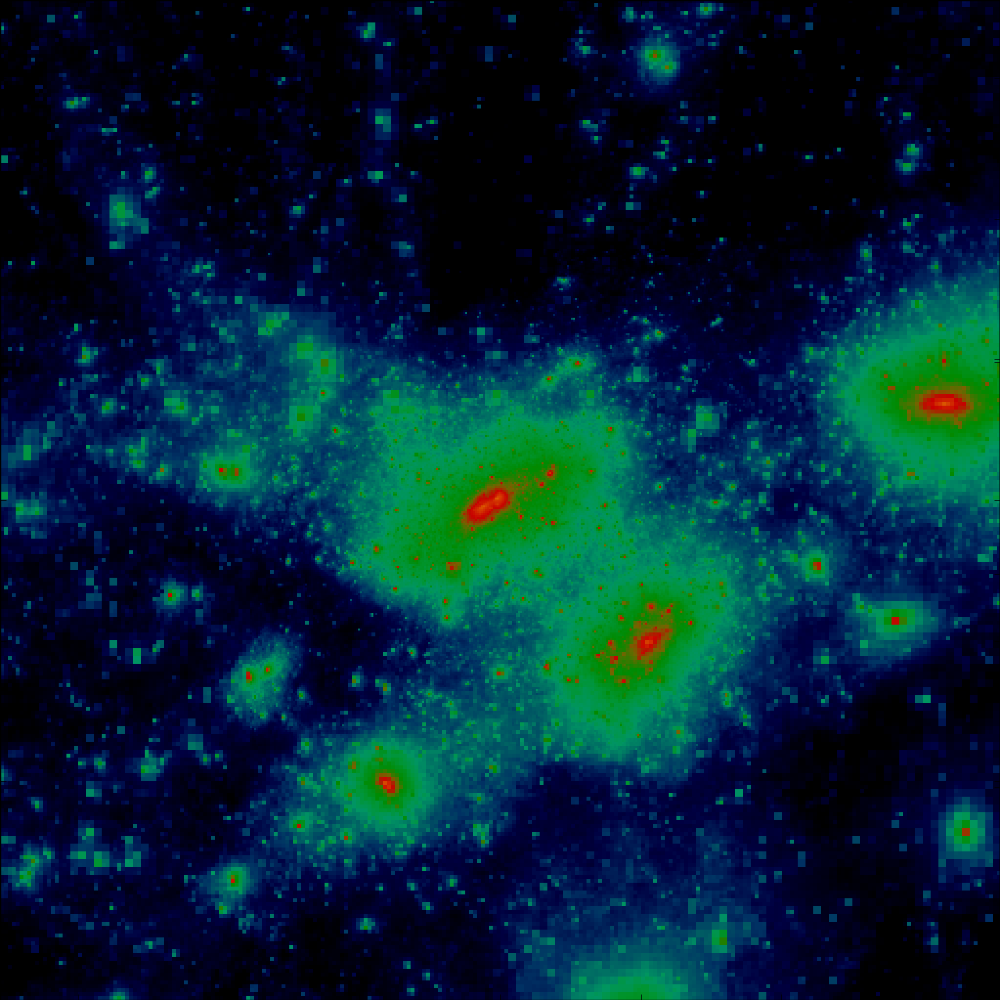}
\label{Fig:DMC06} \caption{C06}
\end{subfigure}

\begin{subfigure}[b]{0.25\textwidth}
\includegraphics[width=\textwidth]{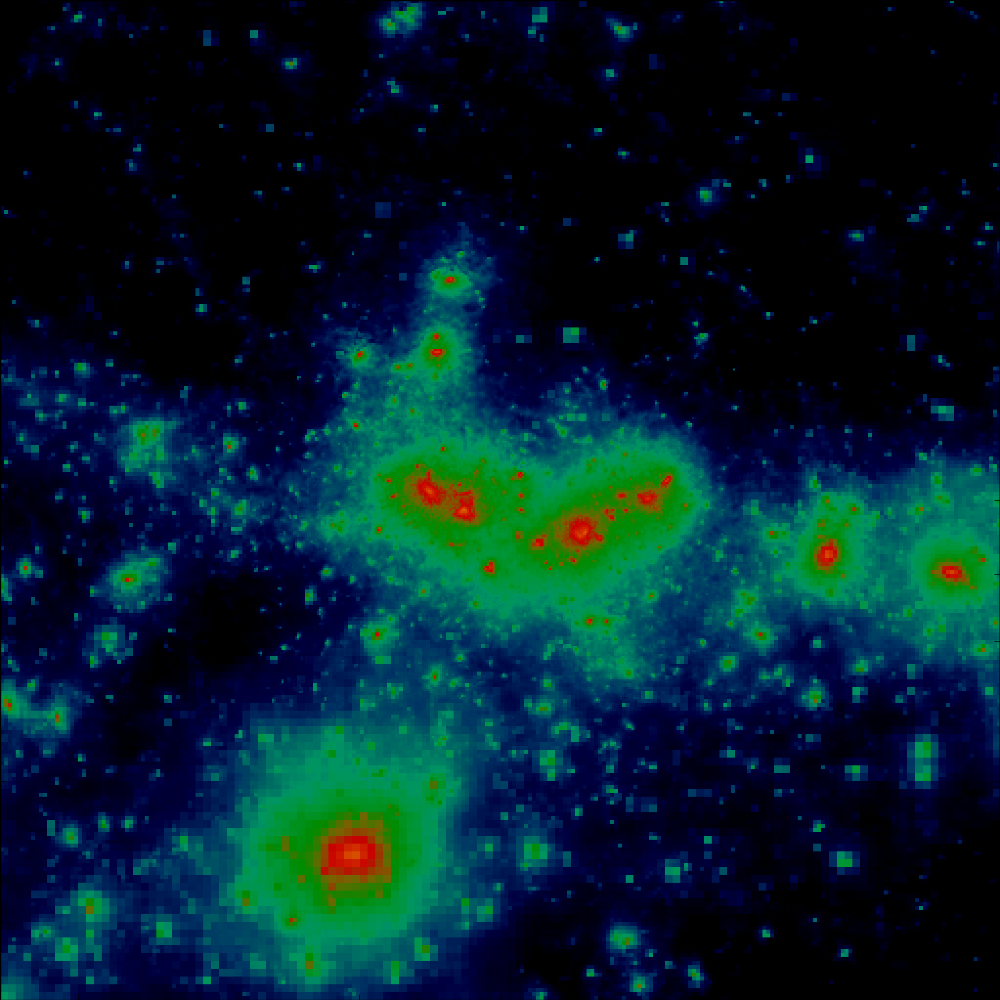}
\label{Fig:DMC07} \caption{C07}
\end{subfigure}
\begin{subfigure}[b]{0.25\textwidth}
\includegraphics[width=\textwidth]{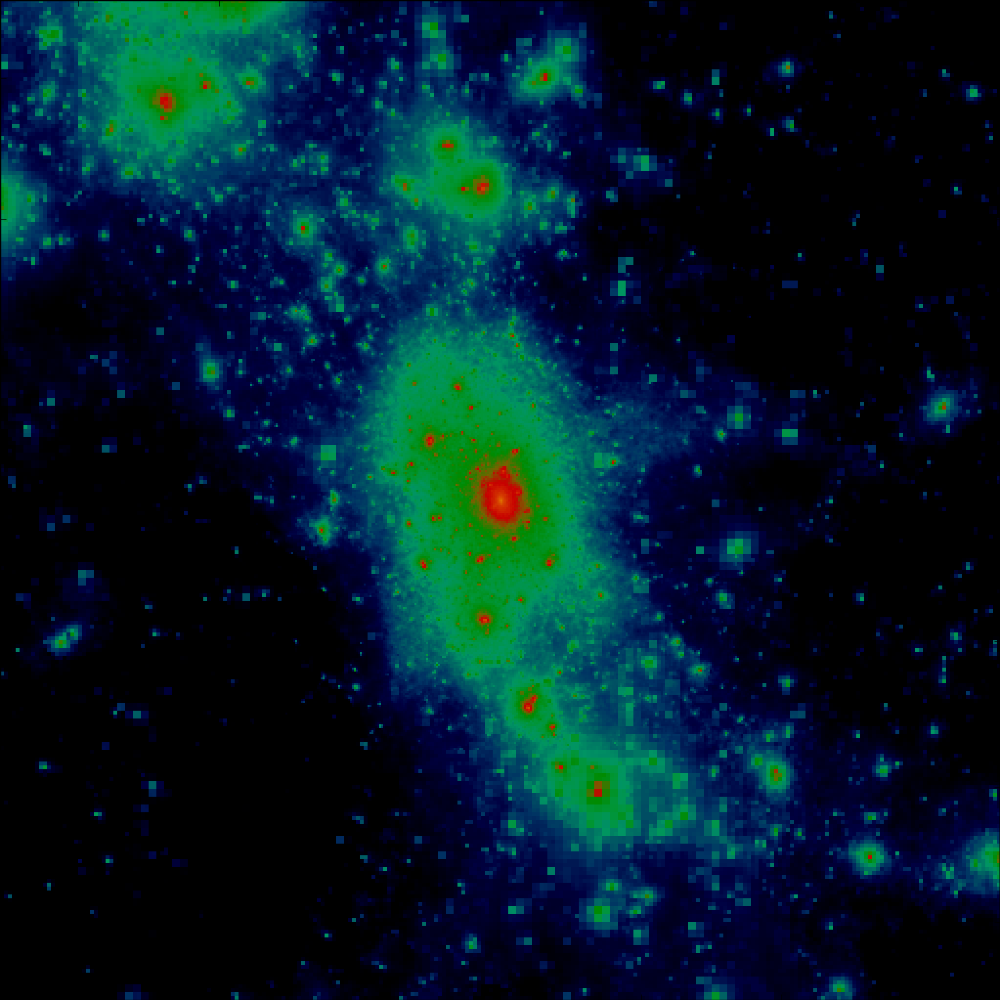}
\label{Fig:DMC08} \caption{C08}
\end{subfigure}
\begin{subfigure}[b]{0.25\textwidth}
\includegraphics[width=\textwidth]{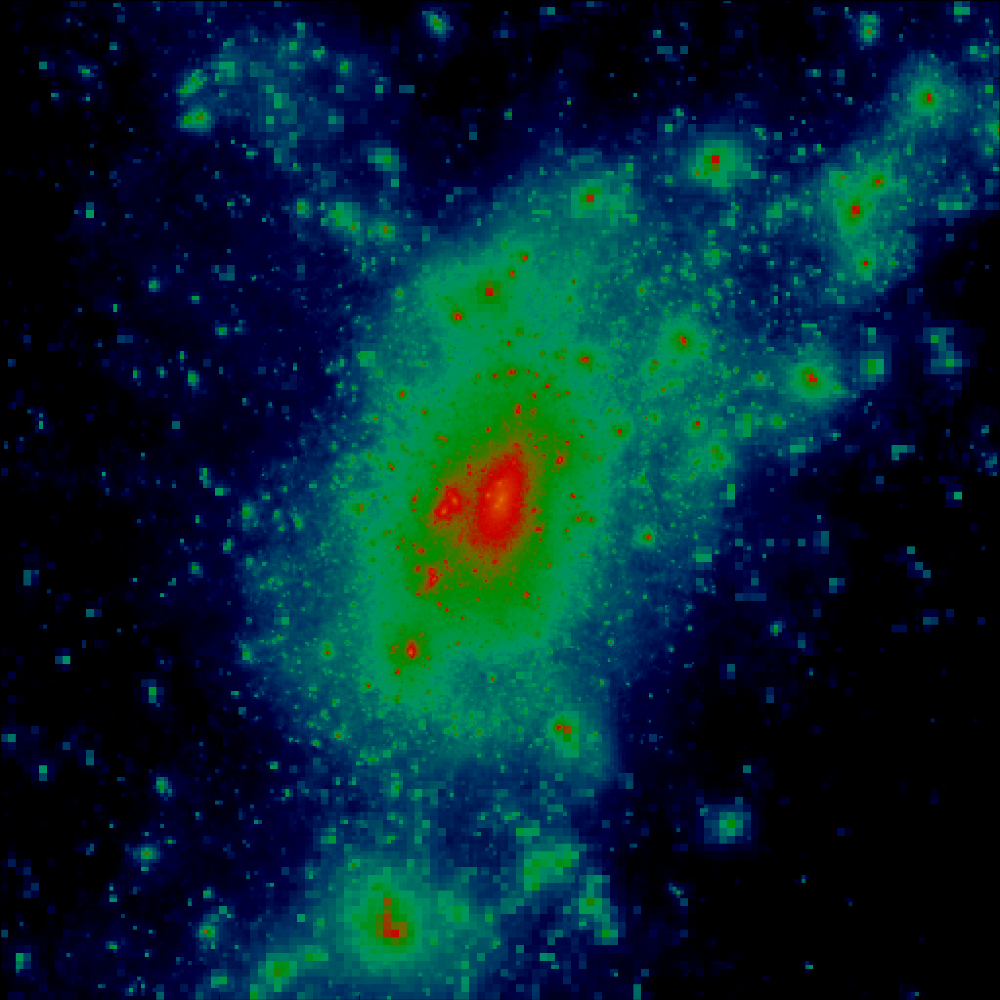}
\label{Fig:DMC09} \caption{C09}
\end{subfigure}

\begin{subfigure}[b]{0.25\textwidth}
\includegraphics[width=\textwidth]{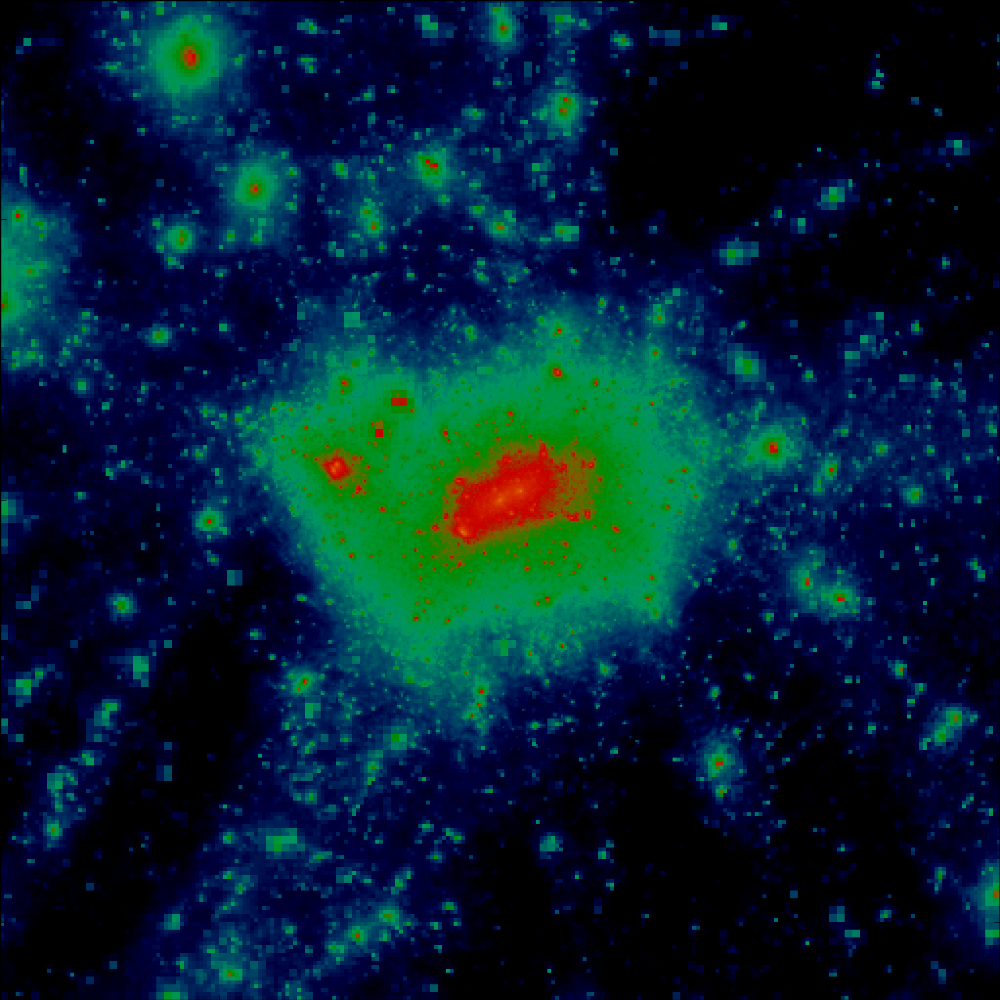}
\label{Fig:DMC10} \caption{C10}
\end{subfigure}
\begin{subfigure}[b]{0.25\textwidth}
\includegraphics[width=\textwidth]{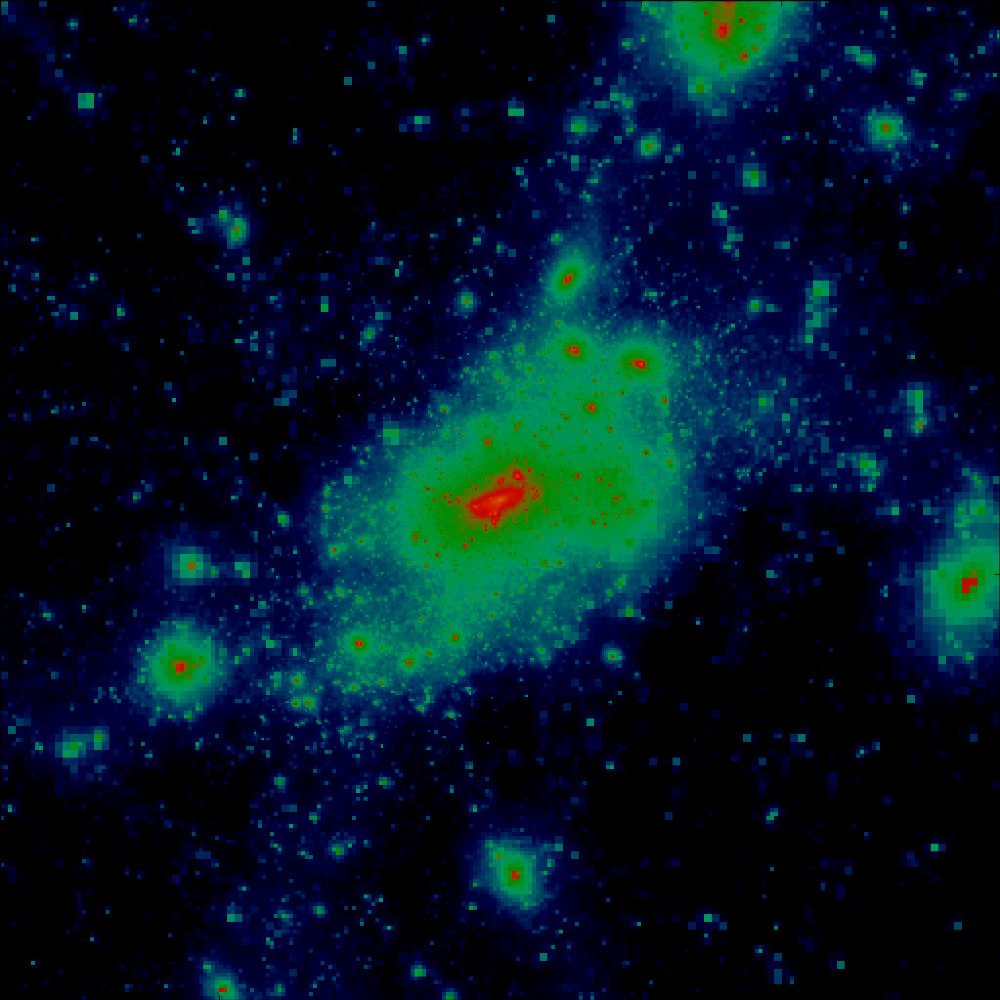}
\label{Fig:DMC11} \caption{C11}
\end{subfigure}
\begin{subfigure}[b]{0.25\textwidth}
\hspace{0.1\textwidth}
\includegraphics[width=1.35\textwidth]{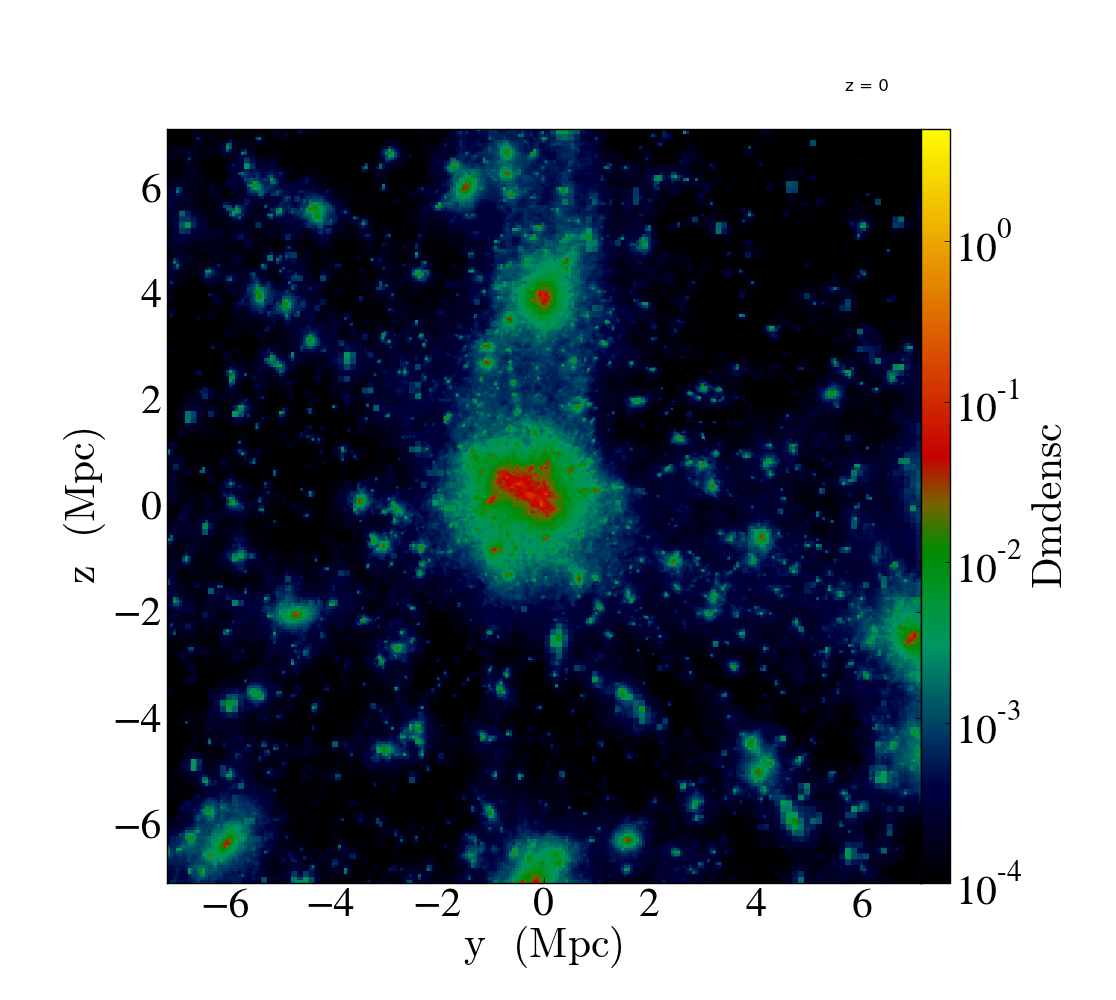}
\vspace{-0.07\textwidth}
\label{Fig:DMC12} \caption{C12}
\end{subfigure}
\caption{Dark matter density $x$-projections in $\mathrm{g}/\mathrm{cm}^2$ of all 12 clusters.
All plots show the inner 5 Mpc/h and use the same color scheme, see Fig. (l). }  \label{Fig:DMall}
\end{figure*}

\begin{figure*}[h!]
\centering
\begin{subfigure}[b]{0.25\textwidth}
\includegraphics[width=\textwidth]{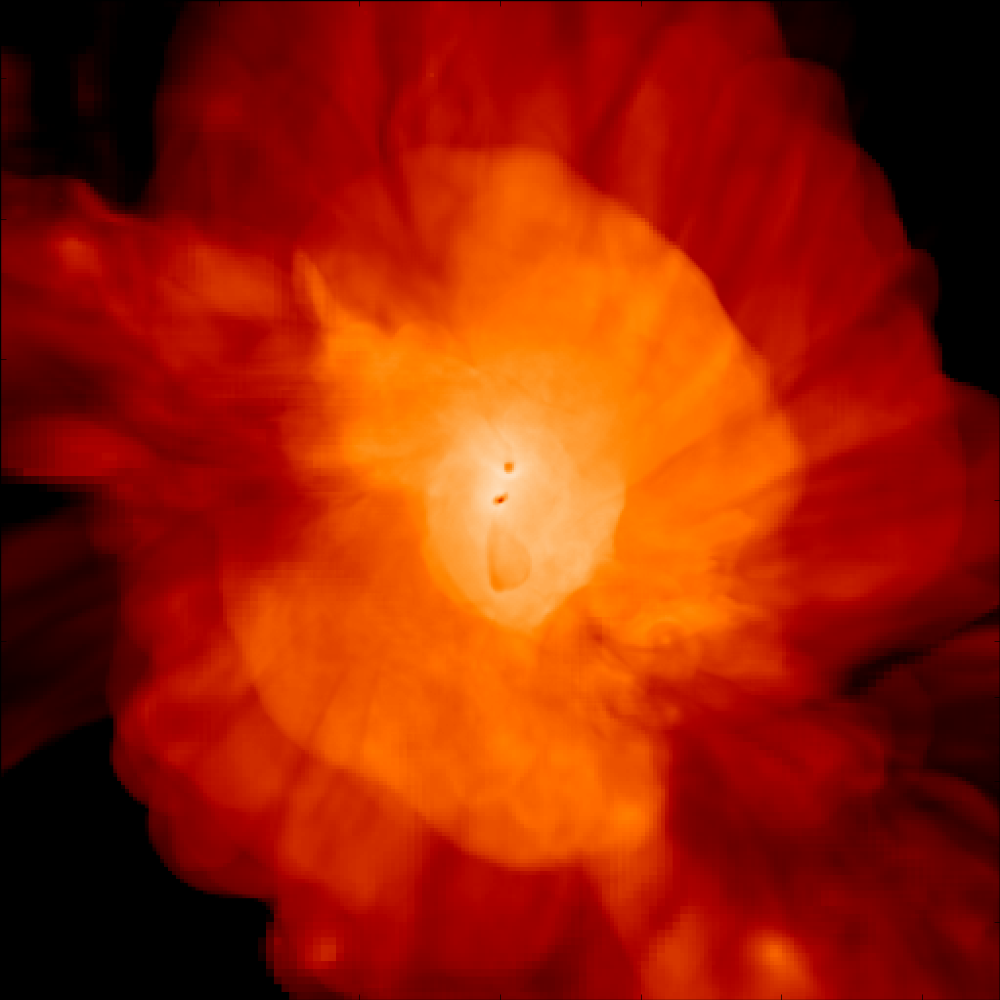}
\label{Fig:TC01} \caption{C01}
\end{subfigure}
\begin{subfigure}[b]{0.25\textwidth}
\includegraphics[width=\textwidth]{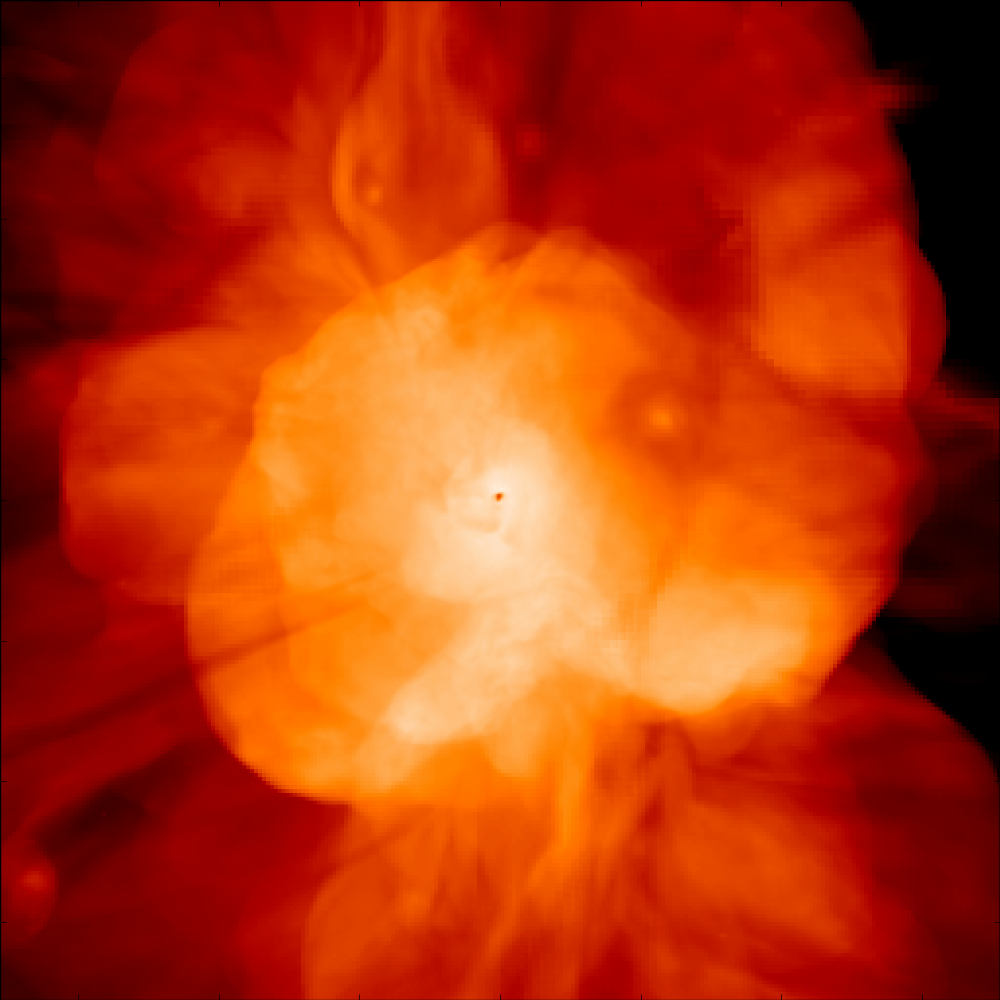}
\label{Fig:TC02} \caption{C02}
\end{subfigure}
\begin{subfigure}[b]{0.25\textwidth}
\includegraphics[width=\textwidth]{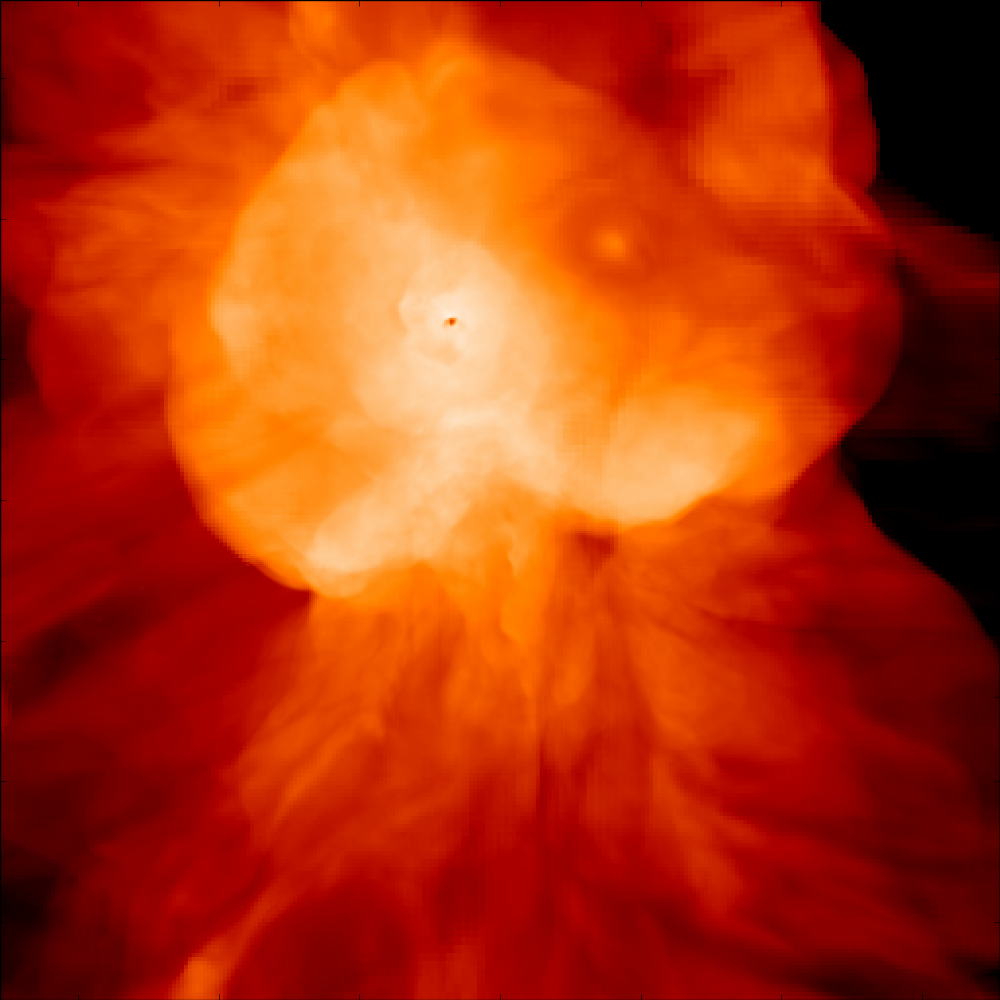}
\label{Fig:TC03} \caption{C03}
\end{subfigure}

\begin{subfigure}[b]{0.25\textwidth}
\includegraphics[width=\textwidth]{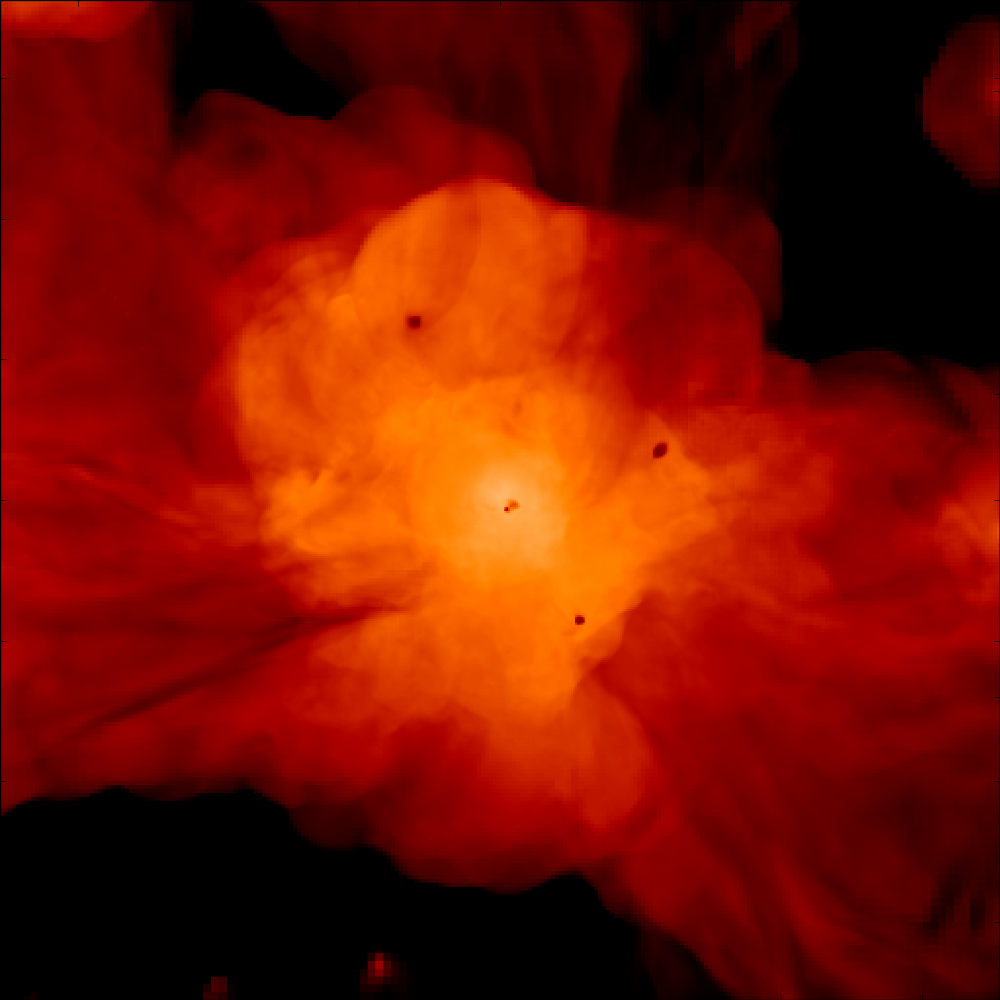}
\label{Fig:TC04} \caption{C04}
\end{subfigure}
\begin{subfigure}[b]{0.25\textwidth}
\includegraphics[width=\textwidth]{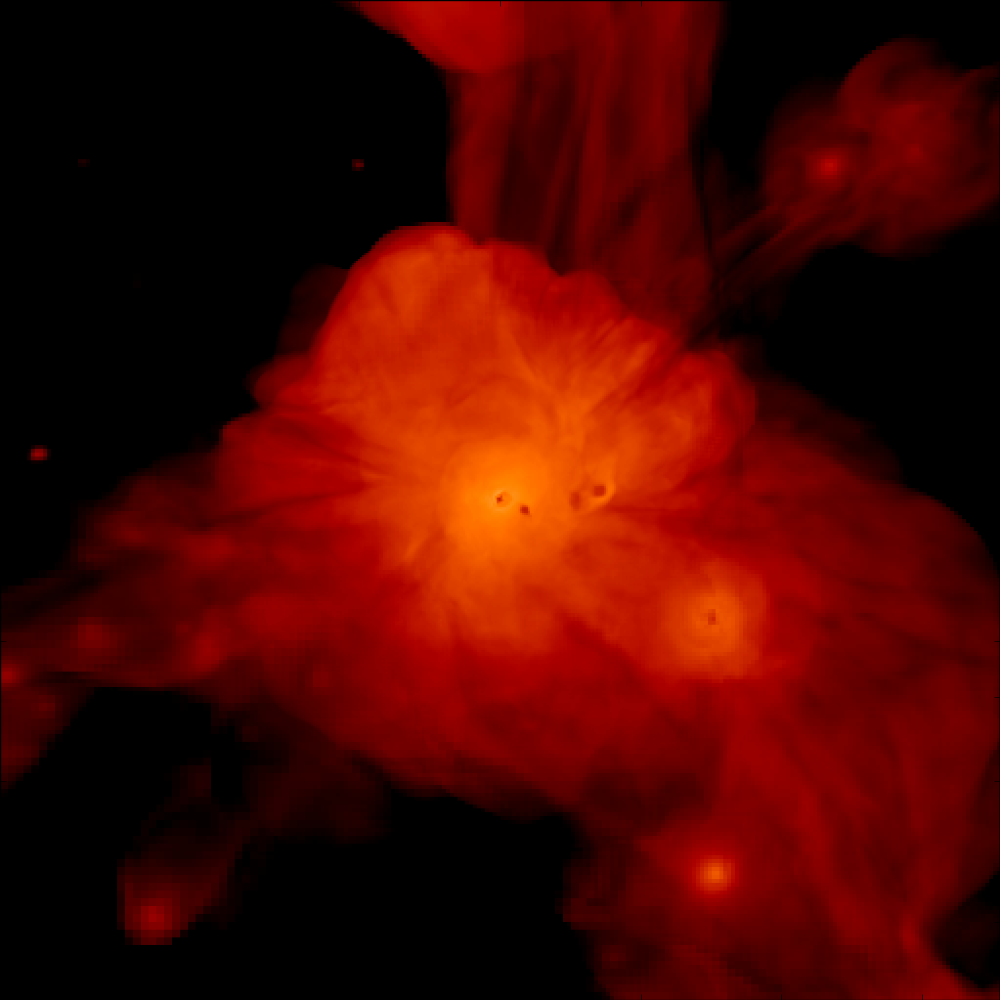}
\label{Fig:TC05} \caption{C05}
\end{subfigure}
\begin{subfigure}[b]{0.25\textwidth}
\includegraphics[width=\textwidth]{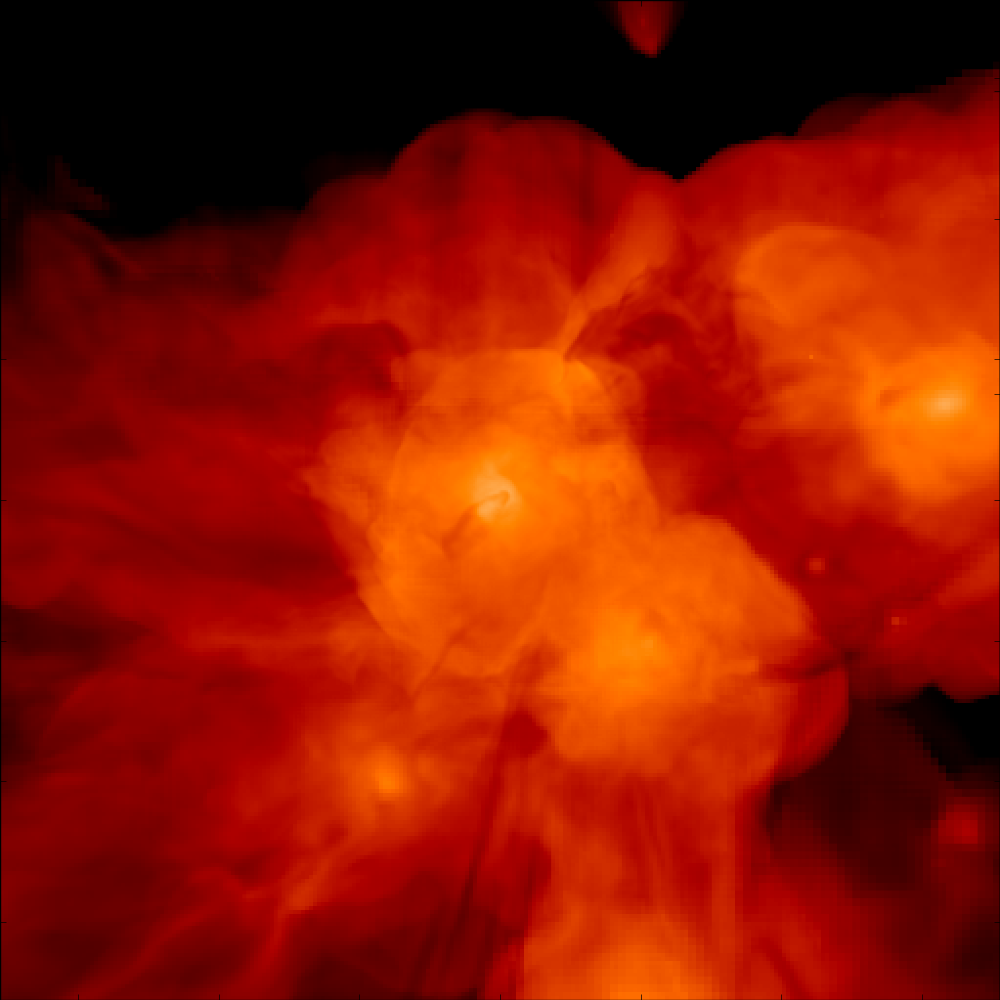}
\label{Fig:TC06} \caption{C06}
\end{subfigure}

\begin{subfigure}[b]{0.25\textwidth}
\includegraphics[width=\textwidth]{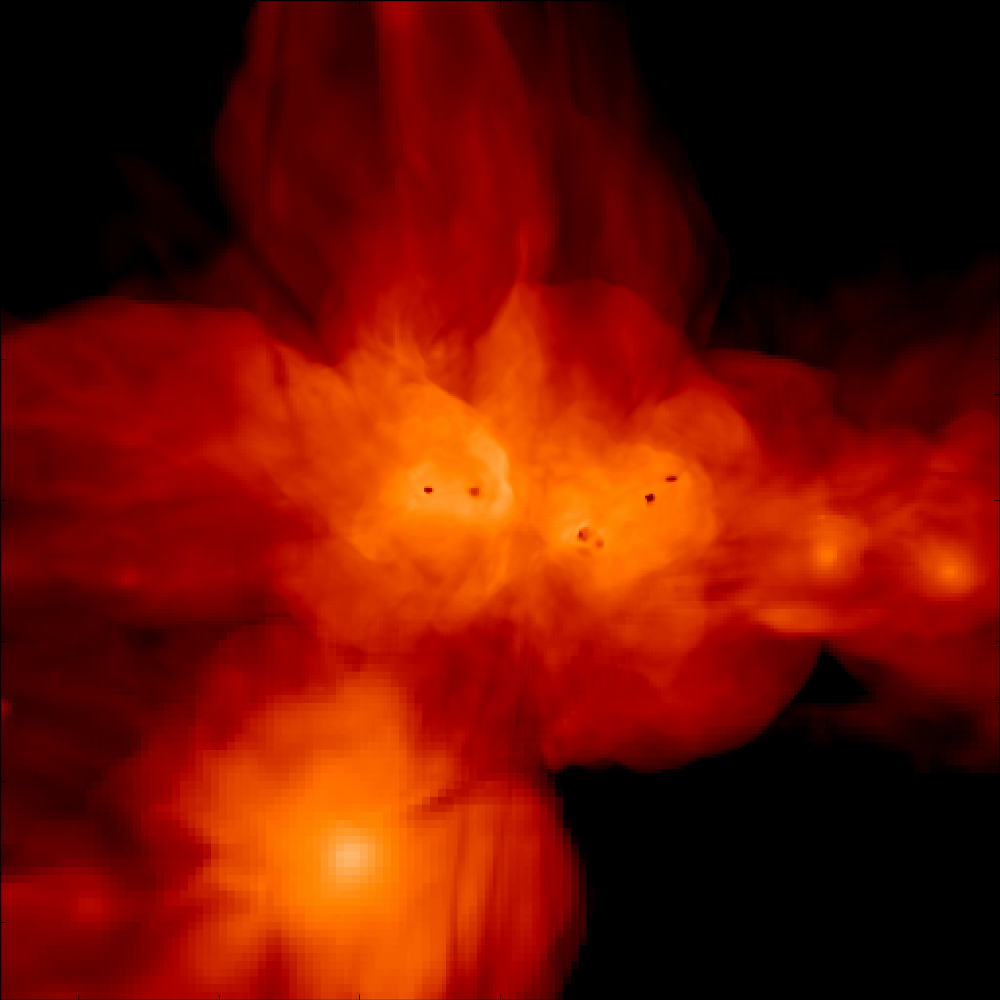}
\label{Fig:TC07} \caption{C07}
\end{subfigure}
\begin{subfigure}[b]{0.25\textwidth}
\includegraphics[width=\textwidth]{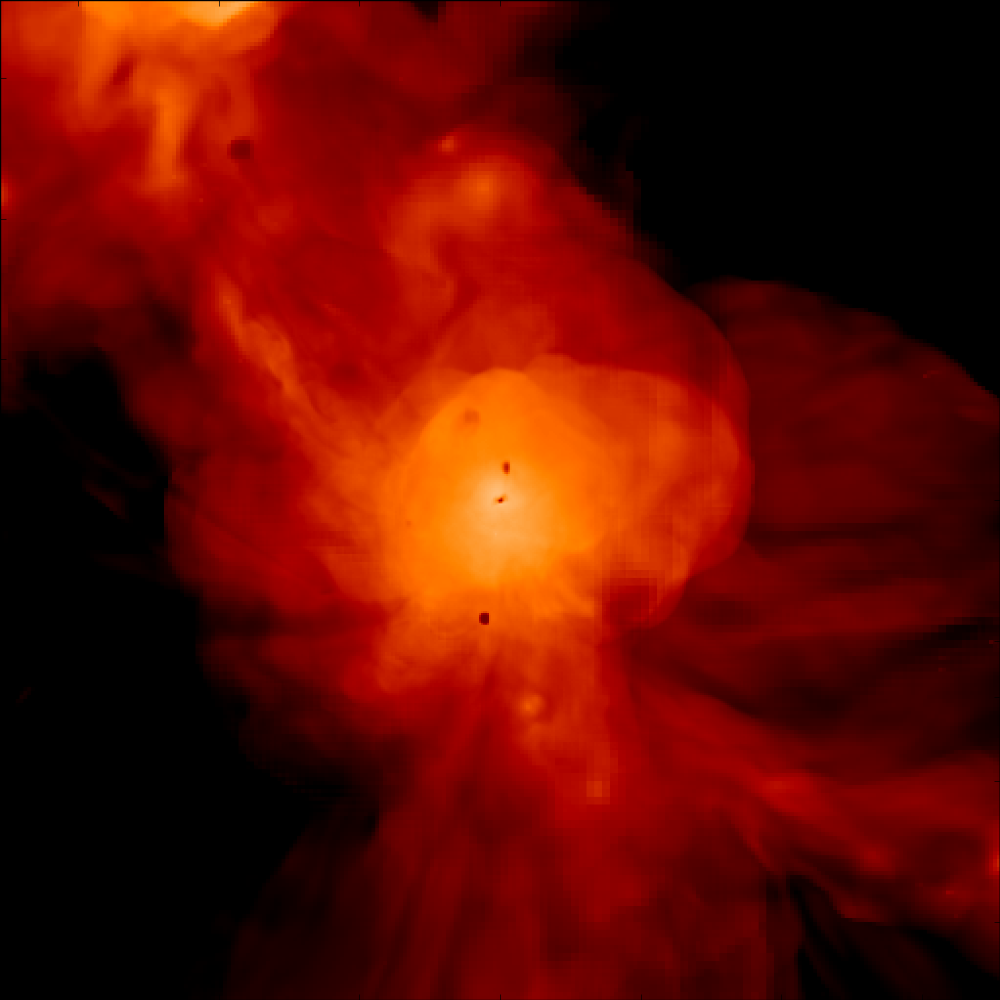}
\label{Fig:TC08} \caption{C08}
\end{subfigure}
\begin{subfigure}[b]{0.25\textwidth}
\includegraphics[width=\textwidth]{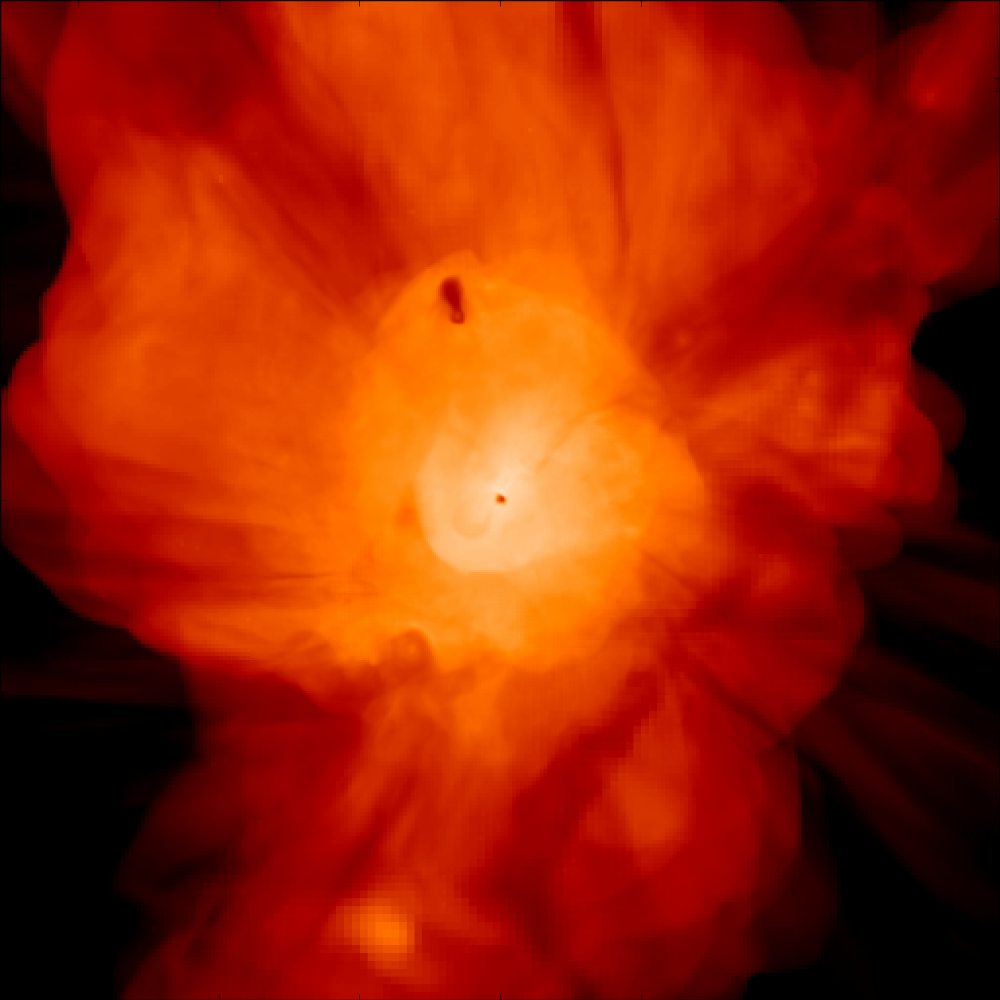}
\label{Fig:TC09} \caption{C09}
\end{subfigure}

\begin{subfigure}[b]{0.25\textwidth}
\includegraphics[width=\textwidth]{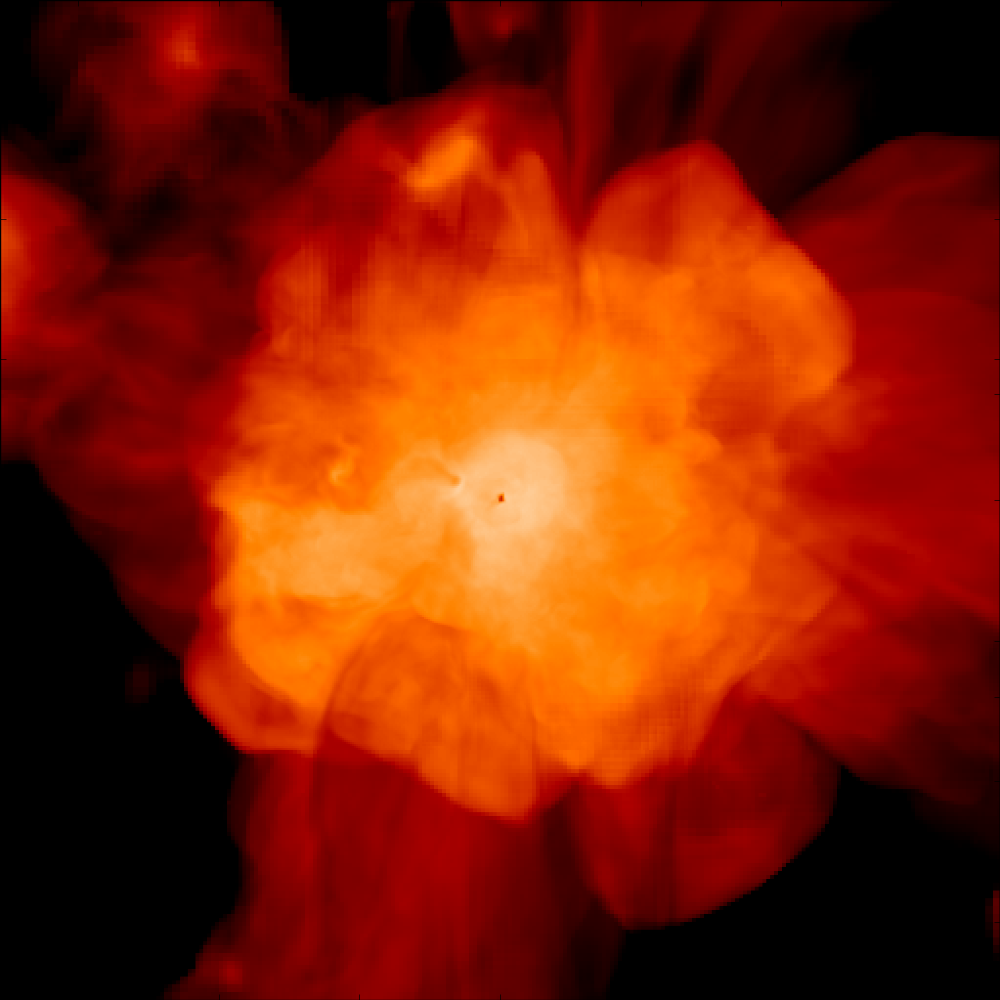}
\label{Fig:TC10} \caption{C10}
\end{subfigure}
\begin{subfigure}[b]{0.25\textwidth}
\includegraphics[width=\textwidth]{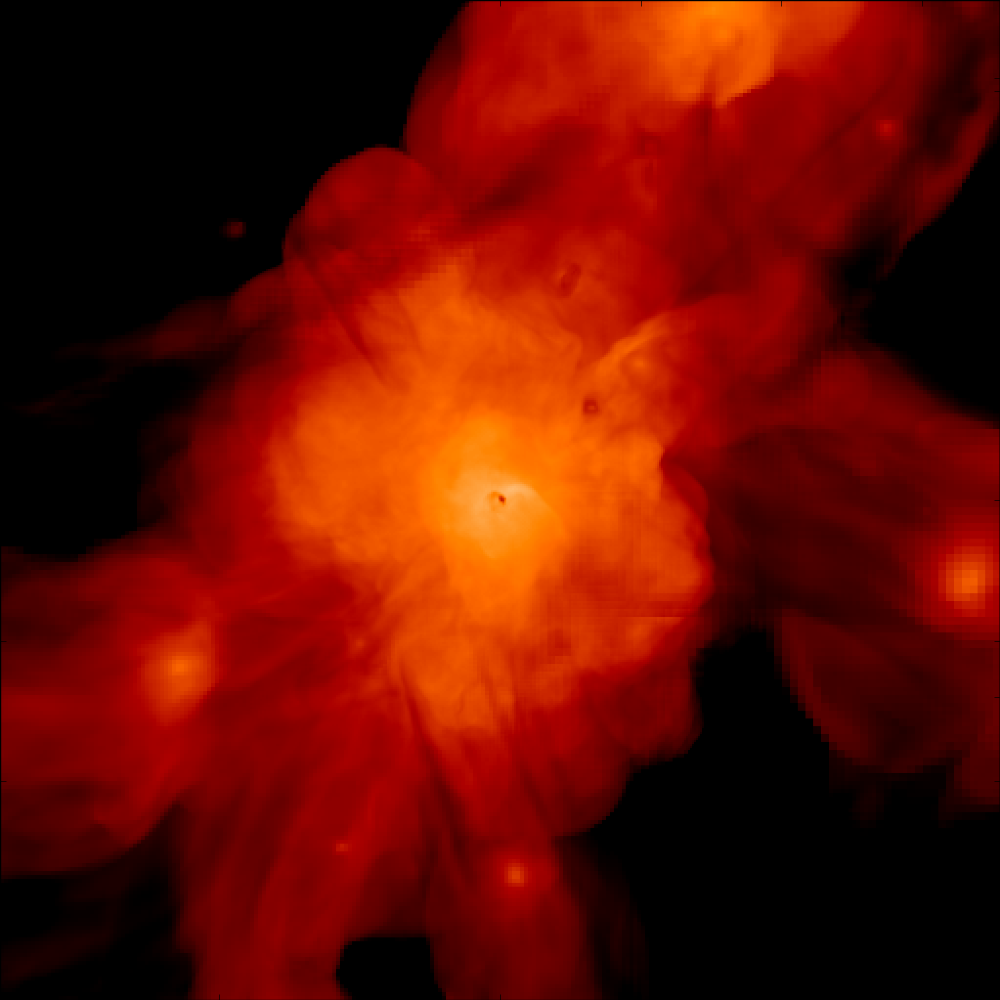}
\label{Fig:TC11} \caption{C11}
\end{subfigure}
\begin{subfigure}[b]{0.25\textwidth}
\hspace{0.1\textwidth}
\includegraphics[width=1.35\textwidth]{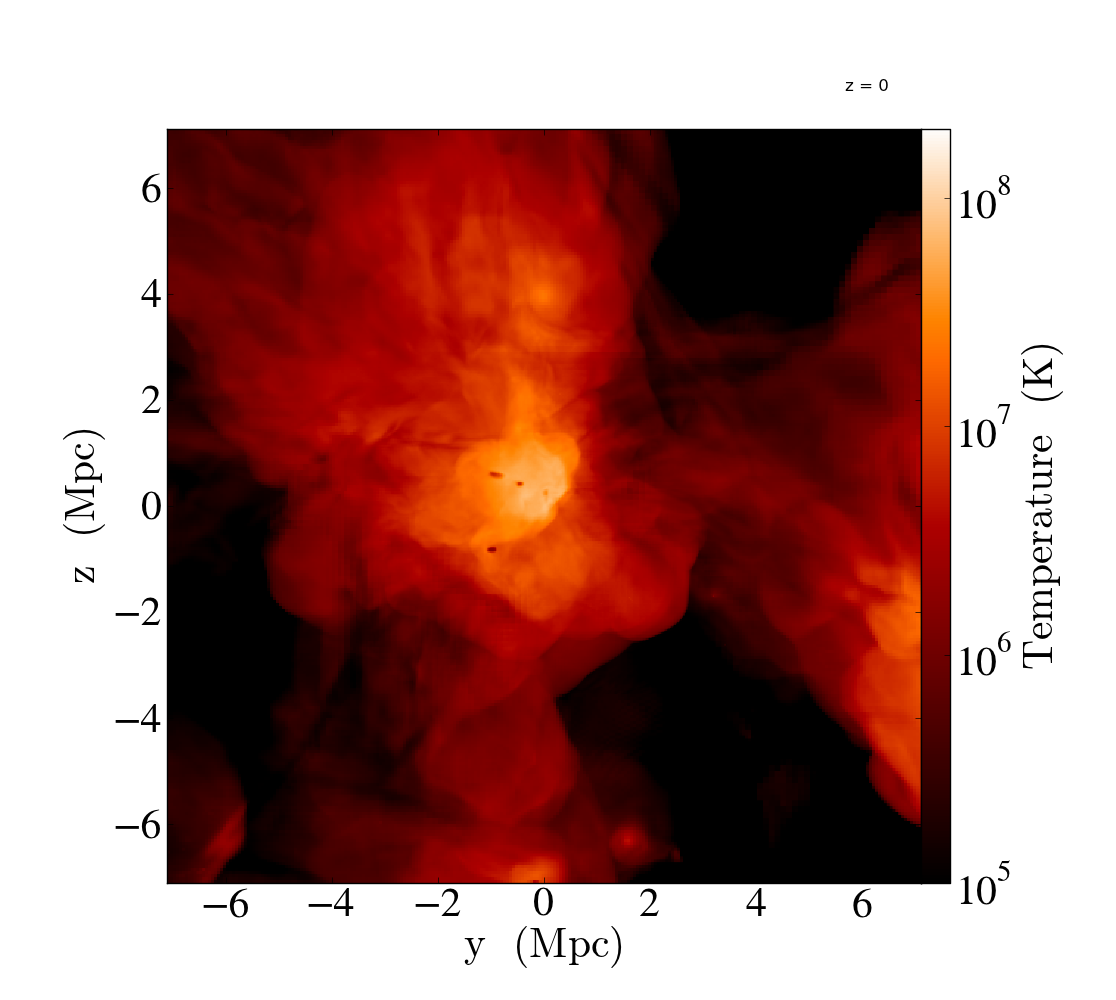}
\vspace{-0.07\textwidth}
\label{Fig:TC12} \caption{C12}
\end{subfigure}
\caption{Temperature $x$-projections of all 12 clusters. All plots show the inner 5 Mpc/h and use 
the same color scheme, see Fig. (l). } \label{Fig:Tempall}
\end{figure*}

\begin{figure*}[h!]
\centering
\begin{subfigure}[b]{0.25\textwidth}
\includegraphics[width=\textwidth]{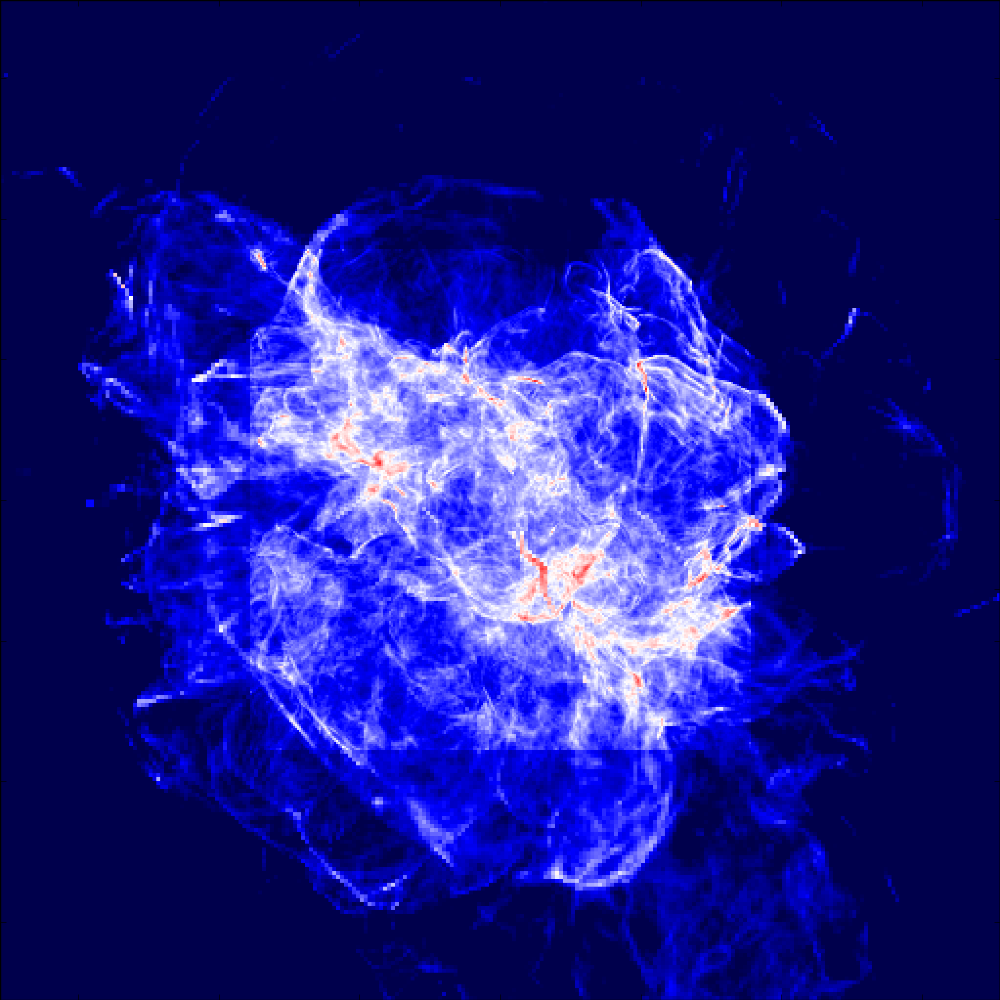}
\label{Fig:BC01} \caption{C01}
\end{subfigure}
\begin{subfigure}[b]{0.25\textwidth}
\includegraphics[width=\textwidth]{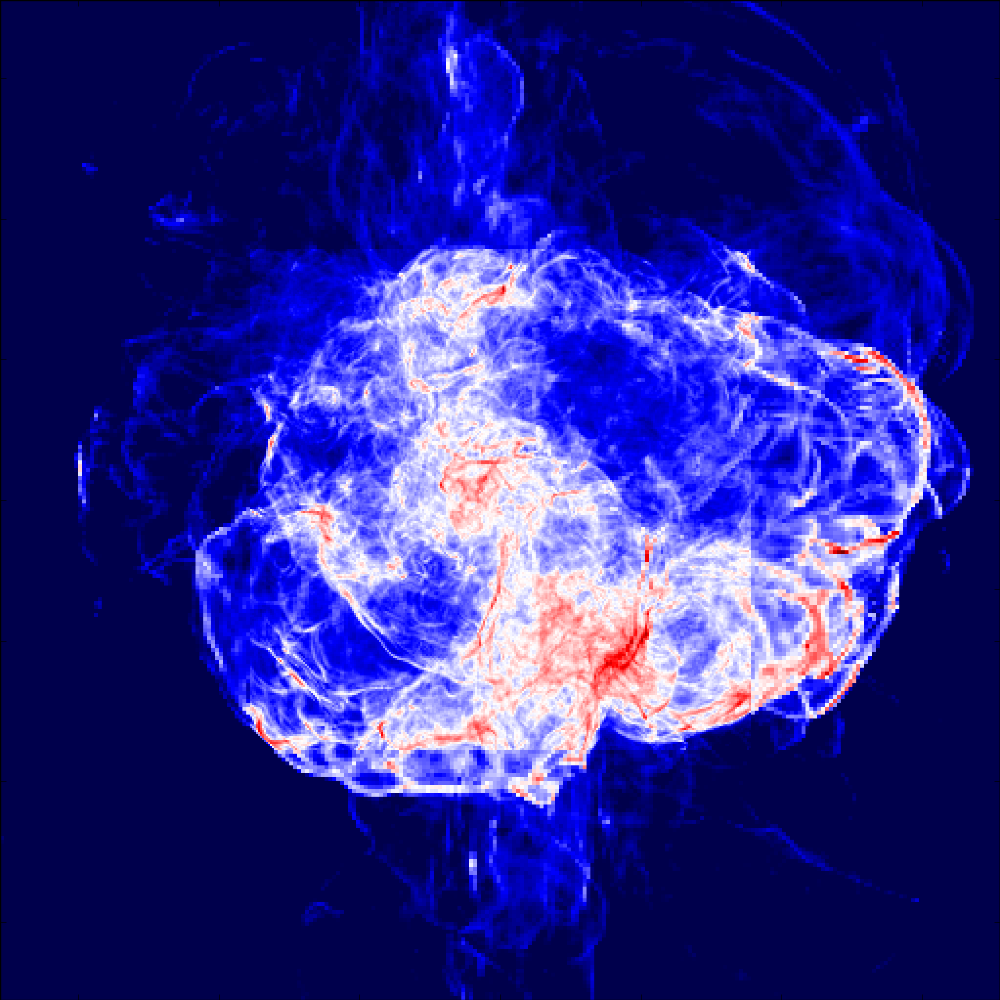}
\label{Fig:BC02} \caption{C02}
\end{subfigure}
\begin{subfigure}[b]{0.25\textwidth}
\includegraphics[width=\textwidth]{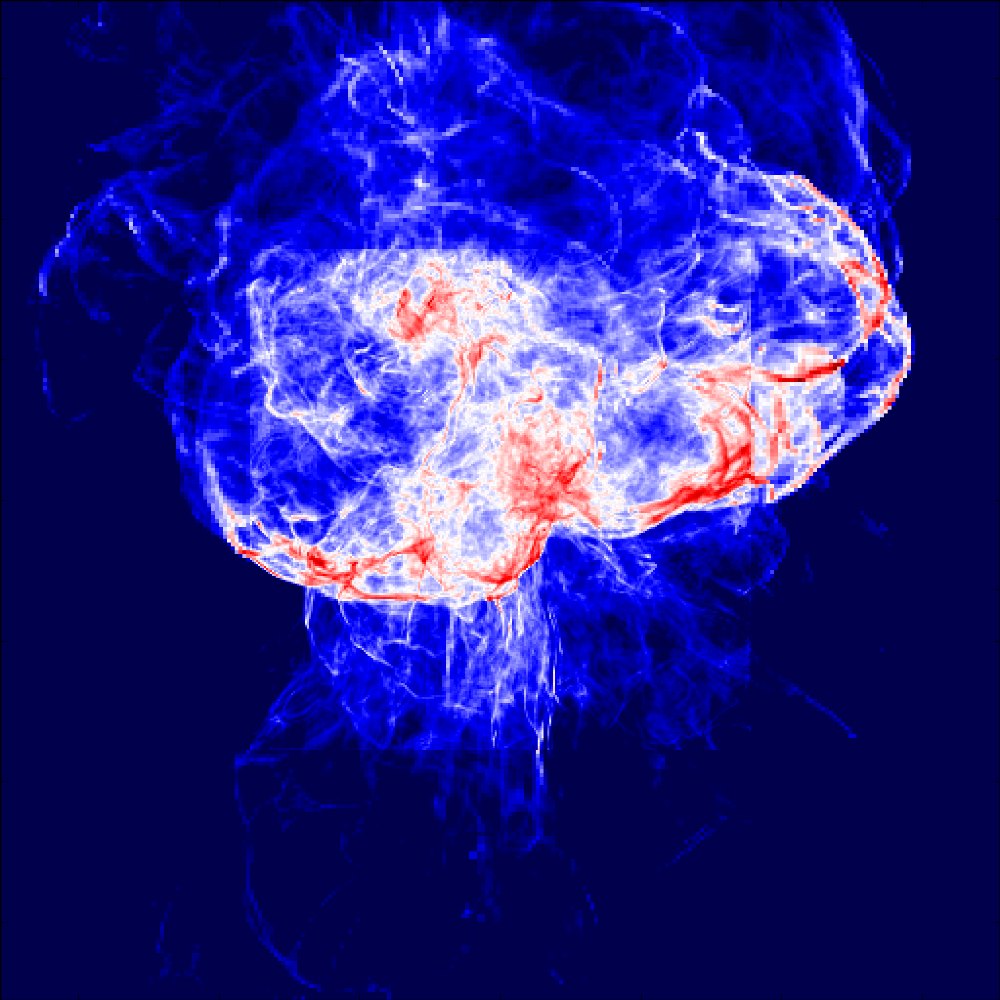}
\label{Fig:BC03} \caption{C03}
\end{subfigure}

\begin{subfigure}[b]{0.25\textwidth}
\includegraphics[width=\textwidth]{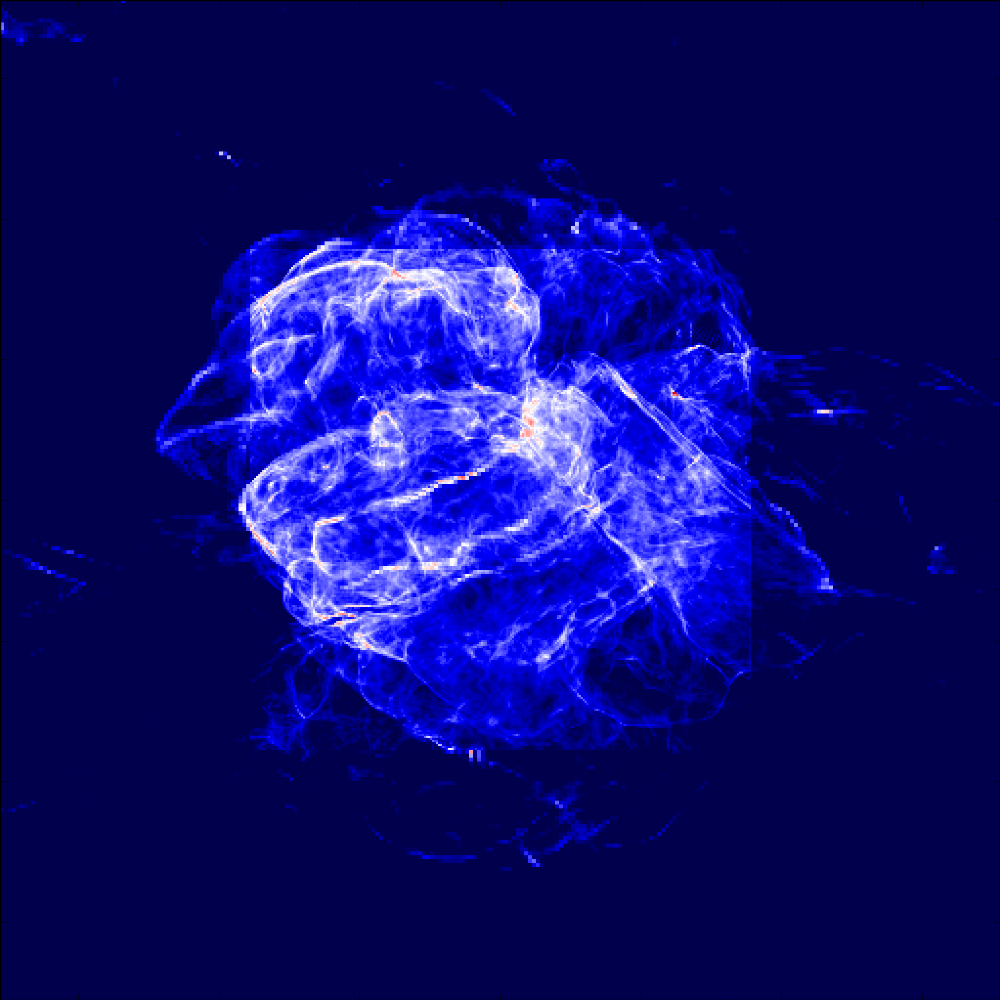}
\label{Fig:BC04} \caption{C04}
\end{subfigure}
\begin{subfigure}[b]{0.25\textwidth}
\includegraphics[width=\textwidth]{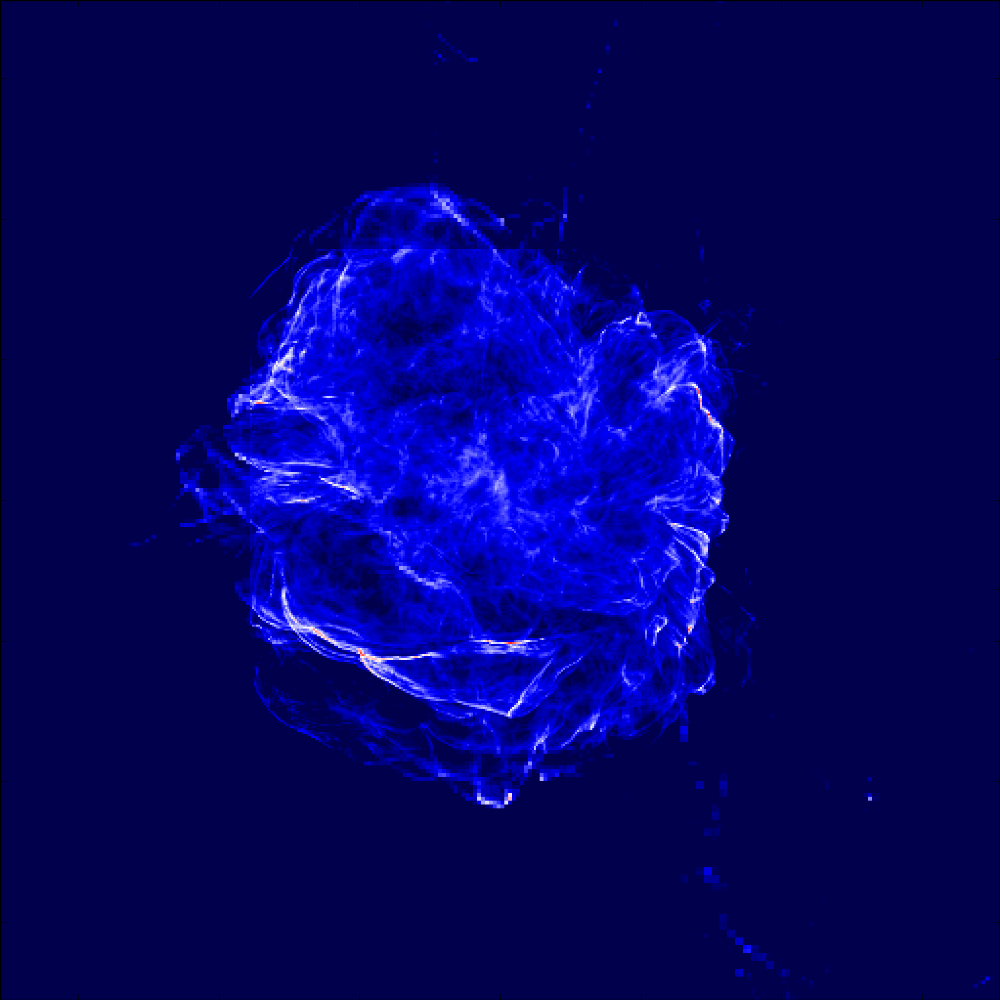}
\label{Fig:BC05} \caption{C05}
\end{subfigure}
\begin{subfigure}[b]{0.25\textwidth}
\includegraphics[width=\textwidth]{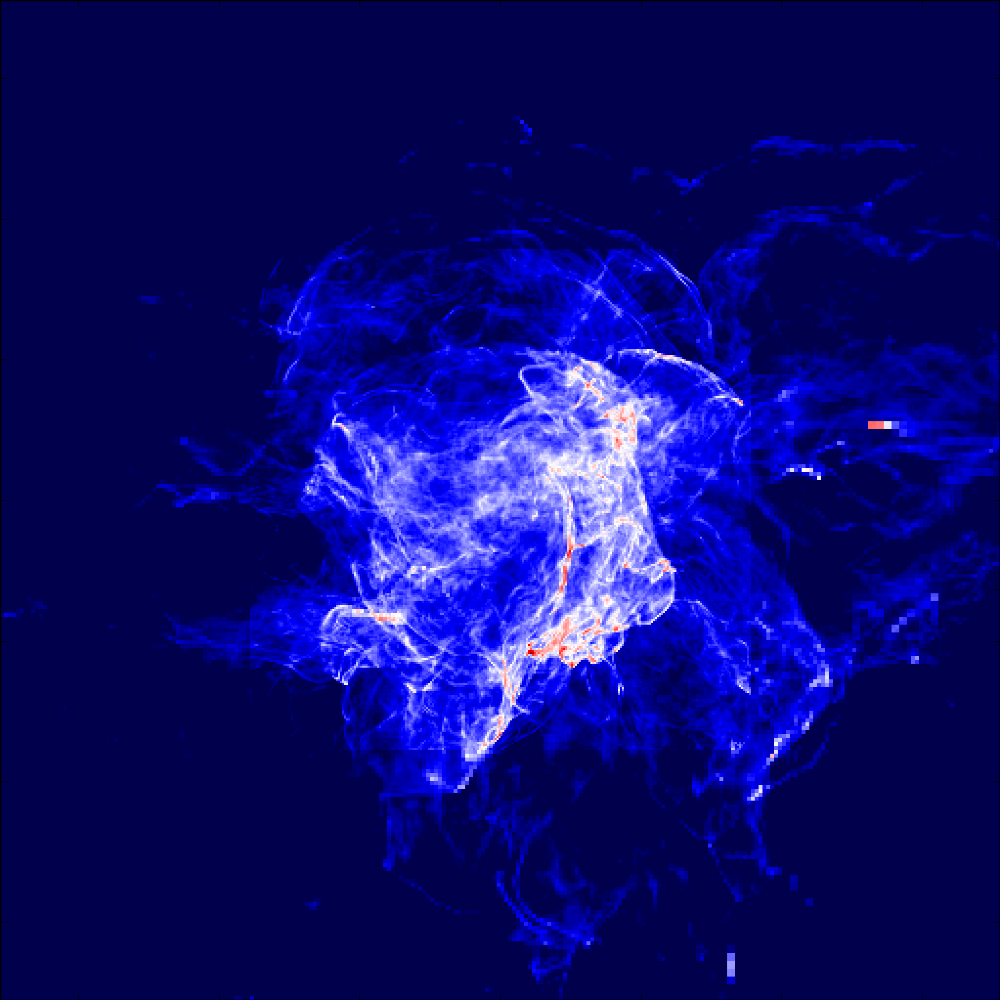}
\label{Fig:BC06} \caption{C06}
\end{subfigure}

\begin{subfigure}[b]{0.25\textwidth}
\includegraphics[width=\textwidth]{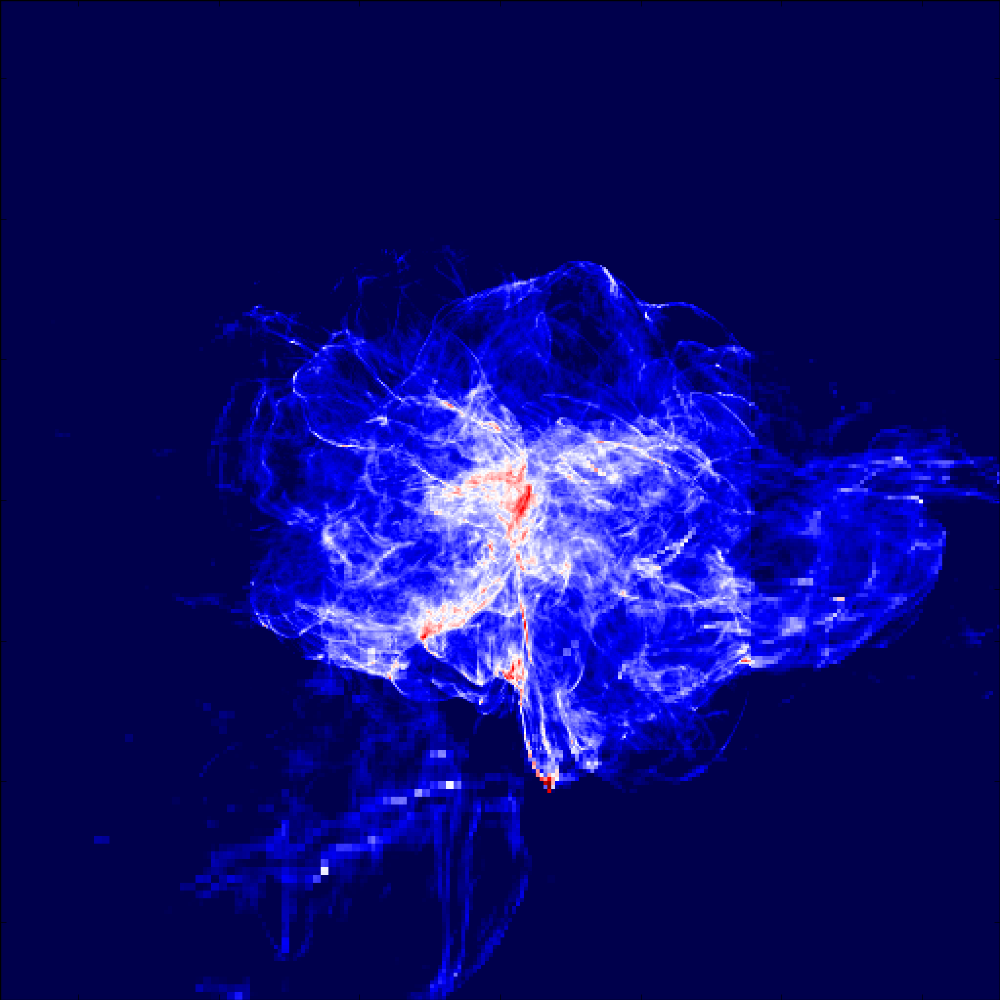}
\label{Fig:BC07} \caption{C07}
\end{subfigure}
\begin{subfigure}[b]{0.25\textwidth}
\includegraphics[width=\textwidth]{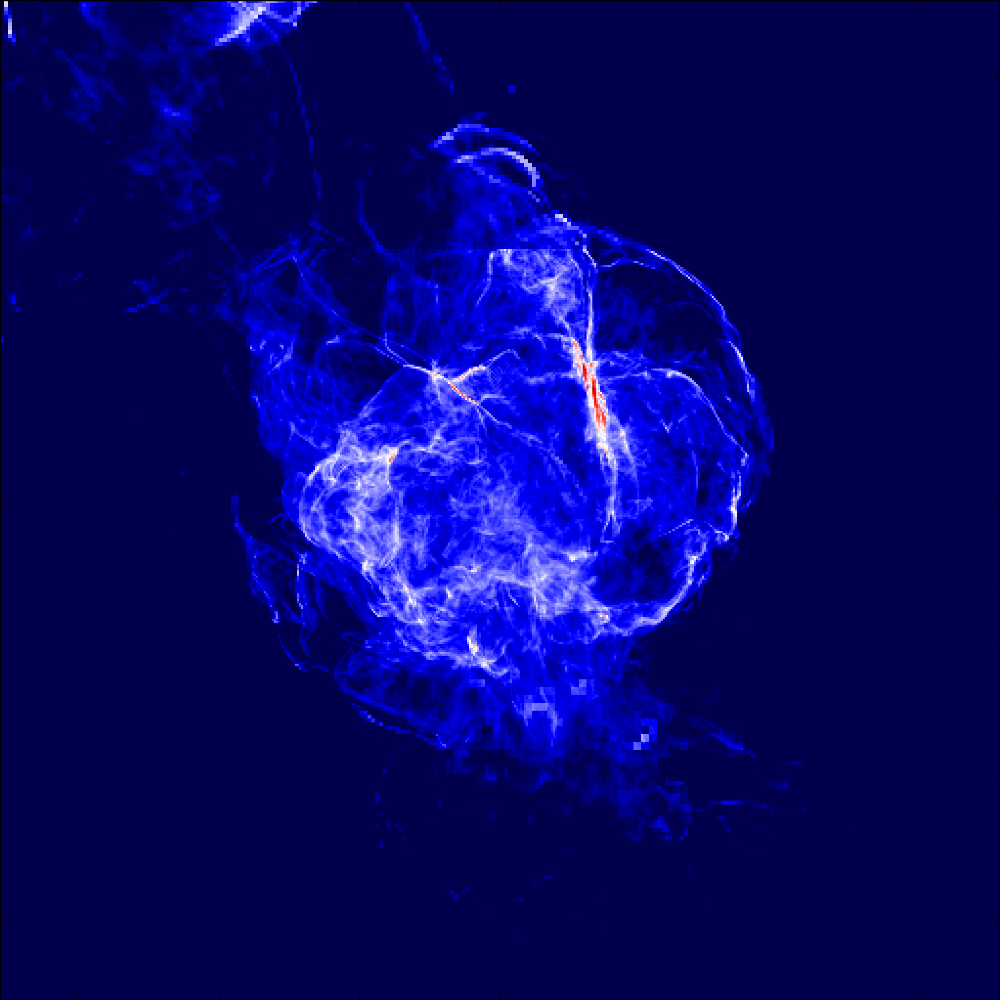}
\label{Fig:BC08} \caption{C08}
\end{subfigure}
\begin{subfigure}[b]{0.25\textwidth}
\includegraphics[width=\textwidth]{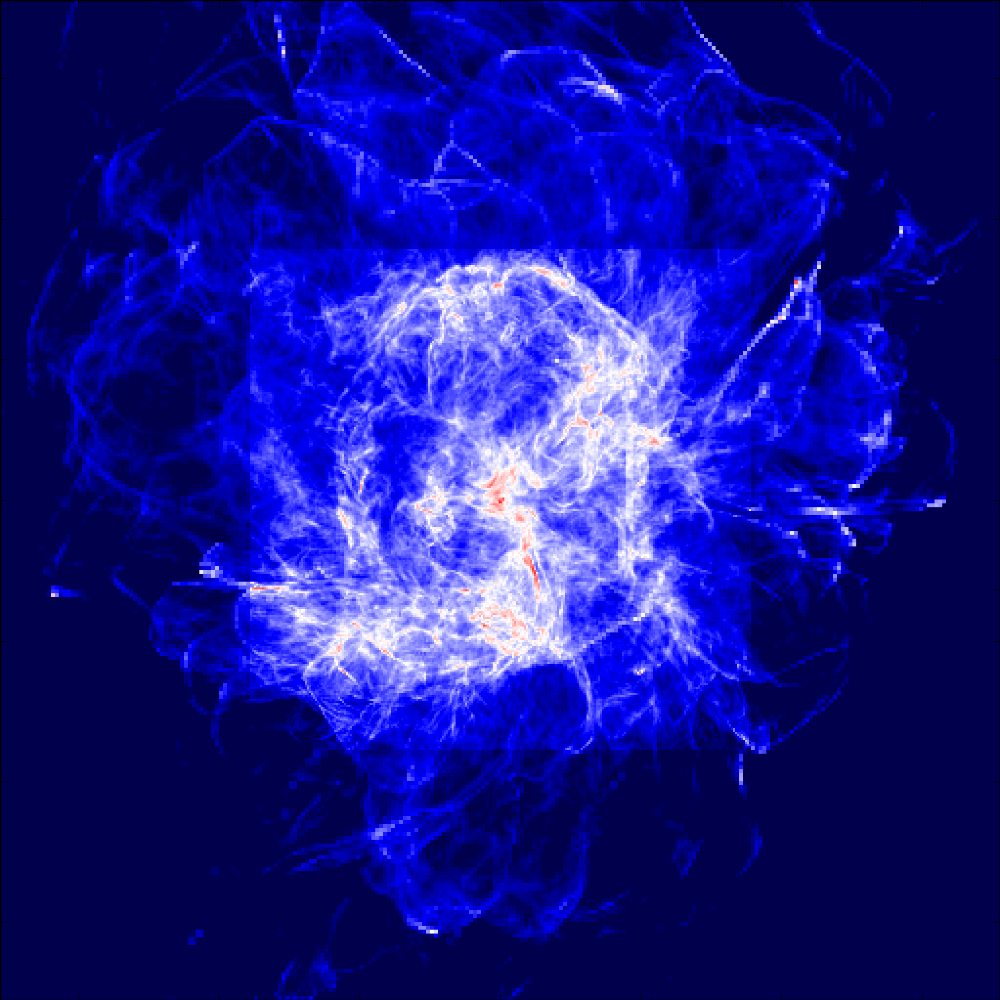}
\label{Fig:BC09} \caption{C09}
\end{subfigure}

\begin{subfigure}[b]{0.25\textwidth}
\includegraphics[width=\textwidth]{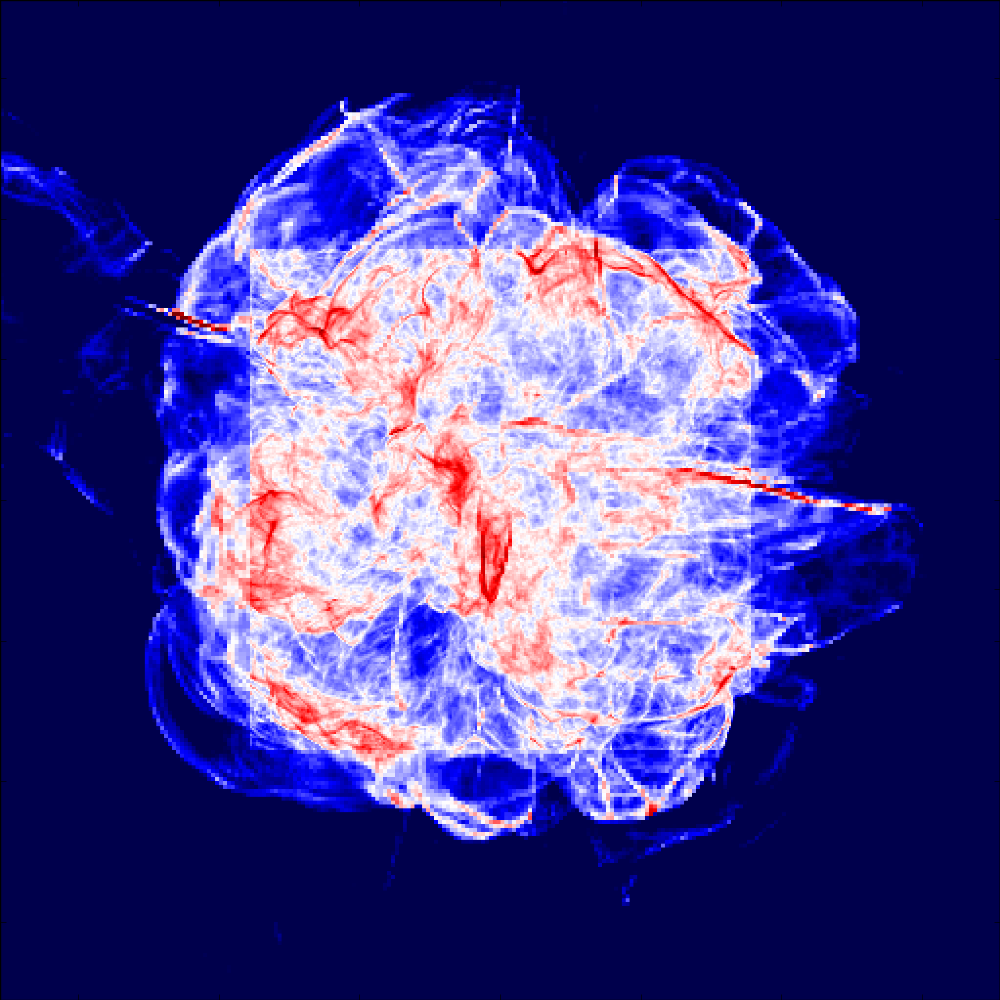}
\label{Fig:BC10} \caption{C10}
\end{subfigure}
\begin{subfigure}[b]{0.25\textwidth}
\includegraphics[width=\textwidth]{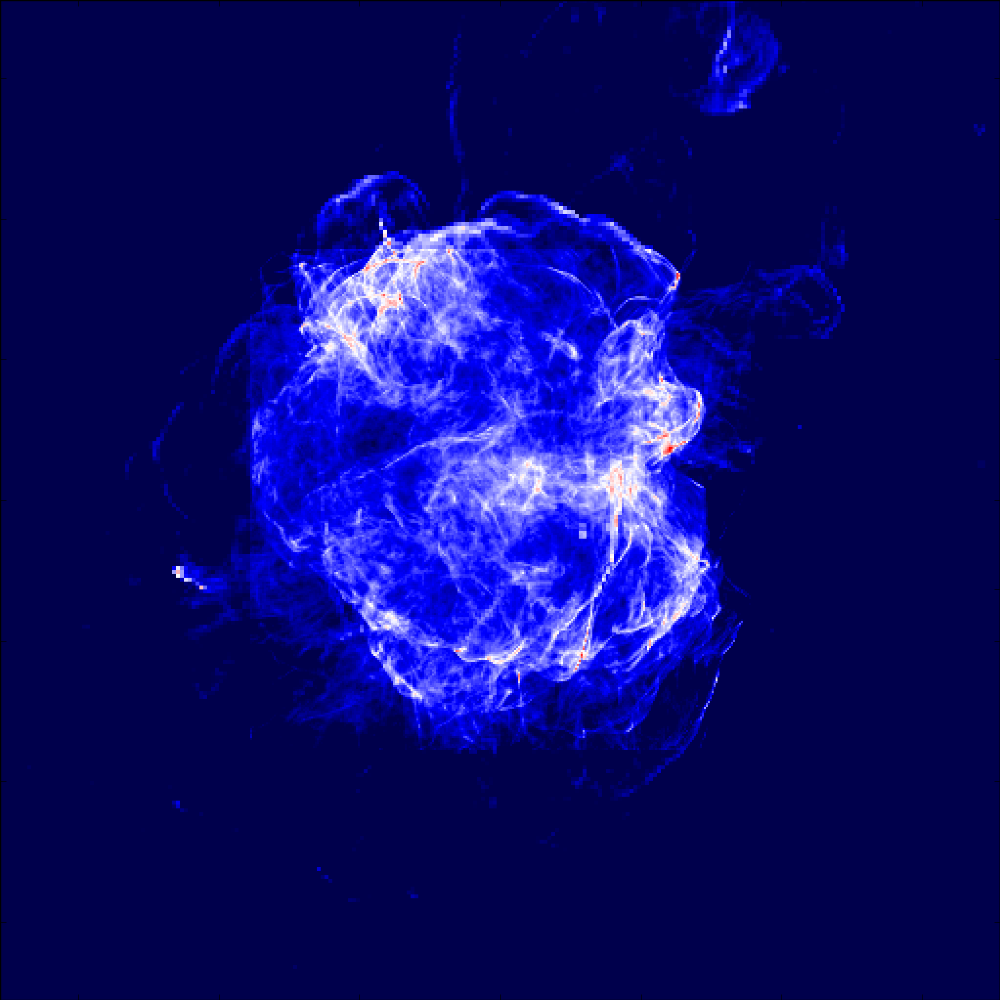}
\label{Fig:BC11} \caption{C11}
\end{subfigure}
\begin{subfigure}[b]{0.25\textwidth}
\hspace{0.1\textwidth}
\includegraphics[width=1.35\textwidth]{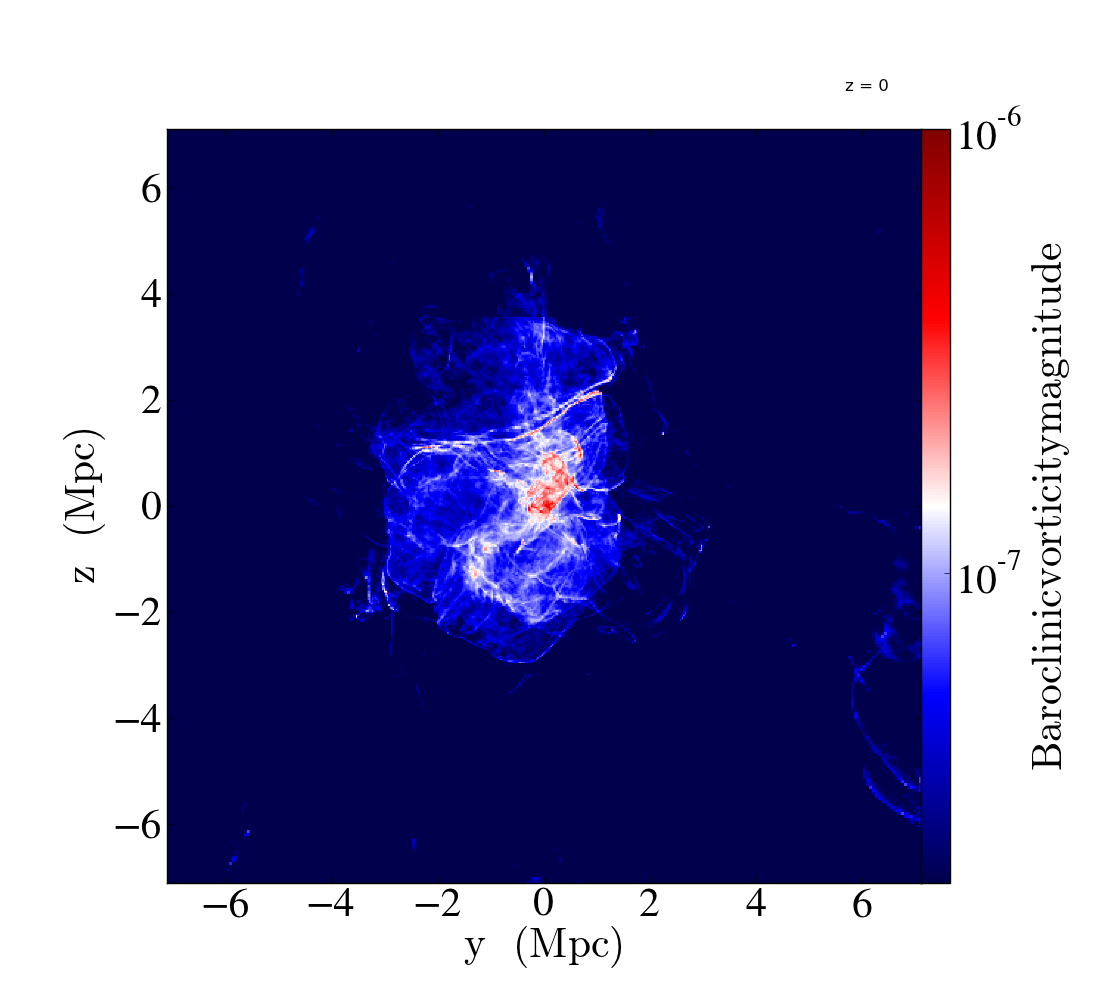}
\vspace{-0.07\textwidth}
\label{Fig:BC12} \caption{C12}
\end{subfigure}
\caption{Baroclinic vorticity magnitude $x$-projections of all 12 clusters. All plots show the inner 5 Mpc/h and use 
the same color scheme, see Fig. (l). The visibility of the grid refinement boundary is due to the usage of a 
common stencil when calculating the gradient or rotation on the grid data. } \label{Fig:BVMall}
\end{figure*}

\begin{figure*}[h!]
\centering
\begin{subfigure}[b]{0.25\textwidth}
\includegraphics[width=\textwidth]{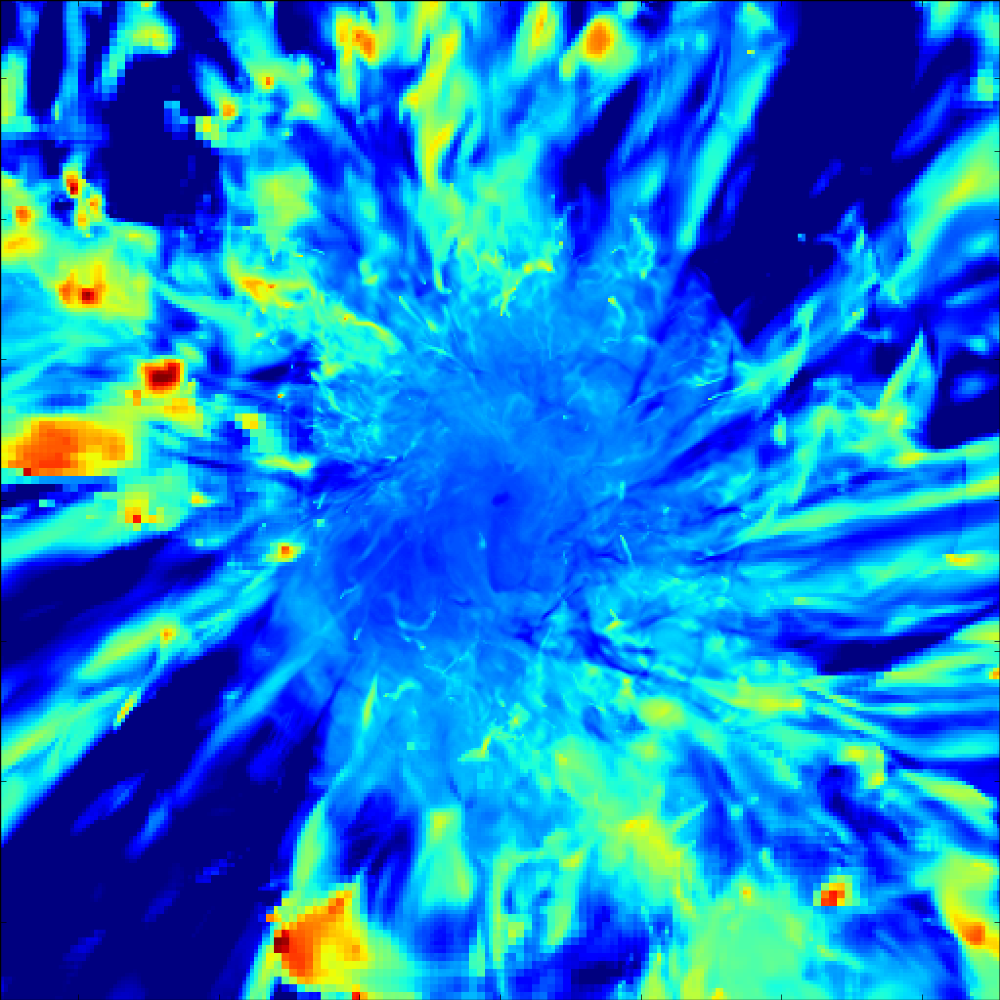}
\label{Fig:WC01} \caption{C01}
\end{subfigure}
\begin{subfigure}[b]{0.25\textwidth}
\includegraphics[width=\textwidth]{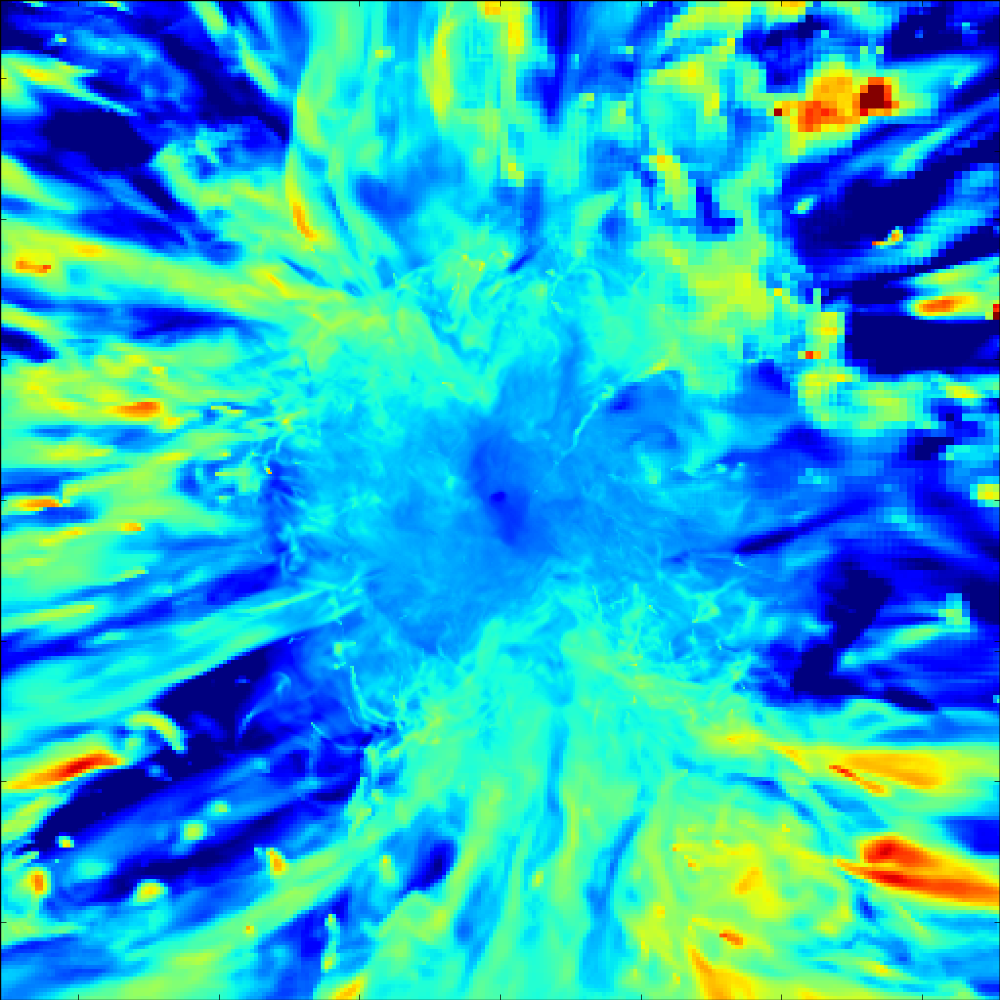}
\label{Fig:WC02} \caption{C02}
\end{subfigure}
\begin{subfigure}[b]{0.25\textwidth}
\includegraphics[width=\textwidth]{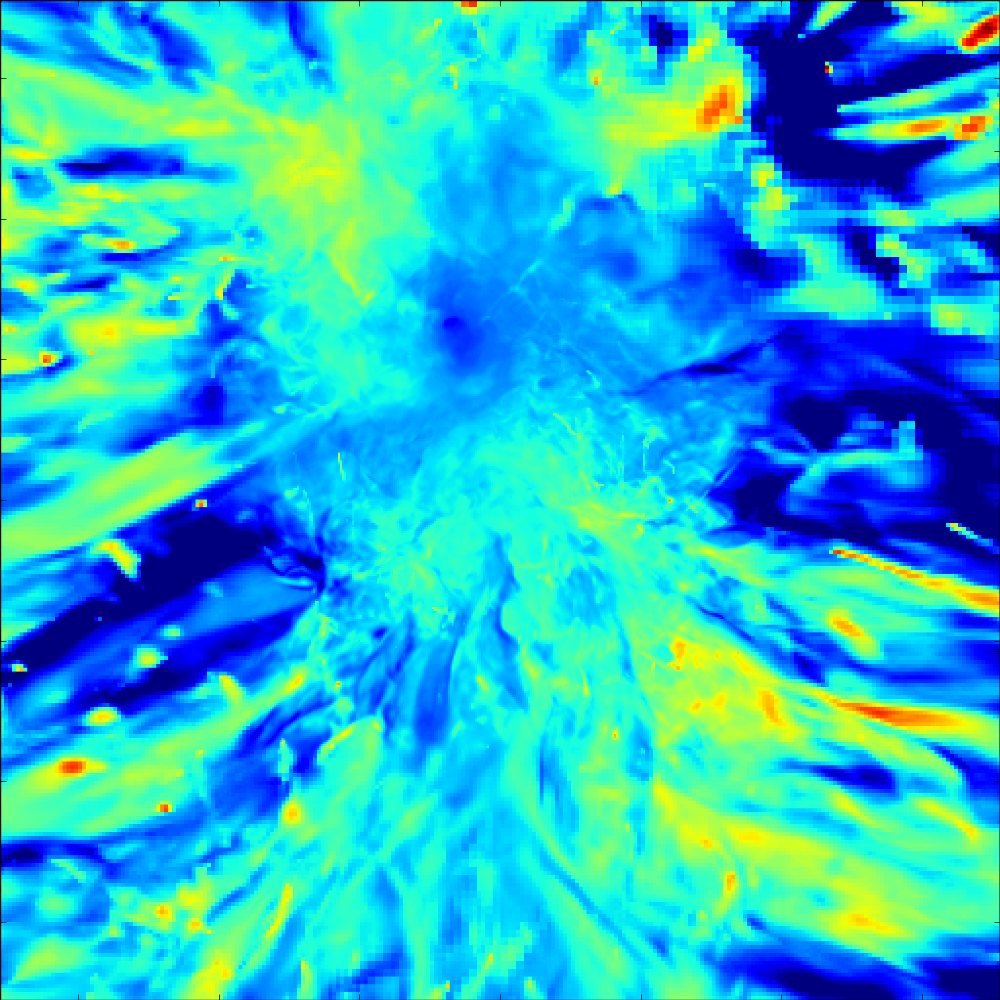}
\label{Fig:WC03} \caption{C03}
\end{subfigure}

\begin{subfigure}[b]{0.25\textwidth}
\includegraphics[width=\textwidth]{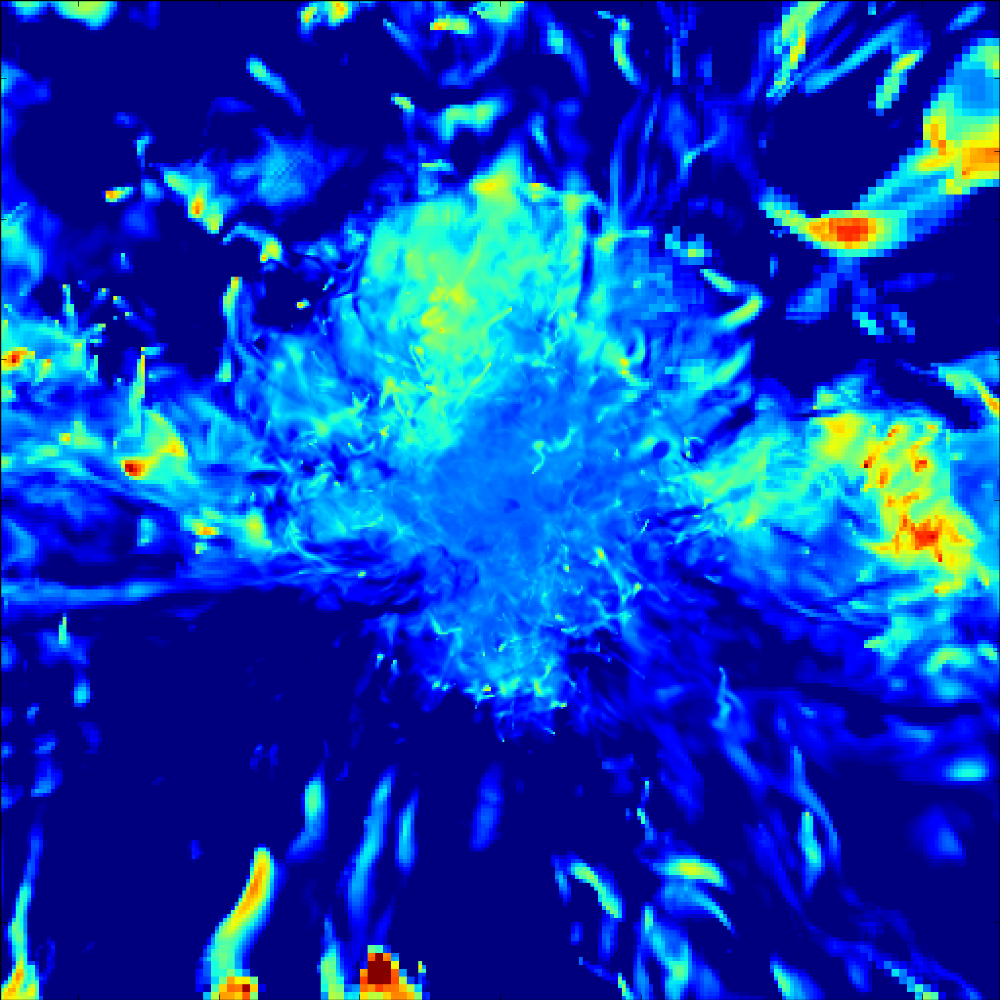}
\label{Fig:WC04} \caption{C04}
\end{subfigure}
\begin{subfigure}[b]{0.25\textwidth}
\includegraphics[width=\textwidth]{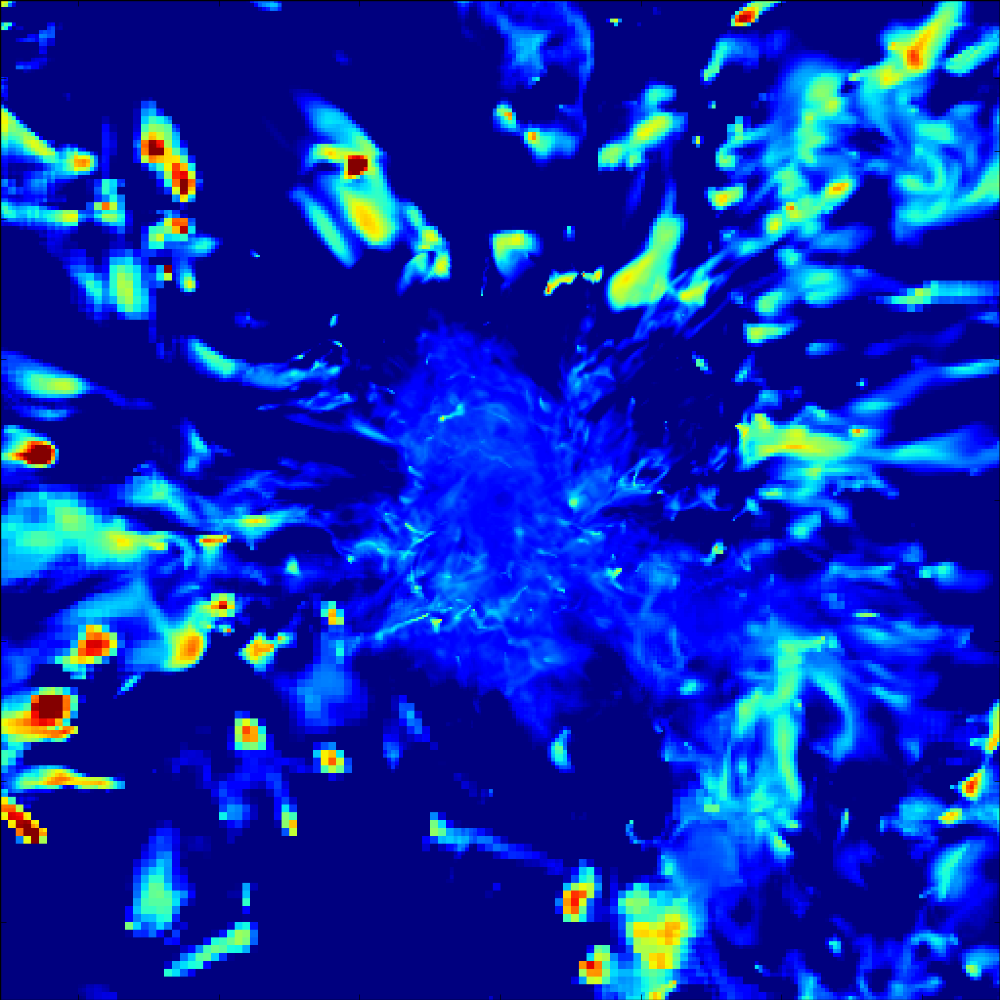}
\label{Fig:WC05} \caption{C05}
\end{subfigure}
\begin{subfigure}[b]{0.25\textwidth}
\includegraphics[width=\textwidth]{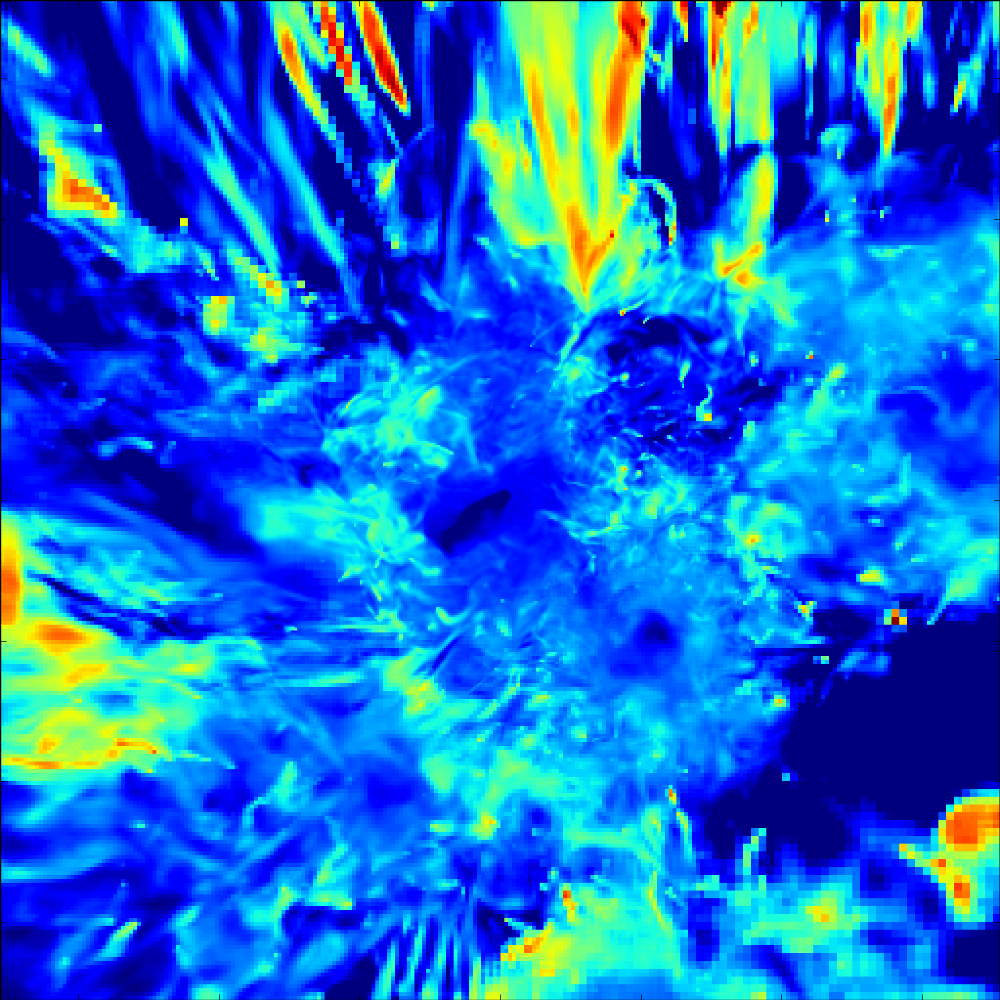}
\label{Fig:WC06} \caption{C06}
\end{subfigure}

\begin{subfigure}[b]{0.25\textwidth}
\includegraphics[width=\textwidth]{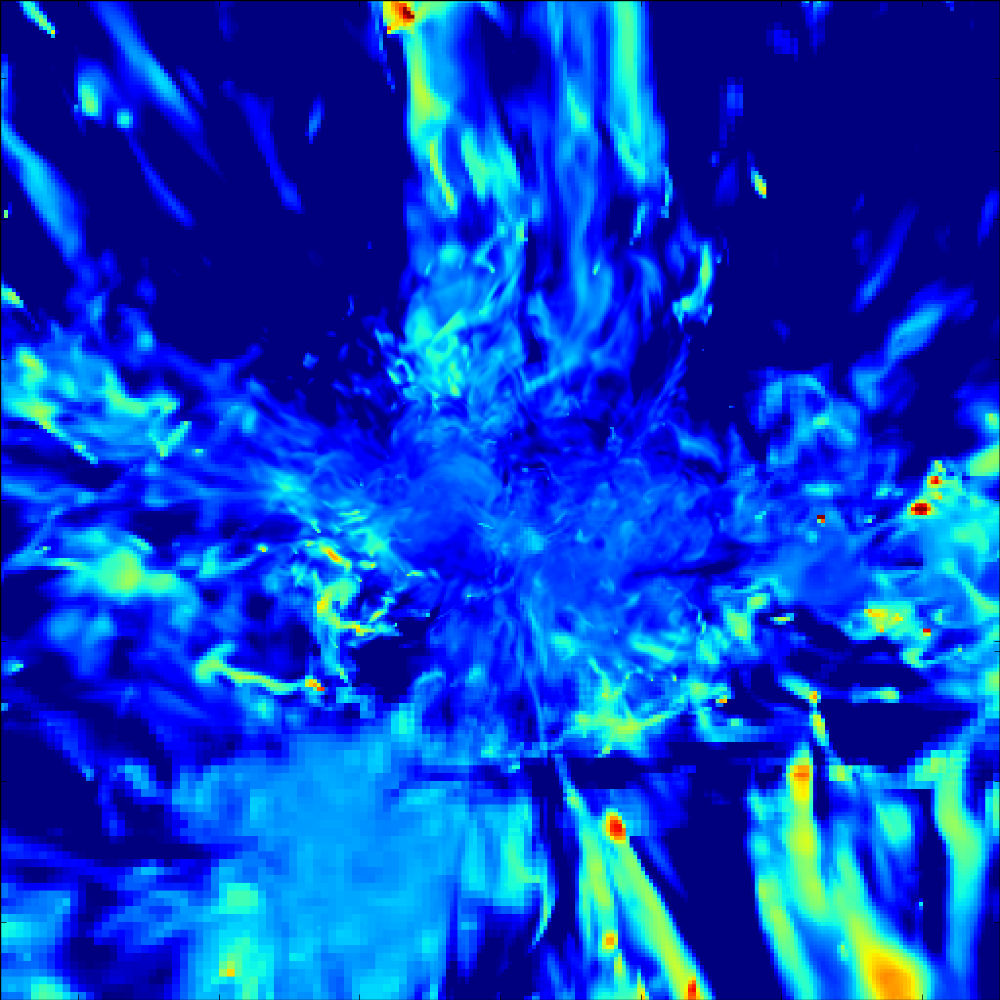}
\label{Fig:WC07} \caption{C07}
\end{subfigure}
\begin{subfigure}[b]{0.25\textwidth}
\includegraphics[width=\textwidth]{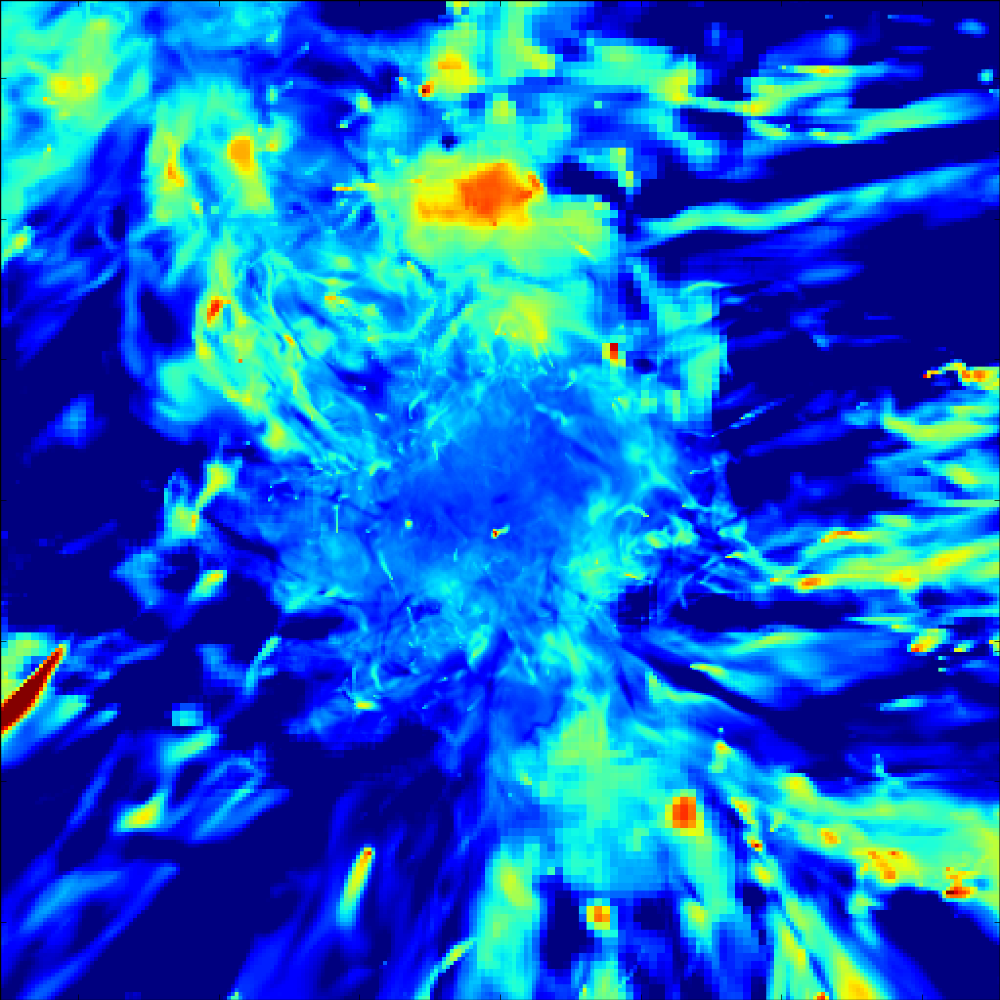}
\label{Fig:WC08} \caption{C08}
\end{subfigure}
\begin{subfigure}[b]{0.25\textwidth}
\includegraphics[width=\textwidth]{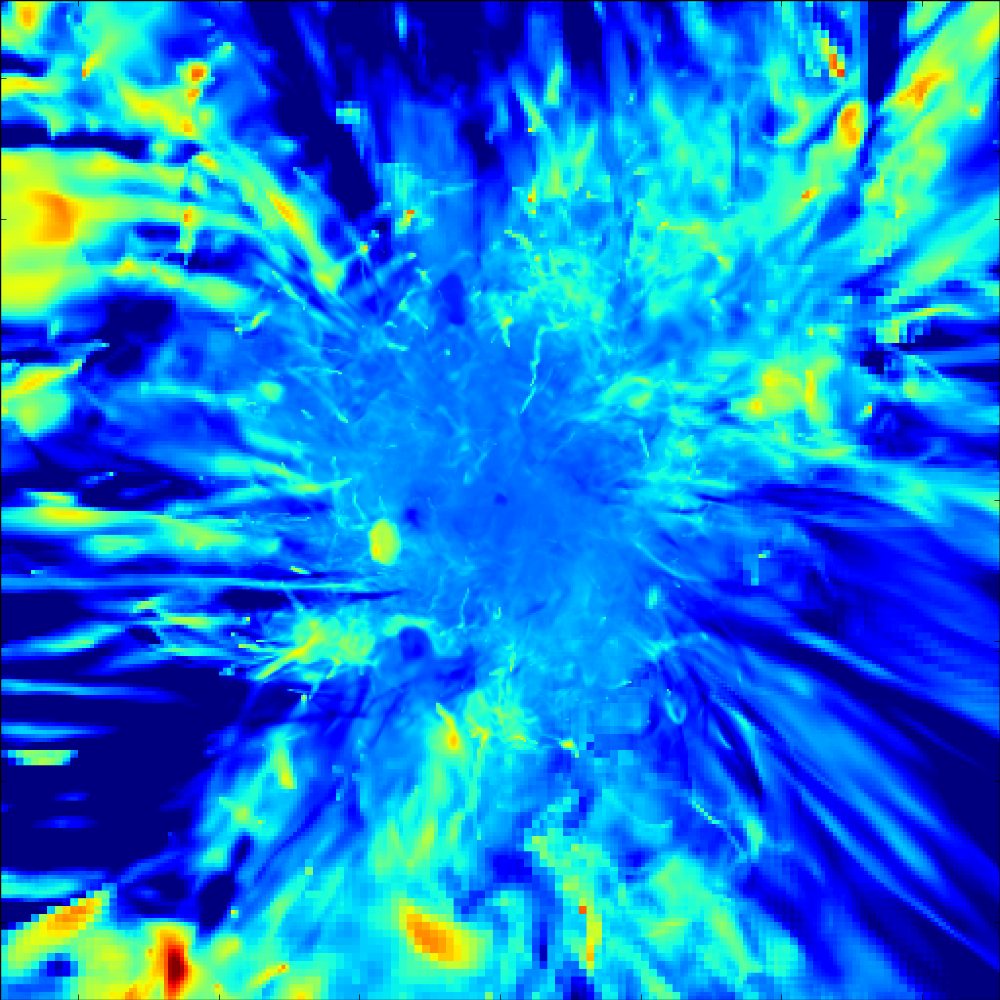}
\label{Fig:WC09} \caption{C09}
\end{subfigure}

\begin{subfigure}[b]{0.25\textwidth}
\includegraphics[width=\textwidth]{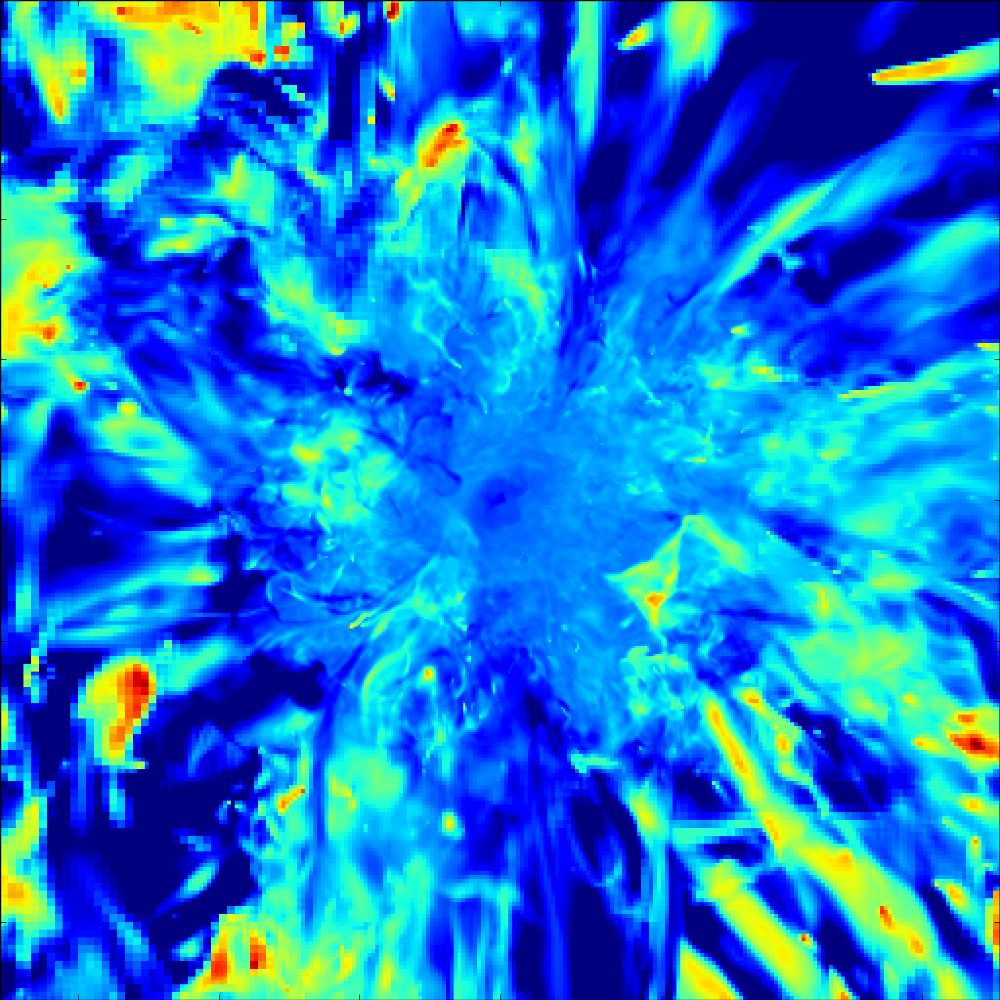}
\label{Fig:WC10} \caption{C10}
\end{subfigure}
\begin{subfigure}[b]{0.25\textwidth}
\includegraphics[width=\textwidth]{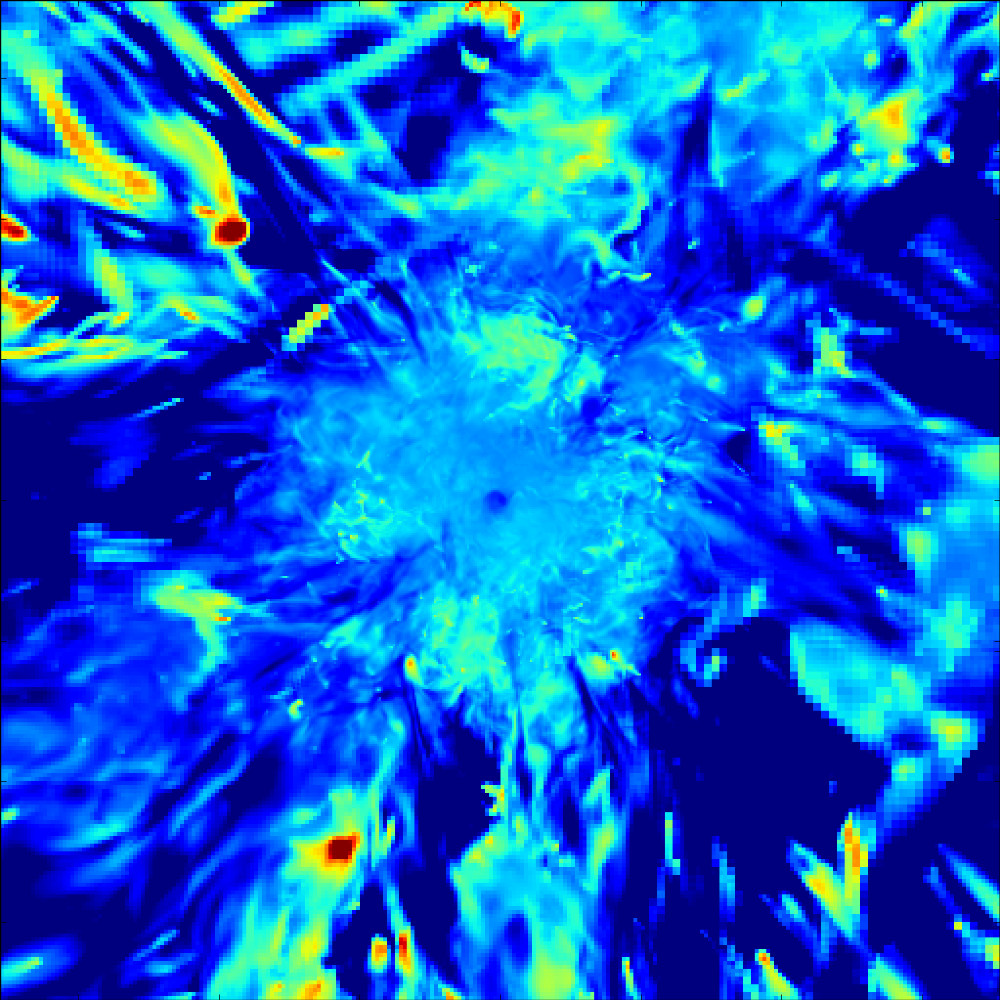}
\label{Fig:WC11} \caption{C11}
\end{subfigure}
\begin{subfigure}[b]{0.25\textwidth}
\hspace{0.1\textwidth}
\includegraphics[width=1.35\textwidth]{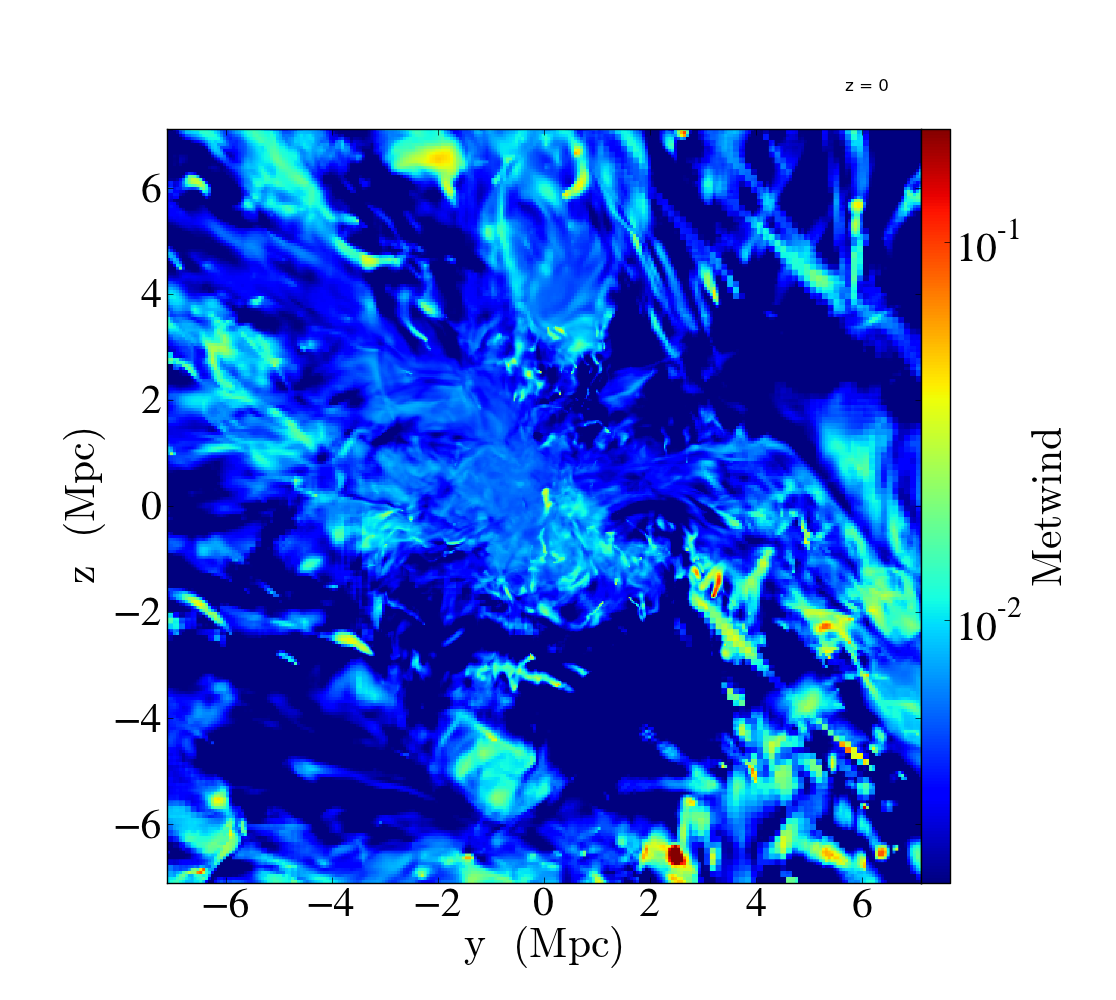}
\vspace{-0.07\textwidth}
\label{Fig:WC12} \caption{C12}
\end{subfigure}
\caption{X-ray emissivity-weighted galactic wind metal mass-fraction $x$-projection of all 12 clusters. All plots show the inner 5
Mpc/h and use
the same color scheme, see Fig. (l). } \label{Fig:Windall}
\end{figure*}

\begin{figure*}[h!]
\centering
\begin{subfigure}[b]{0.25\textwidth}
\includegraphics[width=\textwidth]{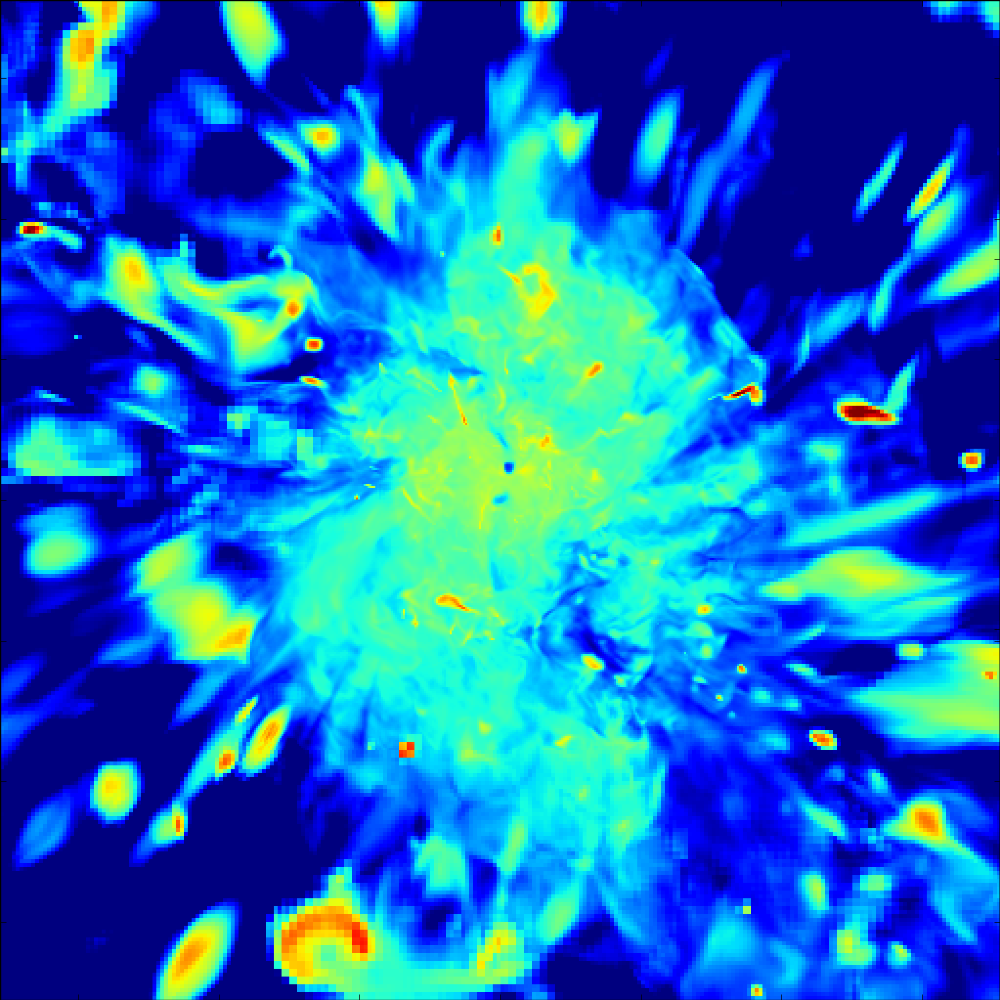}
\label{FigRMC01} \caption{C01}
\end{subfigure}
\begin{subfigure}[b]{0.25\textwidth}
\includegraphics[width=\textwidth]{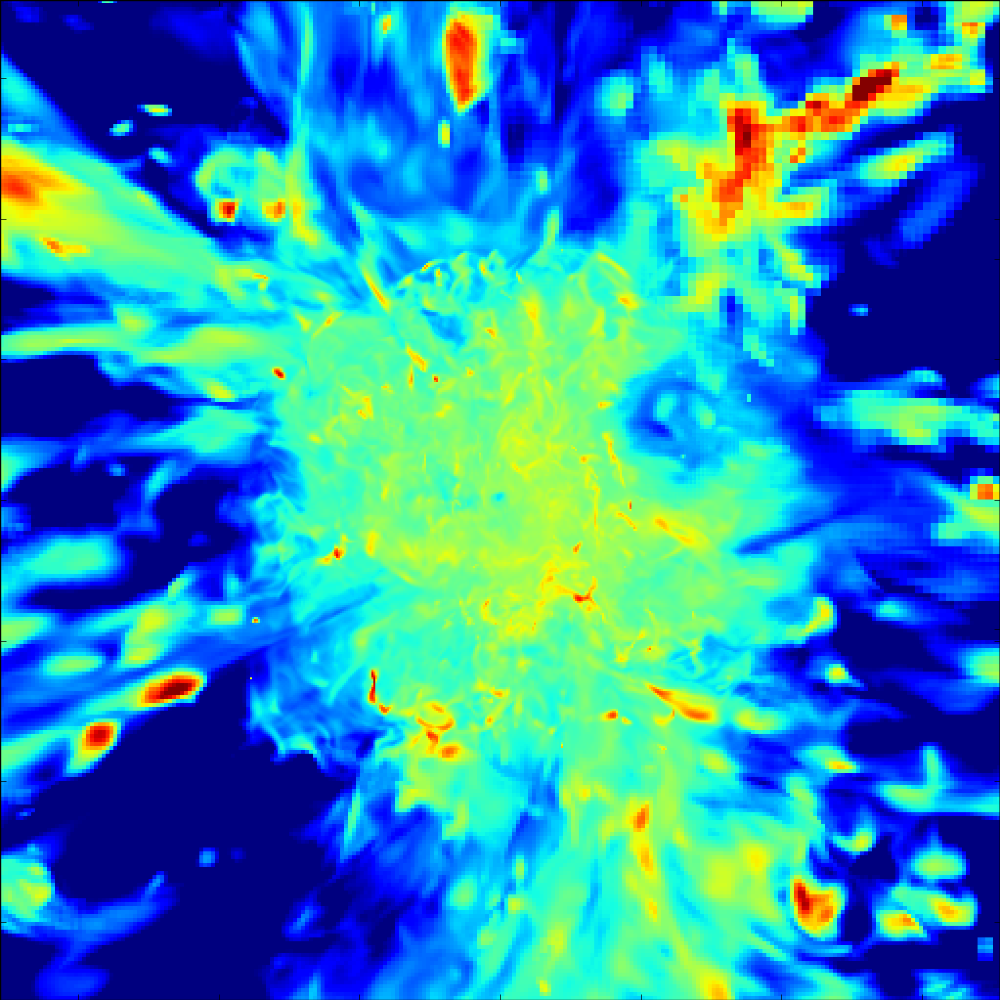}
\label{Fig:RC02} \caption{C02}
\end{subfigure}
\begin{subfigure}[b]{0.25\textwidth}
\includegraphics[width=\textwidth]{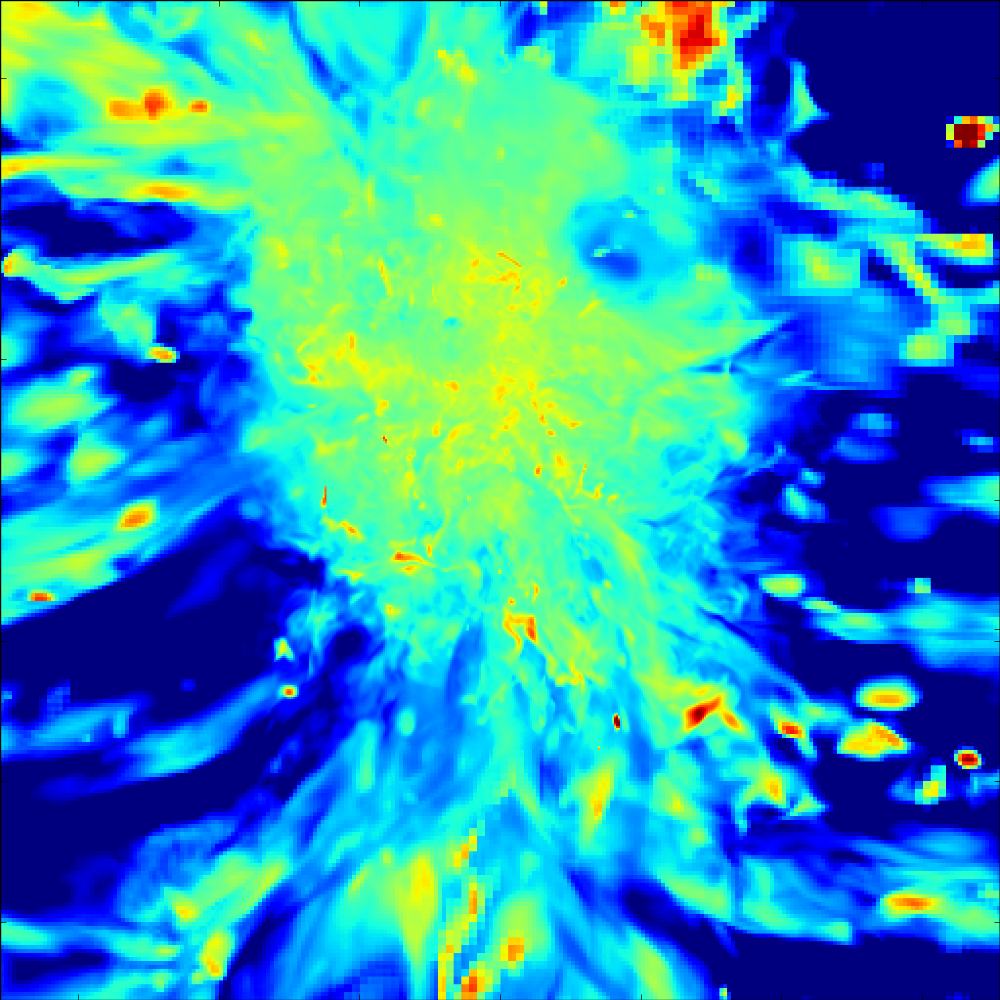}
\label{Fig:RC03} \caption{C03}
\end{subfigure}

\begin{subfigure}[b]{0.25\textwidth}
\includegraphics[width=\textwidth]{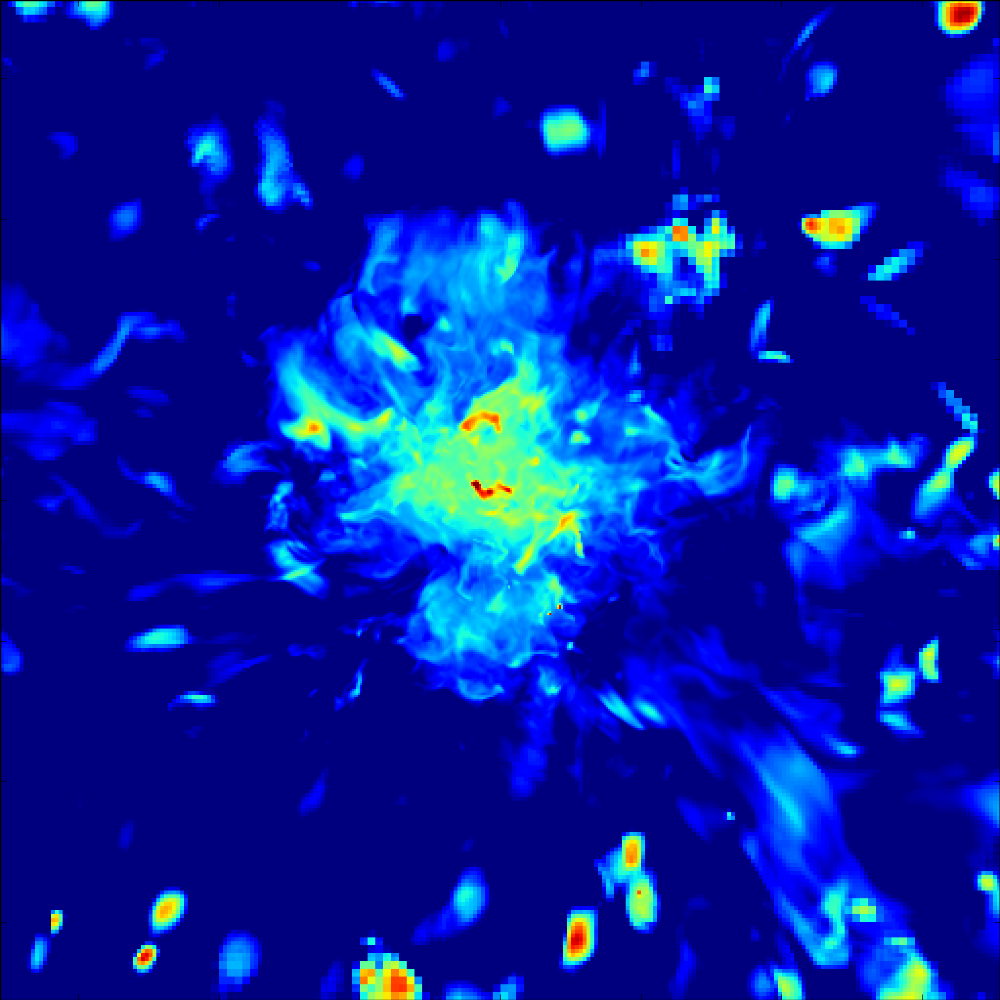}
\label{Fig:RC04} \caption{C04}
\end{subfigure}
\begin{subfigure}[b]{0.25\textwidth}
\includegraphics[width=\textwidth]{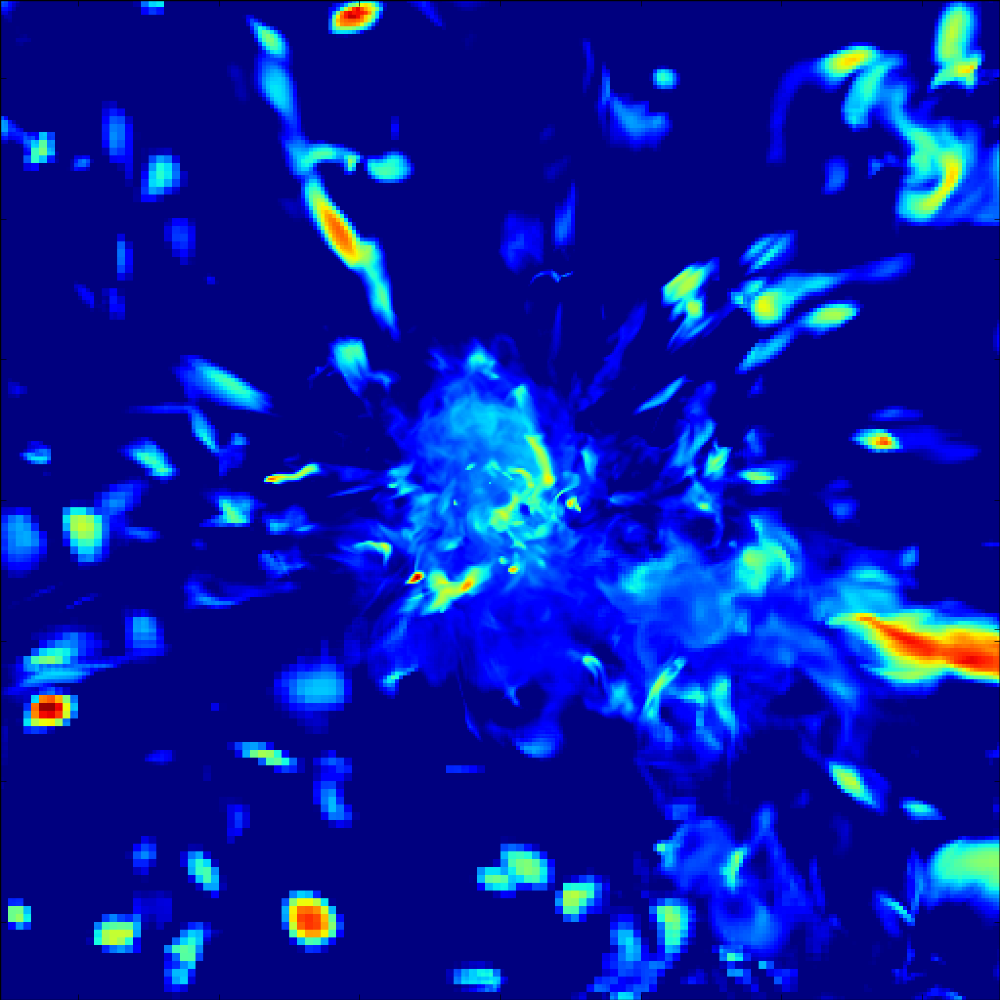}
\label{Fig:RC05} \caption{C05}
\end{subfigure}
\begin{subfigure}[b]{0.25\textwidth}
\includegraphics[width=\textwidth]{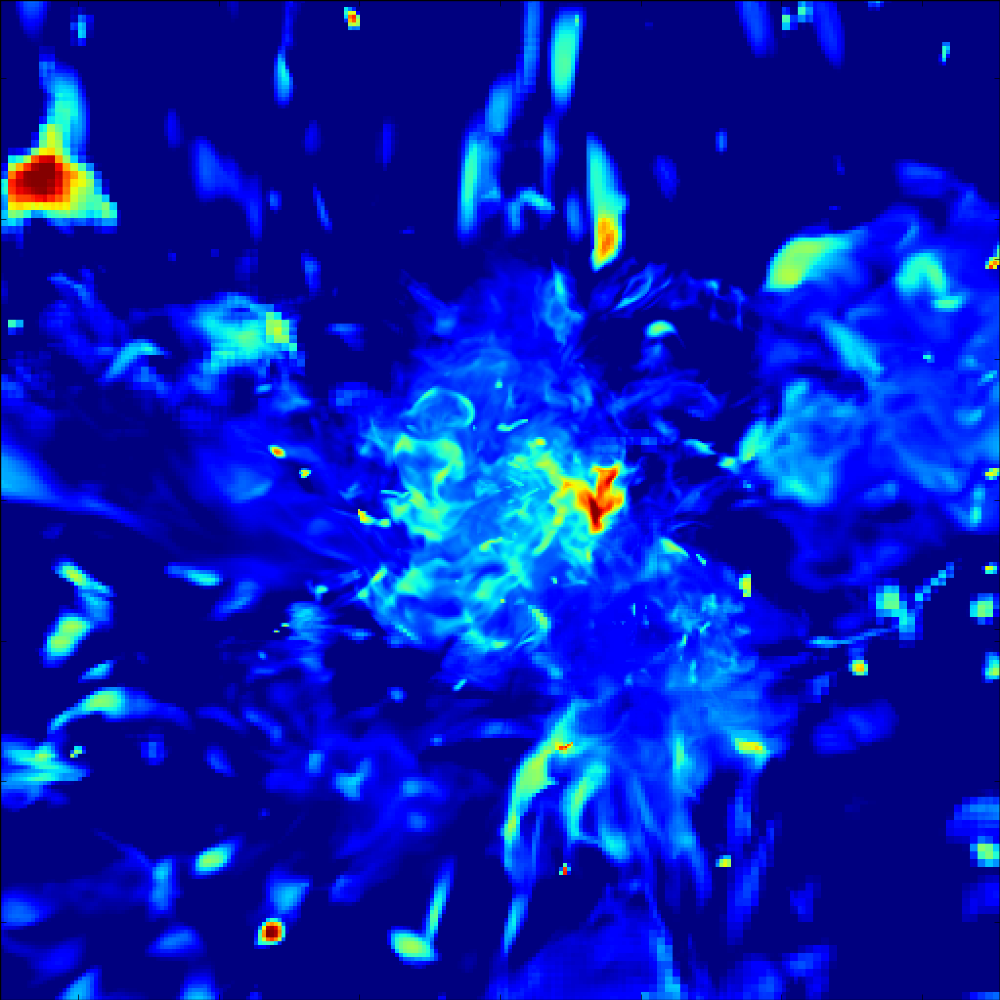}
\label{Fig:RC06} \caption{C06}
\end{subfigure}

\begin{subfigure}[b]{0.25\textwidth}
\includegraphics[width=\textwidth]{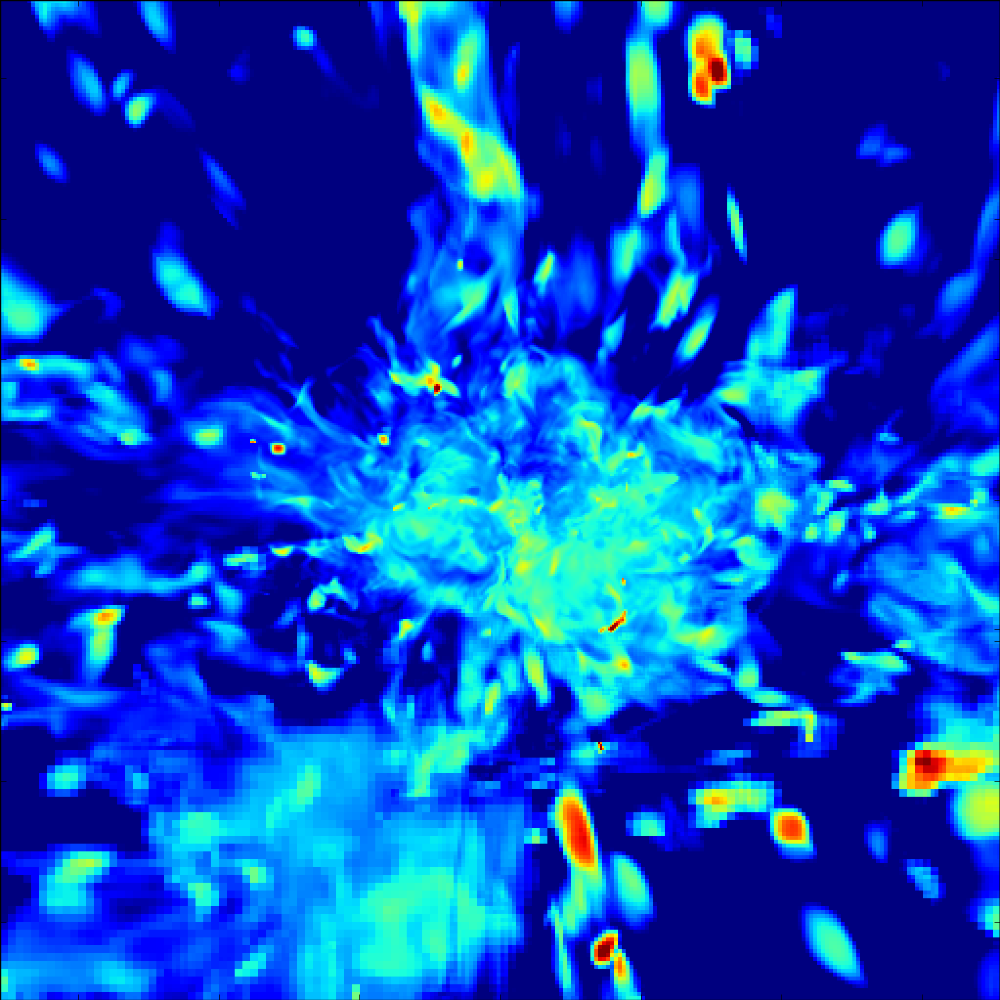}
\label{Fig:RC07} \caption{C07}
\end{subfigure}
\begin{subfigure}[b]{0.25\textwidth}
\includegraphics[width=\textwidth]{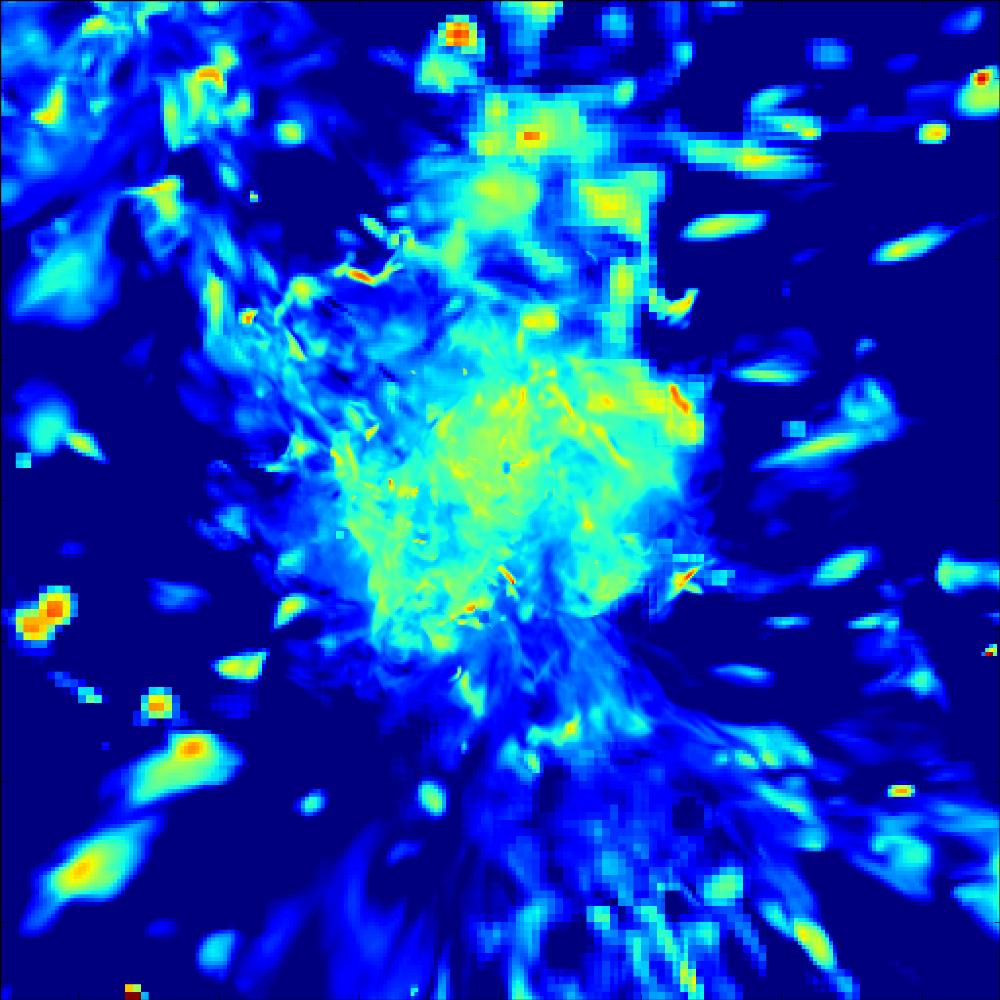}
\label{Fig:RC08} \caption{C08}
\end{subfigure}
\begin{subfigure}[b]{0.25\textwidth}
\includegraphics[width=\textwidth]{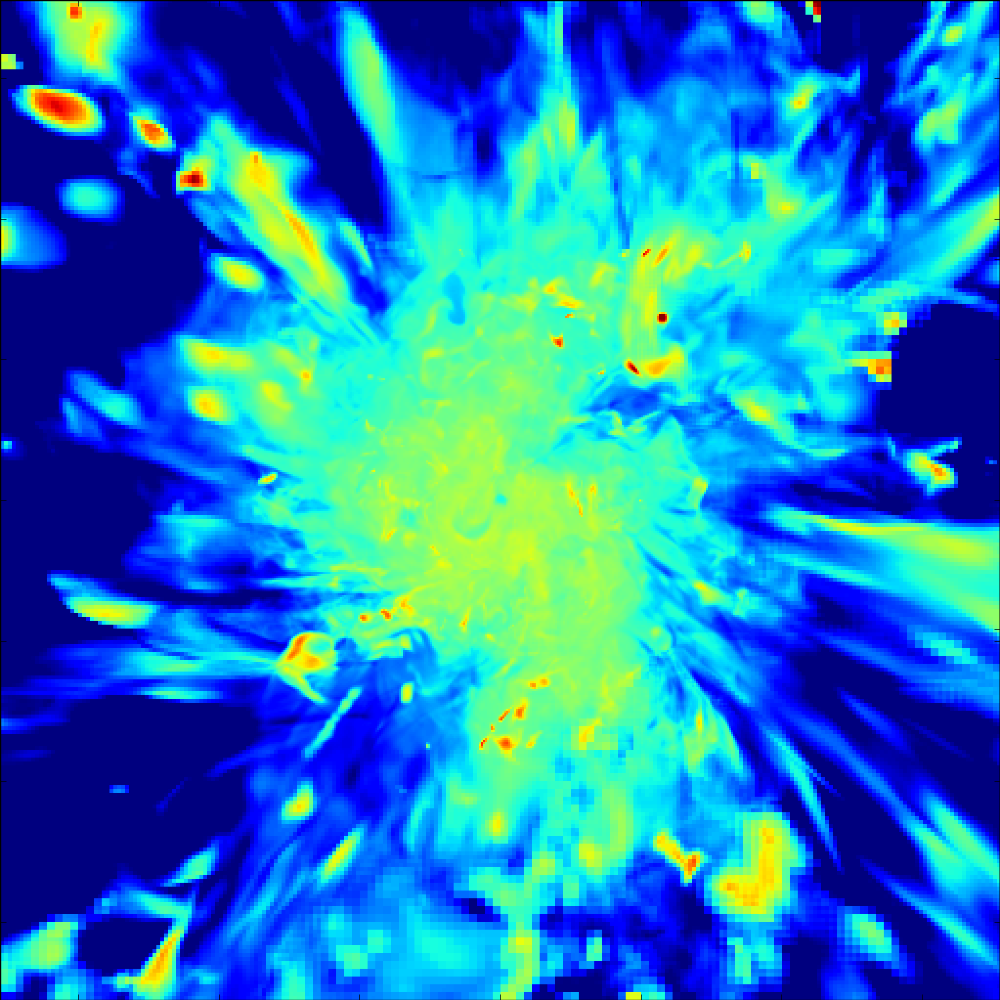}
\label{Fig:RC09} \caption{C09}
\end{subfigure}

\begin{subfigure}[b]{0.25\textwidth}
\includegraphics[width=\textwidth]{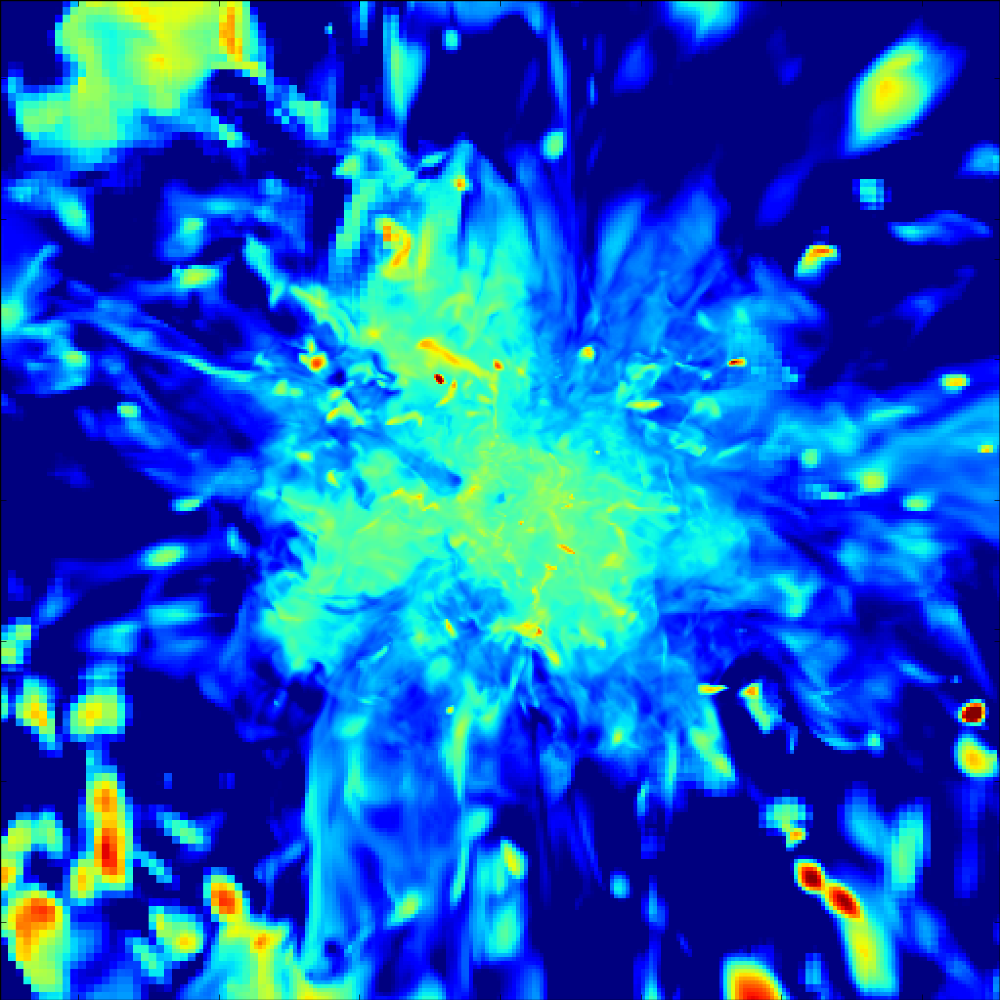}
\label{Fig:RC10} \caption{C10}
\end{subfigure}
\begin{subfigure}[b]{0.25\textwidth}
\includegraphics[width=\textwidth]{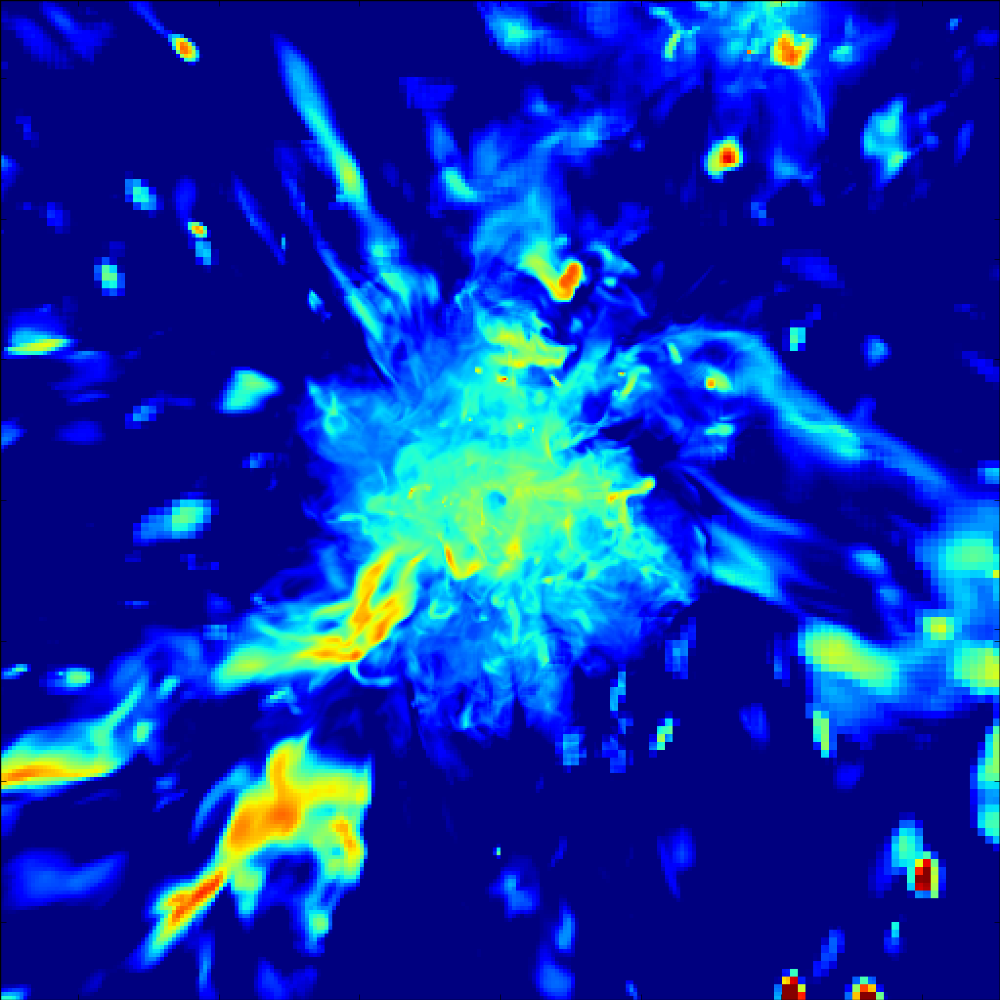}
\label{Fig:RC11} \caption{C11}
\end{subfigure}
\begin{subfigure}[b]{0.25\textwidth}
\hspace{0.1\textwidth}
\includegraphics[width=1.35\textwidth]{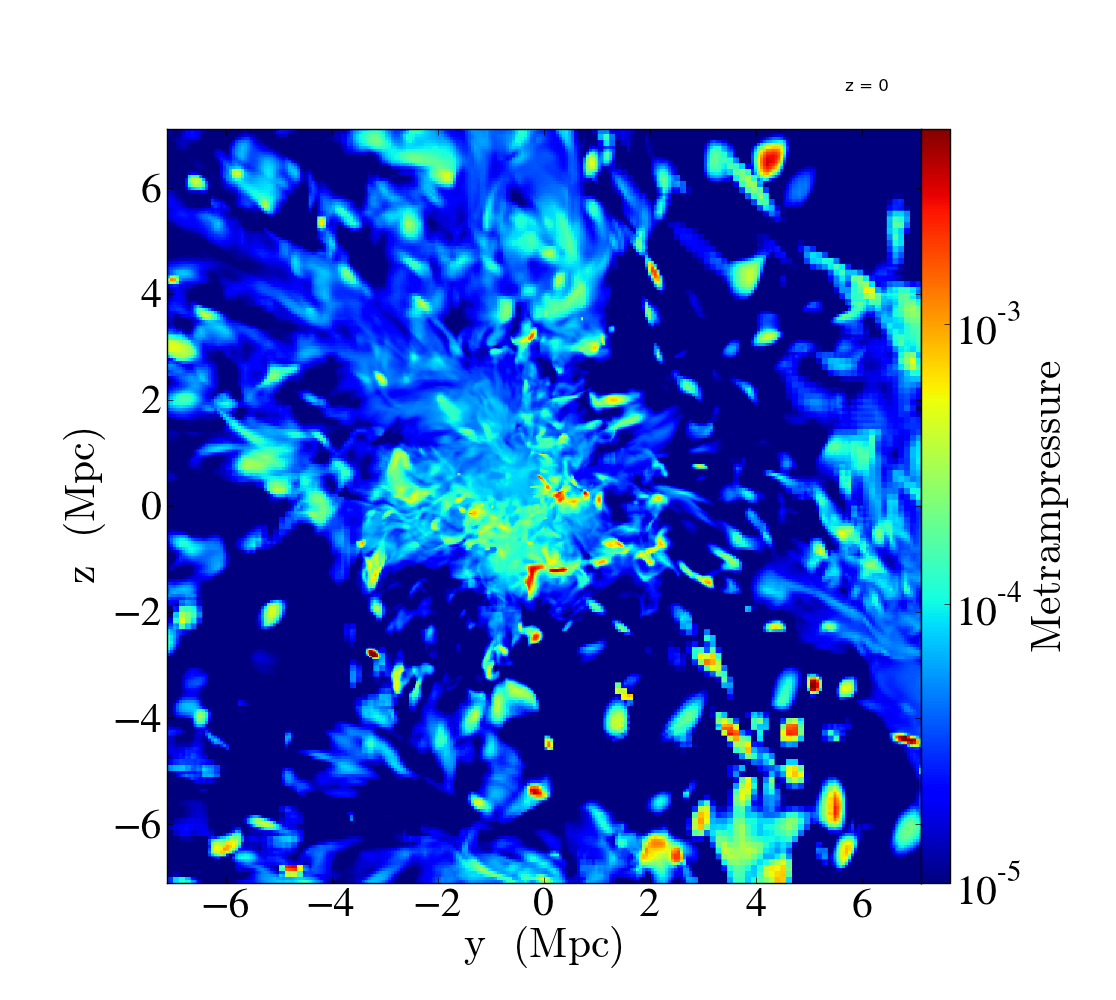}
\vspace{-0.07\textwidth}
\label{Fig:RC12} \caption{C12}
\end{subfigure}
\caption{X-ray emissivity-weighted ram-pressure-stripped metal mass-fraction $x$-projection of all 12 clusters. All plots show the inner 5 Mpc/h and use
the same color scheme, see Fig. (l). } \label{Fig:RPSall}
\end{figure*}

\begin{figure*}[h!]
\centering
\begin{subfigure}[b]{0.25\textwidth}
\includegraphics[width=\textwidth]{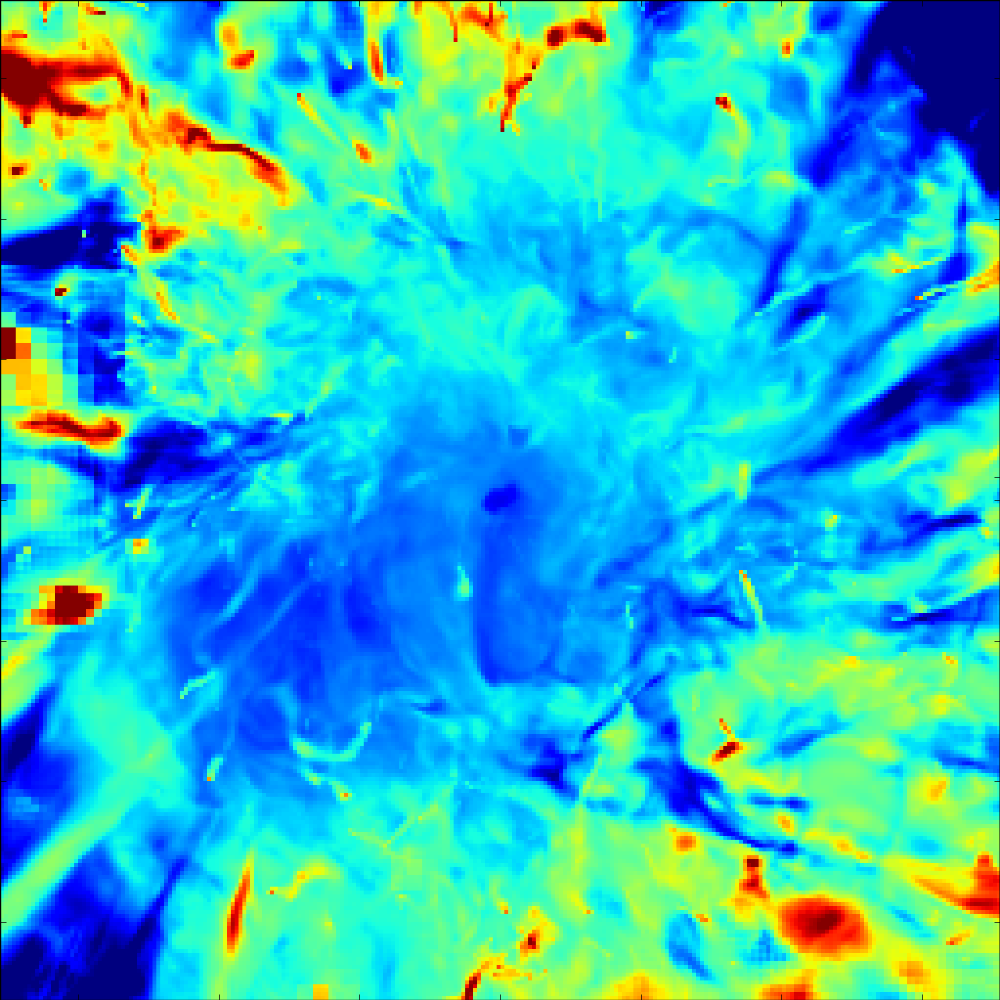}
\label{Fig:CC01} \caption{C01}
\end{subfigure}
\begin{subfigure}[b]{0.25\textwidth}
\includegraphics[width=\textwidth]{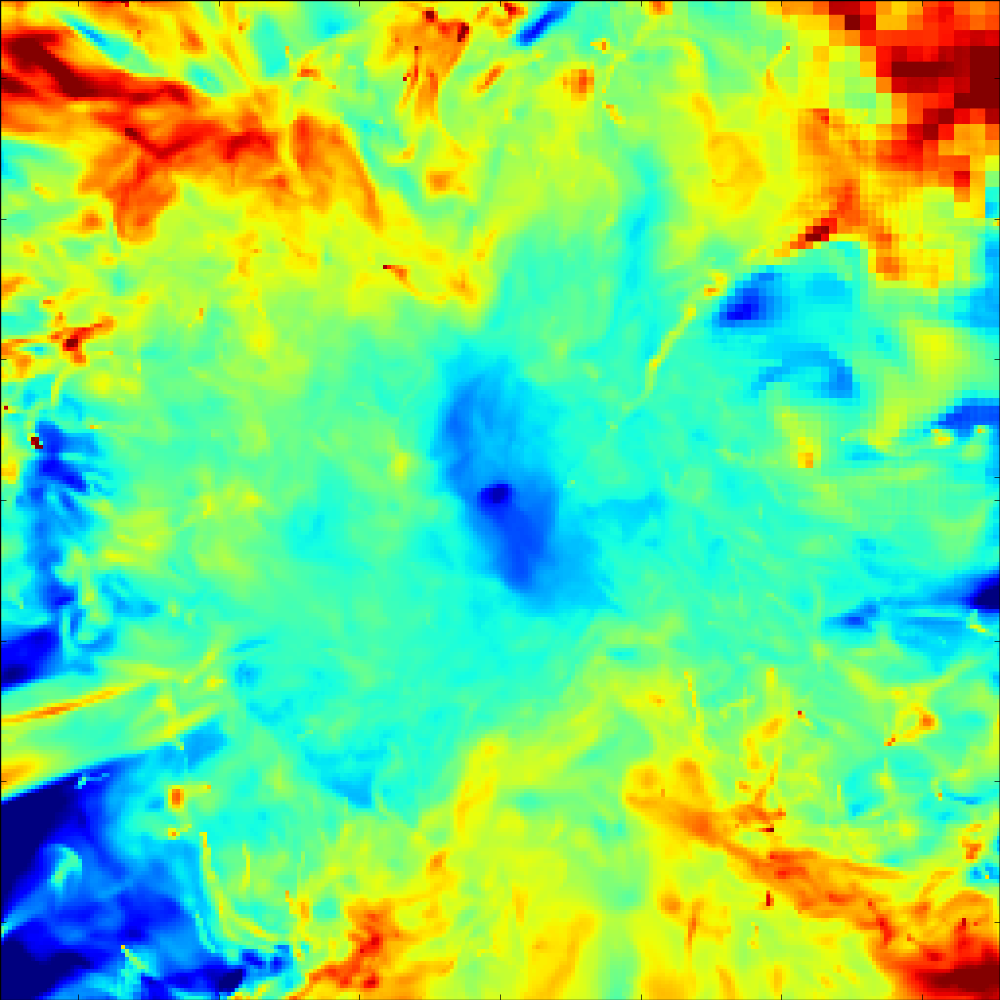}
\label{Fig:CC02} \caption{C02}
\end{subfigure}
\begin{subfigure}[b]{0.25\textwidth}
\includegraphics[width=\textwidth]{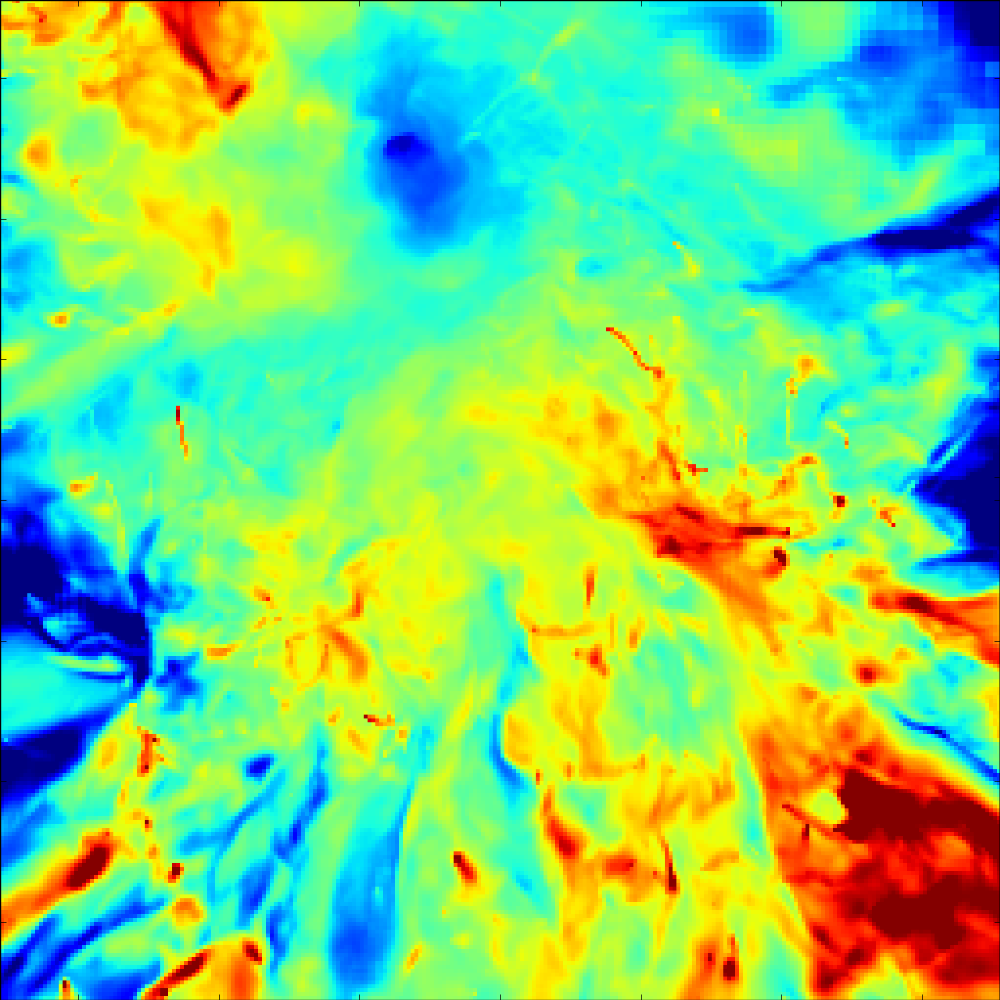}
\label{Fig:CC03} \caption{C03}
\end{subfigure}

\begin{subfigure}[b]{0.25\textwidth}
\includegraphics[width=\textwidth]{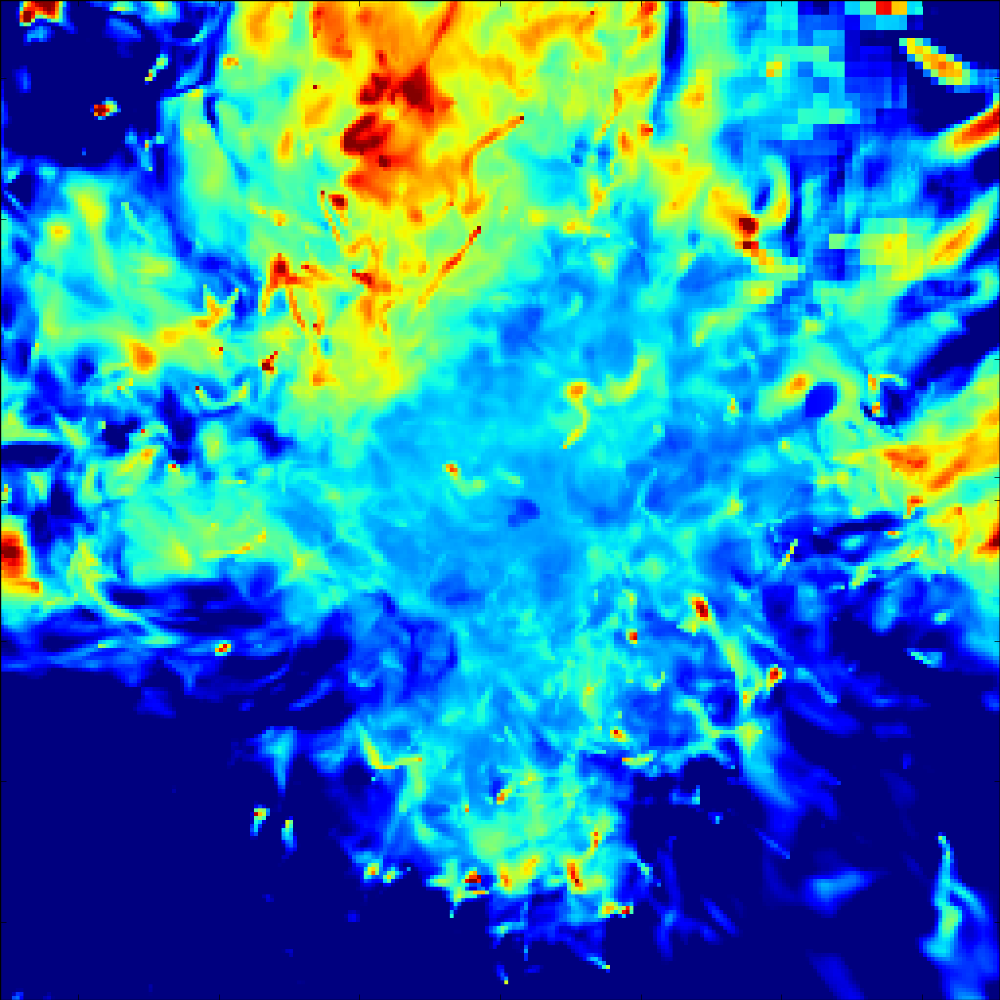}
\label{Fig:CC04} \caption{C04}
\end{subfigure}
\begin{subfigure}[b]{0.25\textwidth}
\includegraphics[width=\textwidth]{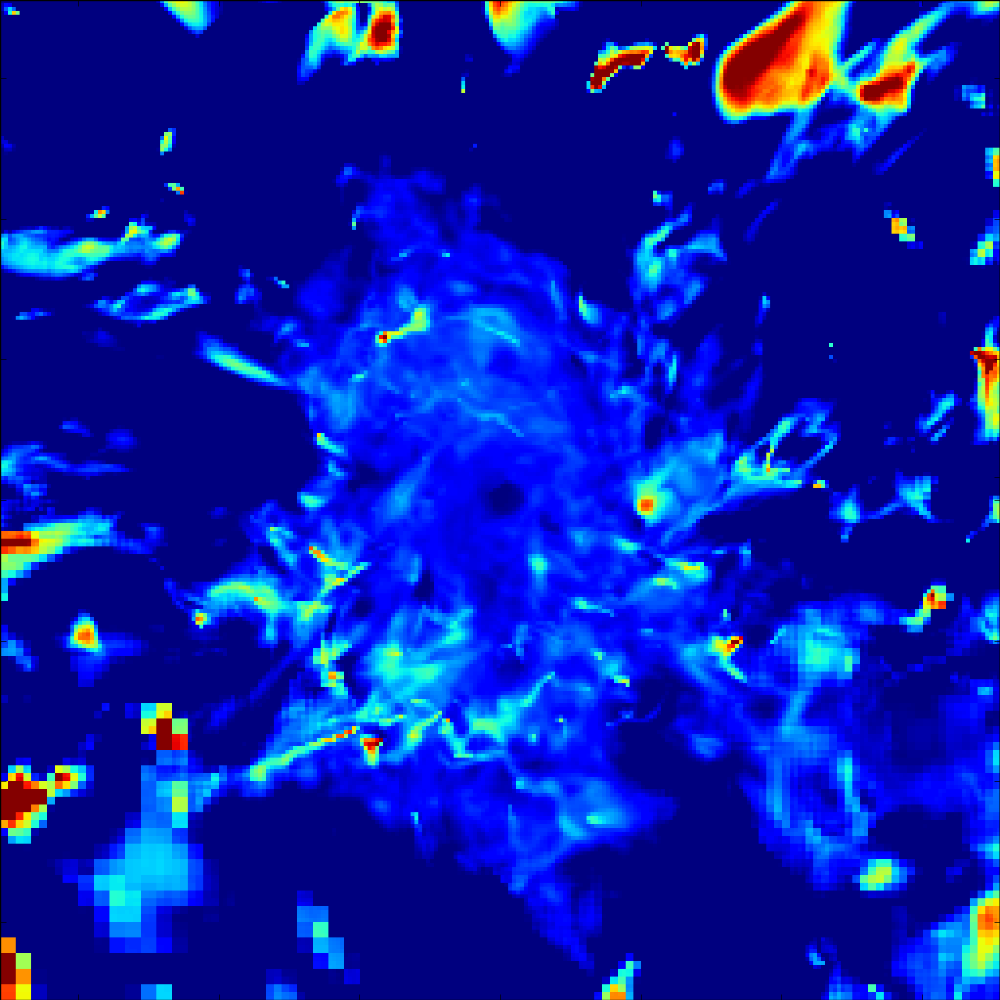}
\label{Fig:CC05} \caption{C05}
\end{subfigure}
\begin{subfigure}[b]{0.25\textwidth}
\includegraphics[width=\textwidth]{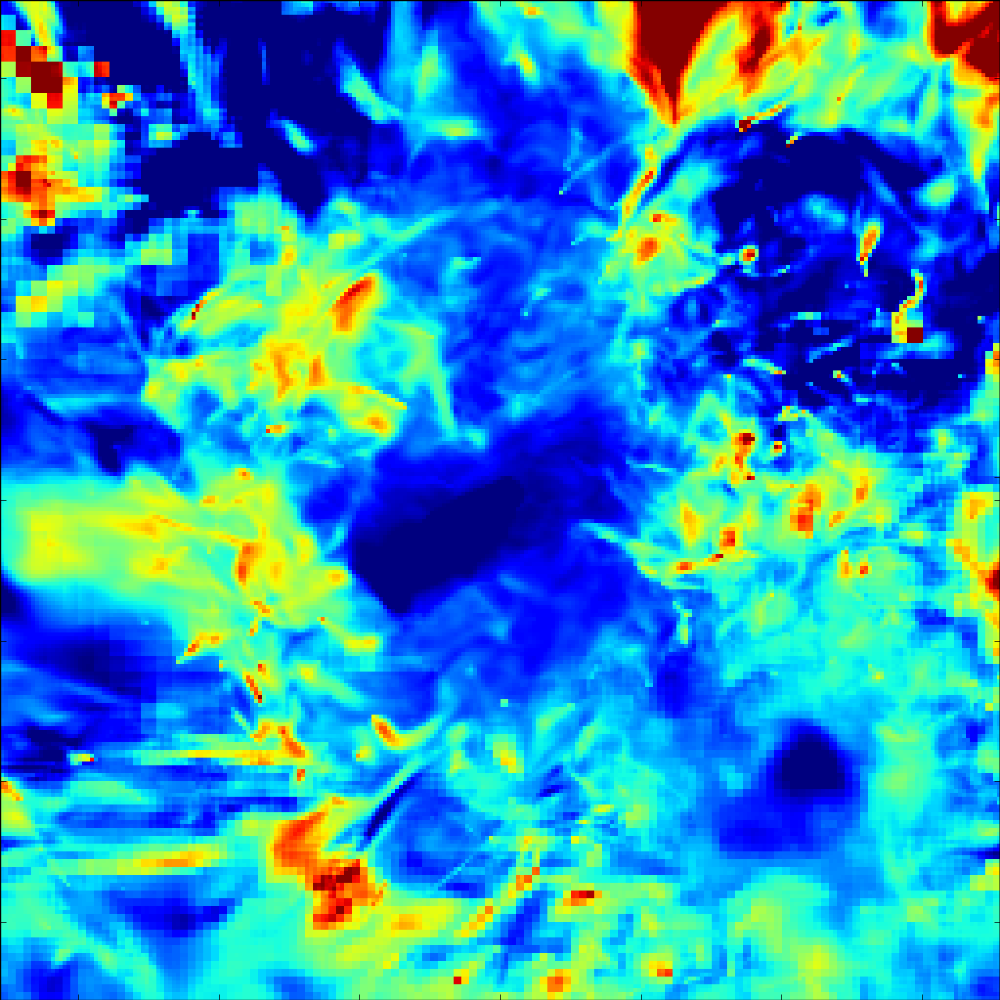}
\label{Fig:CC06} \caption{C06}
\end{subfigure}

\begin{subfigure}[b]{0.25\textwidth}
\includegraphics[width=\textwidth]{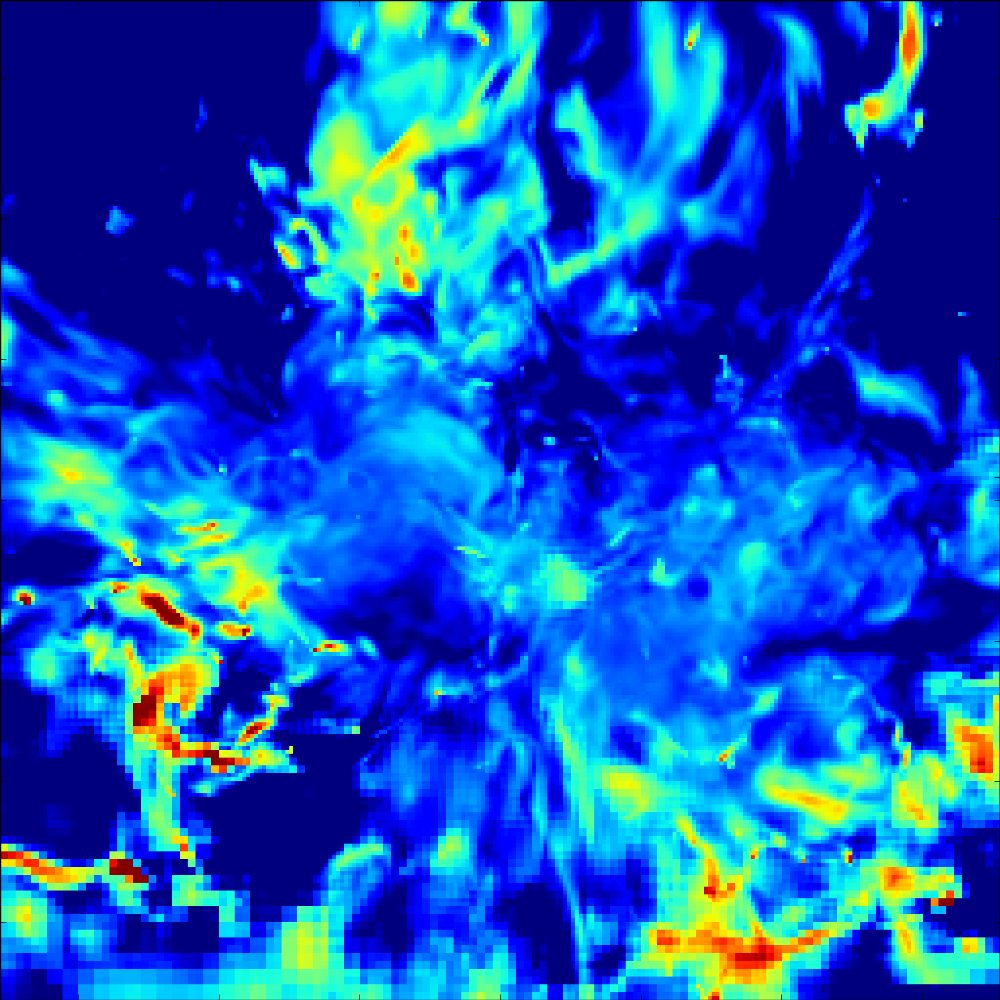}
\label{Fig:CC07} \caption{C07}
\end{subfigure}
\begin{subfigure}[b]{0.25\textwidth}
\includegraphics[width=\textwidth]{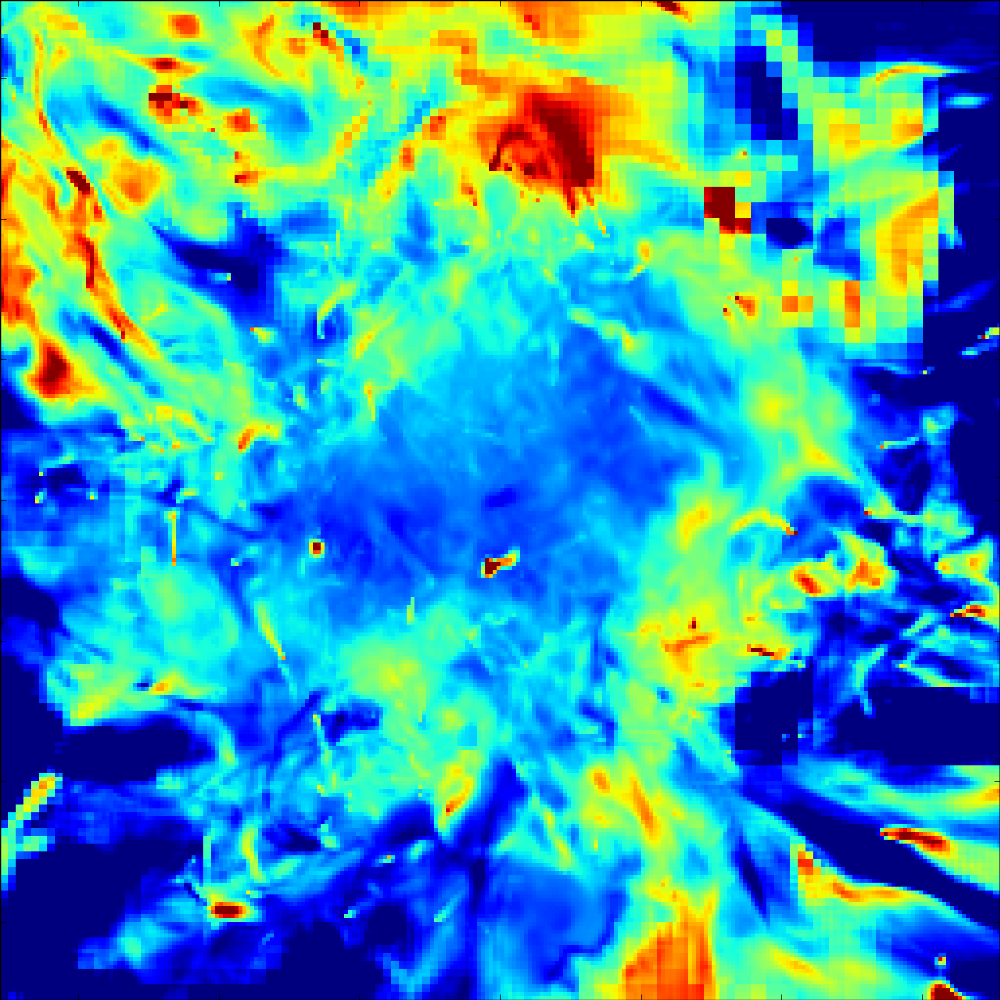}
\label{Fig:CC08} \caption{C08}
\end{subfigure}
\begin{subfigure}[b]{0.25\textwidth}
\includegraphics[width=\textwidth]{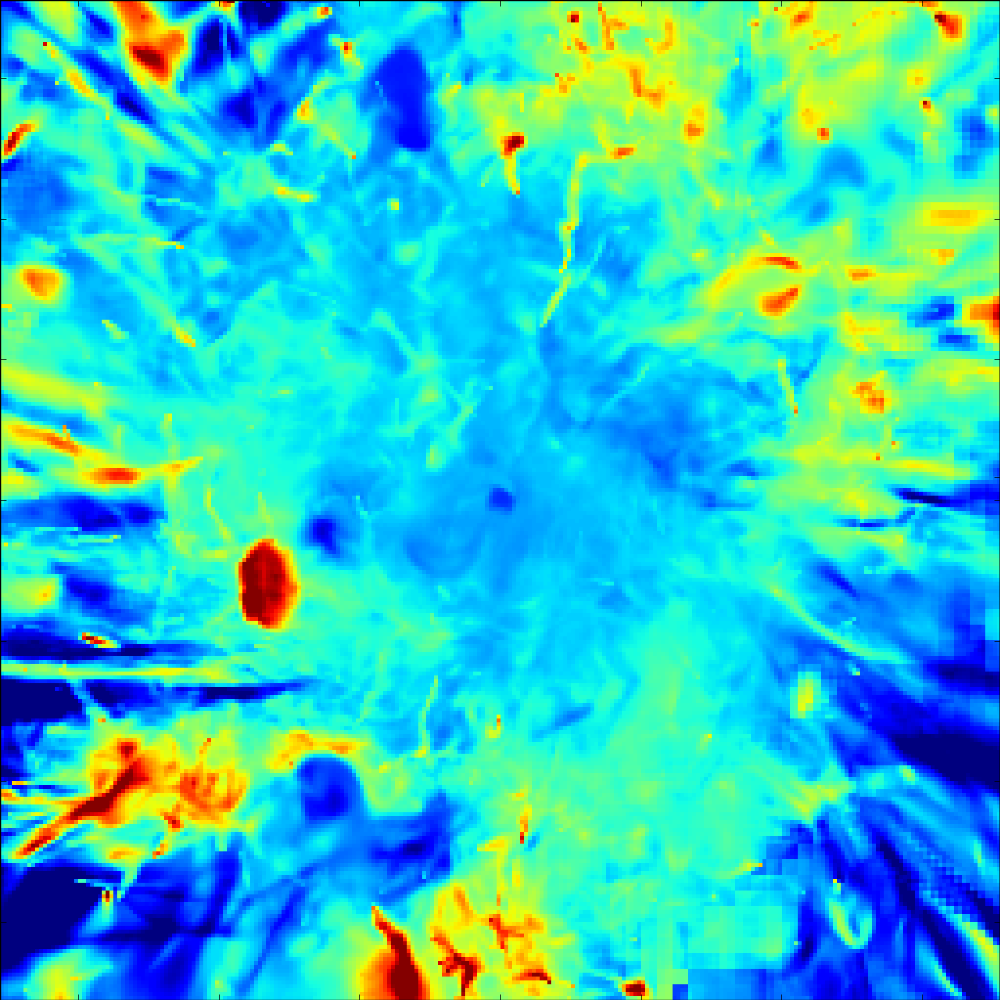}
\label{Fig:CC09} \caption{C09}
\end{subfigure}

\begin{subfigure}[b]{0.25\textwidth}
\includegraphics[width=\textwidth]{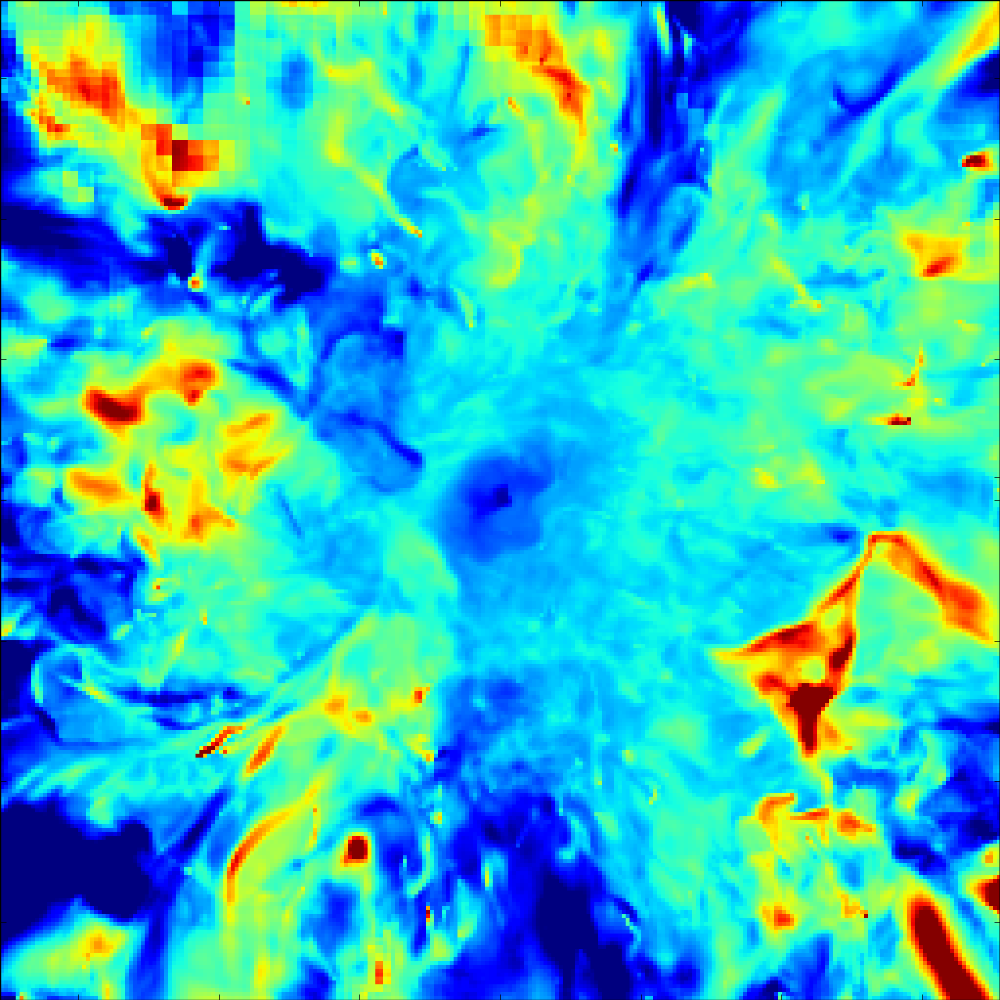}
\label{Fig:CC10} \caption{C10}
\end{subfigure}
\begin{subfigure}[b]{0.25\textwidth}
\includegraphics[width=\textwidth]{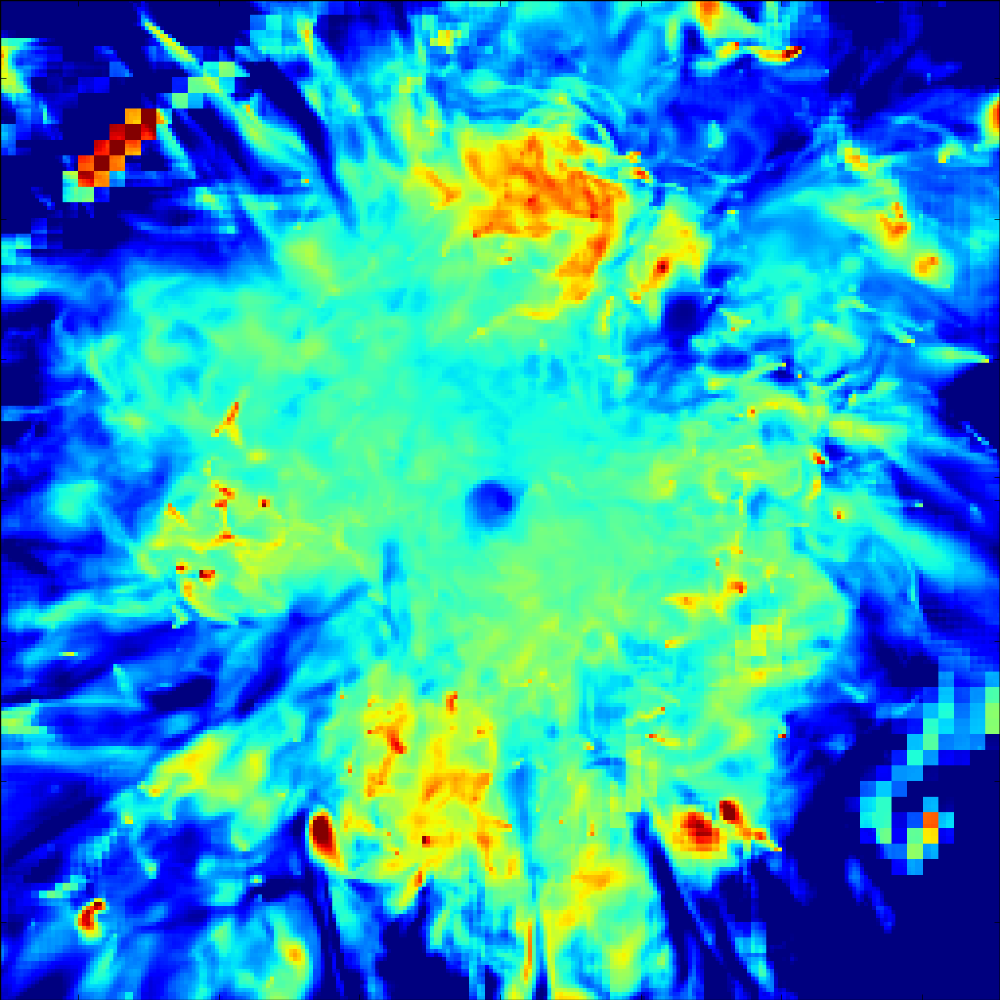}
\label{Fig:CC11} \caption{C11}
\end{subfigure}
\begin{subfigure}[b]{0.25\textwidth}
\hspace{0.1\textwidth}
\includegraphics[width=1.35\textwidth]{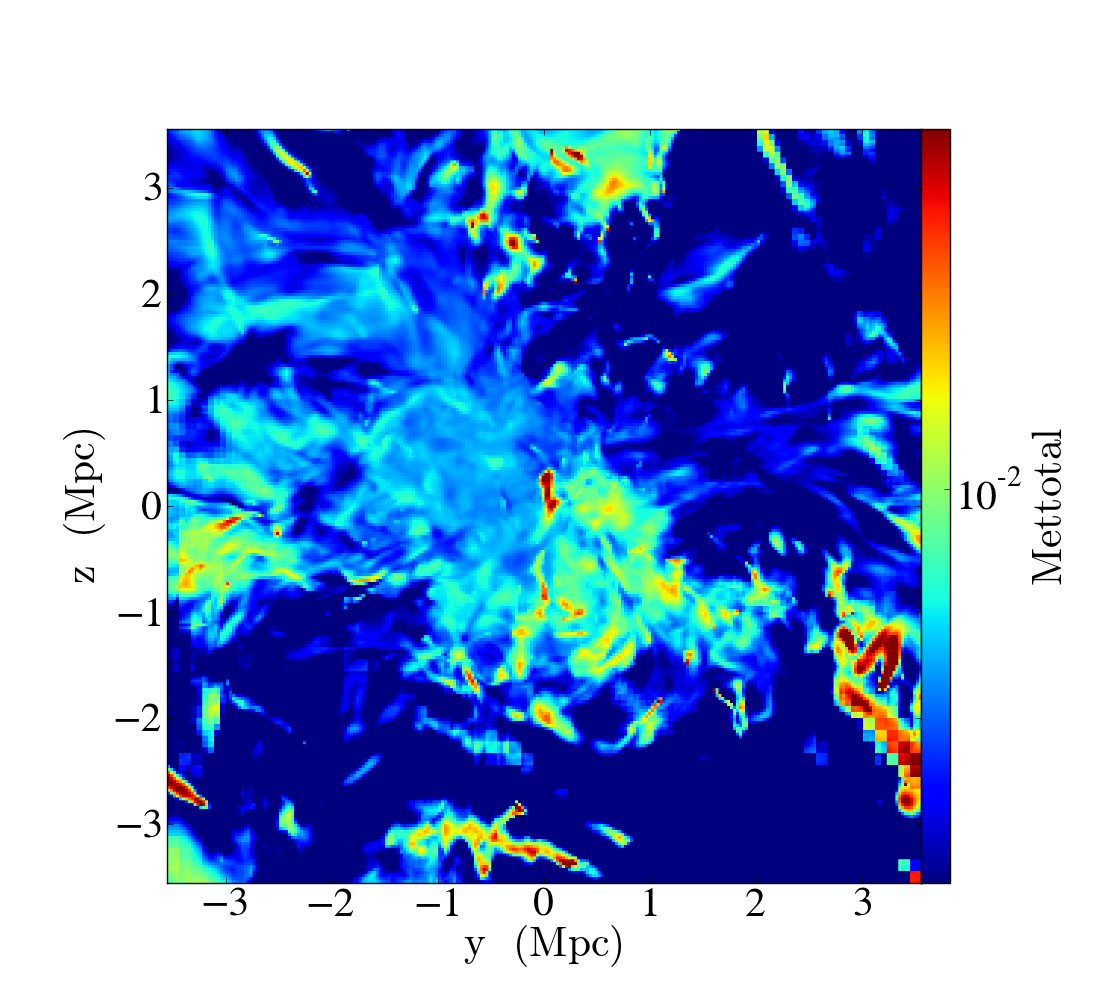}
\vspace{-0.07\textwidth}
\label{Fig:CC12} \caption{C12}
\end{subfigure}
\caption{X-ray emissivity-weighted total metal mass-fraction $x$-projection of all 12 clusters. All plots show the inner 2.5
Mpc/h and use
the same color scheme, see Fig. (l). The color scheme represents a range from $0.2-2\text{Z}_{\odot.}$} \label{Fig:Coreall}
\end{figure*}
\end{appendix}

\end{document}